\documentclass{article}

\usepackage{graphicx}

\usepackage{amsmath,amsfonts,amssymb,latexsym,epsfig}
\usepackage{mathrsfs}
\usepackage{verbatim}
\usepackage{latexsym}
\usepackage{amsthm}
\usepackage{amssymb}
\usepackage{subfig}
\usepackage{graphics}
\usepackage{amsbsy}
\usepackage{fullpage}
\usepackage{enumerate}
\usepackage{times}
\usepackage{multirow}
\usepackage{soul}
\usepackage[normalem]{ulem}

\newtheorem{theorem}{Theorem}[section]

\newtheorem{proposition}[theorem]{Proposition}

\numberwithin{equation}{section}
\usepackage{amscd}
\usepackage{tikz}
\usepackage{xcolor}
\usepackage[english]{babel}
\usepackage[latin1]{inputenc}
\usepackage{times}
\usepackage[T1]{fontenc}
\usepackage{graphicx}
\usepackage{amsmath,amssymb}
\usepackage{slidesec}
\usepackage{verbatim}
\usepackage{url}
\usepackage{epsf}
\usepackage{amsmath}
\usepackage{graphics}
\usepackage{amsfonts}
\usepackage{amsbsy}
\usepackage{lscape}
\usepackage{enumerate}
\usepackage{amsthm}
\usepackage{amssymb}
\usepackage{exscale}
\usepackage{algorithm}
\usepackage{algorithmic}

\begin{document}
\title{Accelerating proximal Markov chain Monte Carlo by using an explicit stabilised method}
\author{Luis Vargas$^{1,2,3}$ \and Marcelo Pereyra$^{2,3}$\and Konstantinos C. Zygalakis$^{1,3}$}

\maketitle

\begin{abstract}
We present a highly efficient proximal Markov chain Monte Carlo methodology to perform Bayesian computation in imaging problems. Similarly to previous proximal Monte Carlo approaches, the proposed method is derived from an approximation of the {Langevin diffusion}. However, instead of the conventional Euler-Maruyama approximation that underpins existing proximal Monte Carlo methods, here we use a state-of-the-art orthogonal Runge-Kutta-Chebyshev stochastic approximation {\cite{abdulleSKROCK}} that combines several gradient evaluations to significantly accelerate its convergence speed, similarly to accelerated gradient optimisation methods. 
The proposed methodology  is demonstrated via a range of numerical experiments, including non-blind image deconvolution, hyperspectral unmixing, and tomographic reconstruction, with total-variation and $\ell_1$-type priors. Comparisons with Euler-type proximal Monte Carlo methods confirm that the Markov chains generated with our method exhibit significantly faster convergence speeds, achieve larger effective sample sizes, and produce lower mean square estimation errors at equal computational budget.
\end{abstract}

\footnotetext[1]{School of Mathematics, University of Edinburgh, Edinburgh, Scotland}
\footnotetext[2]{School of Mathematical and Computer Sciences, Heriot-Watt University, Edinburgh, Scotland}
\footnotetext[3]{Maxwell Institute for Mathematical Sciences, Bayes Centre, 47 Potterrow, Edinburgh, Scotland}

\section{Introduction}
Imaging sciences study theory, methods, models, and algorithms to solve imaging problems, such as image denoising \cite{Houdard2018}, deblurring \cite{Beck2009,Afonso2010}, compressive sensing reconstruction \cite{Lucka2018}, super-resolution \cite{Romano2017}, tomographic reconstruction \cite{Afonso2010}, inpainting \cite{Ye2018}, source separation \cite{hyperspectral1}, and phase retrieval \cite{Elser2018}.

There are currently three main formal paradigms to formulate and solve imaging problems: the variational framework \cite{optimizationAlgImaging}, machine learning \cite{arridge_maass_oktem_schonlieb_2019}, and the Bayesian statistical framework \cite{credibleRegionBook}. In this paper we focus on the Bayesian framework, which is an intrinsically probabilistic paradigm where the data observation process and the prior knowledge available are represented by using statistical models, and where solutions are derived by using inference techniques stemming from Bayesian decision theory {\cite{credibleRegionBook}}. The Bayesian framework is  particularly well equipped to address imaging problems in which uncertainty plays an important role, such as medical imaging or remote sensing problems where it is necessary or desirable to quantify the uncertainty in the delivered solutions to inform decisions or conclusions, see, e.g., \cite{Pereyra2017,Repetti2019}. The framework is also well adapted to blind, semi-blind, and unsupervised problems involving partially unknown models (e.g., unspecified regularisation parameters or observation operators) \cite{Wipf15,Vidal2019}. Bayesian model selection technique also allows the objective comparison of several potential models to analyse the observed imaging data, even in cases where there is no ground truth available \cite{pereyraMYULA}.

In this paper we focus on the computational aspects of performing Bayesian inferences in imaging problems. Modern Bayesian computation methods suitable for imaging sciences can be broadly grouped in three categories: stochastic Markov chain Monte Carlo (MCMC) methods that are computationally expensive but robust, and which can be applied to a wide range of models and inferences; optimisation methods that are significantly more efficient by comparison, but which are only useful for point estimation and some other specific inferences; and deterministic approximation methods such as variational Bayes and message passing methods, which are efficient and support more complex inferences, but can only be applied to specific models, have little theory, and often exhibit convergence issues, {see \cite{mcmcAlgImaging} for a recent survey on the topic}. Recently, there have been significant advances in MCMC methodology for imaging, particularly for Bayesian models that are log-concave w.r.t. the unknown image, for which maximum-a-posteriori estimation is a convex optimisation problem. This paper seeks to further improve MCMC methodology for imaging.

MCMC methods were already actively studied in the imaging literature two decades ago, and have regained attention lately because of their capacity to address challenging imaging problems that are beyond the scope of optimisation-based and machine learning techniques. In particular, the interface between MCMC and optimisation has become a very active research area, especially around the so-called proximal MCMC algorithms \cite{pereyraMYULA} that combine ideas from high-dimensional stochastic simulation with techniques from convex analysis and proximal optimisation to achieve better computational efficiency. Despite being relatively recent, proximal MCMC methods have already been successfully applied to a range of Bayesian inference problems, for example, image deconvolution with total-variation and wavelet priors \cite{pereyraMYULA,Vono_TSP_2019}, inpainting \cite{Vono_TSP_2019}, tomographic reconstruction  \cite{pereyraMYULA}, astronomical imaging \cite{10.1093/mnras/sty2004}, restoration of images corrupted by Poisson  noise \cite{Vono2019}, ultrasound imaging \cite{Corbineau2019}, image coding \cite{Elvira2017}, sparse binary logistic regression \cite{Vono2018}, and graph processing \cite{Bianchi2019}.

This paper seeks to exploit recent developments in stochastic numerical analysis to significantly improve the computational efficiency of proximal MCMC methodology. More precisely, we propose to use a state-of-the-art orthogonal Runge-Kutta-Chebyshev stochastic approximation of the Langevin diffusion process \cite{abdulleSKROCK} that is significantly more computationally efficient than the conventional Euler-Maruyama approximation used by existing proximal MCMC methods. In particular, we present a new proximal MCMC method that applies this approximation to the Moreau-Yosida regularised Langevin diffusion underpinning the Moreau-Yosida unadjusted Langevin algorithm \cite{pereyraMYULA}, and show both theoretically and empirically {that} this leads to dramatic improvements in convergence speed and estimation accuracy.

The remainder of the paper is organised as follows: Section \ref{sec:problemStatement} defines notation, introduces the class of models considered, and recalls the Moreau-Yosida unadjusted Langevin algorithm that is the basis of our method. {In Section \ref{sec:samplingAlgorithms}, we {introduce the proposed proximal MCMC method} and study its convergence properties. For tractability, we focus on the case of Gaussian target densities, which enables the derivation of explicit convergence results in the Wasserstein distance that we compare with those of the conventional Euler-Maruyama approximation.} Section 4 illustrates the methodology in two one-dimensional toy problems, as well as three experiments related to image deconvolution, hyper-spectral unmixing and tomographic reconstruction, {containing $\ell_1$ and $TV$ priors}. Conclusions and perspectives for future work are reported in Section \ref{sec:conclusions}. Proofs are finally reported in Appendices \ref{app:w2analysis} and \ref{app:w2bounds}.

\section{Problem statement}\label{sec:problemStatement}
\subsection{Bayesian inference for imaging inverse problems}
We consider imaging problems involving an unknown image $x \in \mathbb{R}^d$ and some observed data $y \in \mathbb{C}^p$, related to $x$ through {a} statistical model with likelihood function $p(y|x)$. In particular, we are interested in problems where the recovery of $x$ from $y$ is ill-conditioned or ill-posed ({i.e., either the problem does not admit a unique solution that changes continuously with $y$, or there exists a unique solution but it is not stable w.r.t. small perturbations in $y$). For example, problems of the form $y = Ax + w$ with $w \sim \mathcal{N}(0,\sigma^2 \mathbb{I}_p)$ and $\sigma>0$ where the observation operator $A \in  \mathbb{C}^{n \times p}$ is rank deficient, or problems where $A^\top A$ is full rank but has a poor {condition} number}. As mentioned previously, such problems are ubiquitous in imaging sciences and have been the focus of significant research efforts \cite{optimizationAlgImaging, bookKaipioSomersalo}. 

In this paper we adopt a Bayesian approach to regularise the estimation problem and deliver meaningful estimates of $x$, as well as uncertainty quantification for the solutions delivered. More precisely, we represent $x$ as a random quantity with prior distribution $p(x)$ promoting expected properties (e.g., sparsity, piecewise-regularity, smoothness, etc.), and base our inferences on the posterior distribution \cite{bookKaipioSomersalo}
\begin{equation}\label{eqn:piDist}
\pi (x) \triangleq p(x|y)=\frac{p(y|x)p(x)}{\int_{\mathbb{R}^{d}} p(y|x)p(x)dx}\, ,
\end{equation}
henceforth denoted by $\pi$. We focus on {log-concave} models of the following form
\begin{equation}\label{eqn:piDist_U}
\pi (x) = \frac{e^{-f(x)-g(x)}}{\int_{\mathbb{R}^{d}} e^{-f(s)-g(s)}ds},
\end{equation}
where $f: \mathbb{R}^{d} \rightarrow \mathbb{R}$ and $g : \mathbb{R}^{d} \rightarrow\left( -\infty,\infty \right]$ are two lower bounded functions satisfying the following conditions: 
\begin{enumerate}
	\item $f$ is convex and Lipschitz continuously differentiable with constant $L_f$, i.e.,
	\begin{equation*}\label{eqn:lipschitzConstantF}
	\| \nabla f (x) - \nabla f (y) \|_2 \leq L_f \| x-y \|_2, \quad \forall x, y \in \mathbb{R}^d\, ;
	\end{equation*}
	\item $g$ is proper, convex, and lower semi continuous, but potentially non-smooth.
\end{enumerate}
This class of models is widely used in imaging sciences, and includes, for instance, analysis models of the form $f(x) = \|y-A x\|^2 / 2\sigma^2$ and $g(x) = \theta\| \Psi x \|_{\dagger} + \iota_{\mathcal{S}}(x)$ some dictionary or representation $\Psi$ and norm or pseudo-norm $\| \cdot \|_{\dagger}$, and a hard constraint $\mathcal{S} \subset \mathbb{R}^d$ on the solution space\footnote{For any $\mathcal{S} \subset \mathbb{R}^d$, the indicator $\iota_{\mathcal{S}}$ takes value $\iota_{\mathcal{S}}(x) = 0$ if $x \in \mathcal{S}$, and $\iota_{\mathcal{S}}(x) = +\infty$ otherwise.}. {Also note that we do not assume that $p(x|y)$ belongs to the exponential family.}

{As mentioned previously, posterior distributions of the form \eqref{eqn:piDist_U} are log-concave}, which is an important property for Bayesian inference because it guarantees the existence of all posterior moments and hence of moment-based estimators such as the minimum mean square error (MMSE) estimator. Log-concavity also plays a central role in maximum-a-posteriori (MAP) estimation, given by
\begin{equation*}
	\begin{split}
	\hat{x}_{MAP}&=\arg\max\limits_{x} \pi(x)\, ,\\
&= \arg\min\limits_{x} f(x) + g(x) \, ,
\end{split}
\end{equation*}
which is the main estimation strategy in imaging sciences. The popularity of MAP estimation stems from the fact that it is a convex optimisation problem that can be efficiently solved by using modern proximal splitting optimisation techniques \cite{optimizationAlgImaging}. {Some forms of approximate uncertainty quantification can be also computed by using proximal splitting techniques (see  \cite{Repetti2019} and references therein).} 

However, most Bayesian analyses require using specialised computational statistics techniques to calculate expectations and probabilities w.r.t. $\pi$. For example, computing Bayesian estimators (e.g., MMSE estimation), calibrating unknown model parameters (e.g., regularisation parameters), performing Bayesian model selection and predictive model checks, and reporting (exact) credible regions and hypothesis tests. From a Bayesian computation viewpoint, this typically requires using a high-dimensional Markov chain Monte Carlo (MCMC) method to simulate samples from $x|y$ followed by Monte Carlo integration \cite{mcmcAlgImaging}. Unfortunately, this approach has been traditionally {computationally too expensive} for wide adoption in imaging sciences, limiting the impact of Bayesian statistics in this field. {Alternatively, one can also perform approximate inferences by using deterministic surrogate methods \cite{mcmcAlgImaging}. However, these can exhibit convergence issues and have little theoretical guarantees, and hence they have not been widely adopted either.}

Recent works have sought to addressed these limitation{s} of Bayesian computation methodology by developing new and highly efficient MCMC methods tailored for imaging sciences, particularly by using techniques from proximal optimisation that are already widely adopted in the field. These so-called proximal MCMC methods \cite{pereyraMYULA,Vono_TSP_2019,Bianchi2019} have been an important step towards promoting Bayesian imaging techniques, as they are easy to implement, have significantly reduced computing times, and improve theoretical guarantees on the solutions delivered. However, there remain some fundamental features of modern optimisation methodology that have not yet been replicated in proximal MCMC approaches. In particular, modern optimisation methods rely strongly on acceleration techniques to achieve faster convergence rates and improve their robustness to poor conditioning \cite{WibisonoE7351}. In this paper, we accelerate proximal MCMC methods to improve their convergence properties.

\subsection{Bayesian computation for imaging inverse problems}
\subsubsection{Langevin Markov chain Monte Carlo methods}
Proximal MCMC methods are derived from the overdamped Langevin diffusion, which we recall bellow. For clarity we first introduce the approach for models that are smooth, and then explain the generalisation to non-smooth models.

Suppose the need to sample from a high-dimensional density $\bar{\pi}$ that is continuously differentiable on $\mathbb{R}^{d}$. Langevin MCMC methods address this task by using the overdamped Langevin stochastic differential equation (SDE), given by 
\begin{equation}\label{eqn:LangevinSDE}
\textrm{dX}_t = \nabla \log \bar{\pi}(\textrm{X}_t) \textrm{d}t + \sqrt{2} \textrm{d}W_t \, ,
\end{equation}
where $(W_t)_{t\geq 0}$ is a $d$-dimensional Brownian motion. Under mild regularity assumptions, this SDE has an unique strong solution and admits $\bar{\pi}$ as unique invariant distribution. Consequently, if we could solve \eqref{eqn:LangevinSDE} and let $t \rightarrow \infty$, this would provide Monte Carlo samples from $\bar{\pi}$ useful for Bayesian computation. This strategy is particularly computationally efficient when $\bar{\pi}$ is log-concave because in that case $\textrm{X}_t$ converges in distribution to $\bar{\pi}$ exponentially fast with a good rate \cite{convergenceULA}.

Unfortunately, it is generally not possible to exactly solve \eqref{eqn:LangevinSDE}, and discrete approximations of $\textrm{X}_t$ need to be considered instead. In particular, {most algorithms use the Euler-Maruyama (EM) discretization {\cite{KP92}}:}
\begin{equation}\label{eqn:eulerMaruyDisc}
X_{n+1} = X_{n} + \delta \nabla  \log \bar{\pi} (X_{n}) + \sqrt{2 \delta} Z_{n+1},
\end{equation}
where $\delta >0$ is a given stepsize and $(Z_{n})_{n\geq 1}$ is a sequence of i.i.d. \textit{d}-dimensional standard Gaussian random variables. This MCMC method is known as the unadjusted Langevin algorithm (ULA) \cite{ULAMALAPaper}.

Under some regularity assumptions, namely $\bar{L}$-Lipschitz continuity of $\nabla \log \bar{\pi}$ and $\delta < 2/\bar{L}$, the Markov chain $(X_{n})_{n\geq 0}$ is ergodic with stationary distribution $\bar{\pi}_{\delta}(x)$ close to $\bar{\pi}$ \cite{convergenceULA}. Additionally, when $\bar{\pi}$ is log-concave, ULA inherits the favourable properties of \eqref{eqn:LangevinSDE} and converges to $\bar{\pi}_{\delta}(x)$ geometrically fast with good convergence rates, offering an efficient Bayesian computation methodology for large problems \cite{convergenceULA}.

The estimation bias {\cite{AVZ14}} associated with targeting $\bar{\pi}_{\delta}(x)$ instead of $\bar{\pi}$ can be reduced by decreasing $\delta$, and vanishes as $\delta \rightarrow 0$. However, decreasing $\delta$ deteriorates the convergence properties of the chain and amplifies the associated non-asymptotic bias and variance. Therefore, to apply ULA to large problems in a computationally efficient way it is necessary to use values of $\delta$ that are close to the stability limit $2/\bar{L}$, at the expense of some asymptotic bias. Notice that it is also possible to remove the asymptotic bias by combining ULA with a Metrolopolis Hastings correction step targeting $\bar{\pi}$, leading to the so-called Metropolis adjusted Langevin algorithm (MALA) \cite{ULAMALAPaper}. This strategy is widely used in computational statistics for medium-sized problems. However, in large problems such as imaging problems, using a Metropolis-Hastings correction may dramatically deteriorate {the} convergence speed \cite{pereyraMYULA}.

\subsubsection{Proximal Markov chain Monte Carlo methods}
We now consider the class of models $\pi$ given by \eqref{eqn:piDist_U}, which are not smooth. Unfortunately, ULA and MALA cannot be directly applied to such models, as they require Lipschitz differentiability of $\log \pi$. Proximal MCMC methods address this difficulty by carefully constructing a smooth approximation $\pi_\lambda$ that by construction satisfies all the regularity conditions required by ULA and MALA, and which can be made arbitrarily close to the original model $\pi$ by tuning a regularisation parameter $\lambda > 0 $. This strategy, originally proposed in \cite{proxMCMC}, can be implemented in different ways. In particular, \cite{pereyraMYULA} replaces the non-smooth term $g$ in \eqref{eqn:piDist_U} with its Moreau-Yosida (MY) envelope
\begin{equation*}
g^{\lambda}(x)=\min\limits_{y\in\mathbb{R}^{d}} \left\lbrace g(y) + \frac{1}{2\lambda} \| x-y \|^{2} \right\rbrace ,
\end{equation*}
to construct the approximation
\begin{equation*}
\pi^\lambda (x) = \frac{e^{-f(x)-g^{\lambda}(x)}}{\int_{\mathbb{R}^{d}} e^{-f(s)-g^{\lambda}(s)}ds} \, ,
\end{equation*}
which has the following key properties that are useful for Bayesian computation \cite{pereyraMYULA}:
\begin{itemize}
	\item For all $\lambda > 0$, $\pi^\lambda$ defines a proper density on $\mathbb{R}^d$.
	\item For all $\lambda > 0$, $\pi^\lambda$ is log-concave and Lipschitz continuously differentiable with
\begin{equation}\label{eqn:gradientMoreauYoshida}
\begin{split}
\nabla \log \pi ^\lambda &= -\nabla f(x) - \nabla g^{\lambda}(x) \, ,\\
&= -\nabla f(x) - \frac{1}{\lambda} \left( x - \text{prox}_{g}^{\lambda}(x) \right)\, ,
\end{split}
\end{equation}	
with Lipschitz constant $L = L_f + 1/\lambda$, and where for all $x \in \mathbb{R}^d$
$$
\text{prox}_{g}^{\lambda}(x)=\underset{u\in\mathbb{R}^{d}}{\arg\min} \,\, g(u) + \frac{1}{2\lambda} \| x-u \|^{2}.
$$
	\item The approximation $\pi^\lambda$ converges to $\pi$ in total-variation norm; i.e.,
	$$
	\lim_{\lambda \rightarrow 0} \|\pi^\lambda - \pi\|_{TV} = 0\, .
	$$
\end{itemize} 

Given the smooth approximation $\pi^\lambda$, we define the auxiliary Langevin SDE 
\begin{equation} \label{eq:MYLeq}
\textrm{dX}_{t}= \nabla \log \pi^{\lambda}(\textrm{X}_{t})\textrm{d}t+\sqrt{2}\textrm{d}W_{t}\, ,
\end{equation}
and derive the MYULA Markov chain by discretising this SDE by the EM method
\begin{equation}\label{eqn:MYULADisc}
X_{n+1} = X_{n} - \delta \nabla f (X_n)-\frac{\delta}{\lambda}\left(X_{n}-\text{prox}_{g}^{\lambda}(X_{n})\right)+ \sqrt{2\delta} Z_{n+1} \,.
\end{equation}
If necessary, the asymptotic bias can then be removed by complementing MYULA with a Metropolis-Hastings step \cite{proxMCMC}, which is useful for benchmarking purposes \cite{pereyraMYULA, 10.1093/mnras/sty2004}.
Notice that one can also consider other approximations constructed by applying the Moreau-Yosida envelope directly to $f+g$ \cite{proxMCMC}, or by replacing the Moreau-Yosida envelope with a forward-backward envelope \cite{proxMCMC,Atchade2018}. It is also possible to apply the Moreau-Yosida envelope separately to $f$ and $g$ and integrate MYULA (with or without Metropolisation) within an auxiliary-variable Gibbs sampling scheme{,} see \cite{Vono_TSP_2019}.

As mentioned previously, despite being relatively recent, proximal MCMC methods have already been successfully applied to a many large-scale inference problems related to imaging sciences \cite{pereyraMYULA,10.1093/mnras/sty2004,Vono_TSP_2019,Corbineau2019}, and machine learning  \cite{Elvira2017,Vono2018,Bianchi2019,Brosse2017}.  
	
\subsubsection{Limitations of proximal Markov chain Monte Carlo methods and recent improvements}
A main limitation of ULA, MALA and their proximal variants is that they are all derived from the EM approximation \eqref{eqn:eulerMaruyDisc} of the Langevin SDE. This approximation is mainly used because it is computationally efficient in high-dimensions, it is easy to implement, and it can be rigorously theoretically analysed. However, the EM approximation is not particularly suitable for problems that are ill-conditioned or ill-posed as its performance is very sensitive to the anisotropy of the target density, which is a common feature of imaging problems.  More precisely, in order to be useful for Bayesian computation, the EM approximation of the Langevin SDE \eqref{eq:MYLeq} has to be numerically stable.
For MYULA, this requires using a stepsize $\delta < 2/L$ with $L = L_f + 1/\lambda$, where we recall that $L_f$ is the Lipschitz constant of $\nabla f$ and that $\lambda$ controls the quality of the approximation $\pi_\lambda$ of $\pi$. This restriction essentially guarantees that the chain moves slowly enough to follow changes in $\nabla \log \pi_\lambda$ in a numerically stable manner, particularly along directions of fast change. However, this is problematic when $\pi_\lambda$ has some directions or regions of the parameter space that change relatively very slowly, as the chain will struggle to properly explore the solution space and will require a very large number of iterations to converge. In imaging models, this typically arises when the likelihood $p(y|x)$ has identifiability issues (e.g, if it involves an observation operator $A$ for which $A^\top A$ is badly conditioned or rank deficient), or if we seek to use a small value of $\lambda$ to bring $\pi_\lambda$ close to $\pi$.

\begin{figure}
	\centering
	\subfloat[MYULA, $\mathcal{N}(\mu_1,\Sigma_1)$]{
		\label{subfig:2dGauss_ULA_wellCond_pdf}
		\includegraphics[scale=.269]{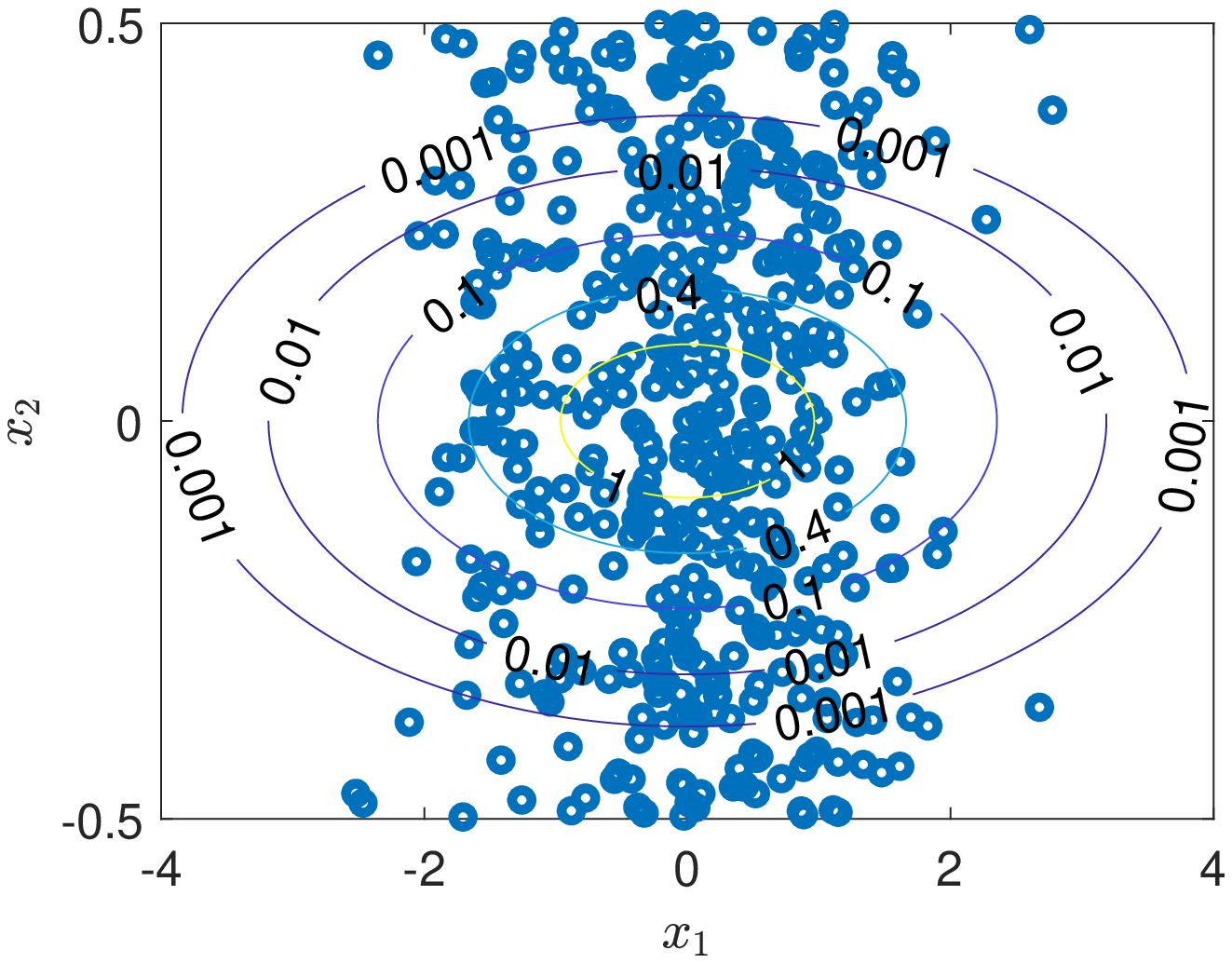}
	}
	\subfloat[ACF, $x_1$]{
		\label{subfig:2dGauss_ULA_wellCond_acf_1stcomp}
		\includegraphics[scale=.269]{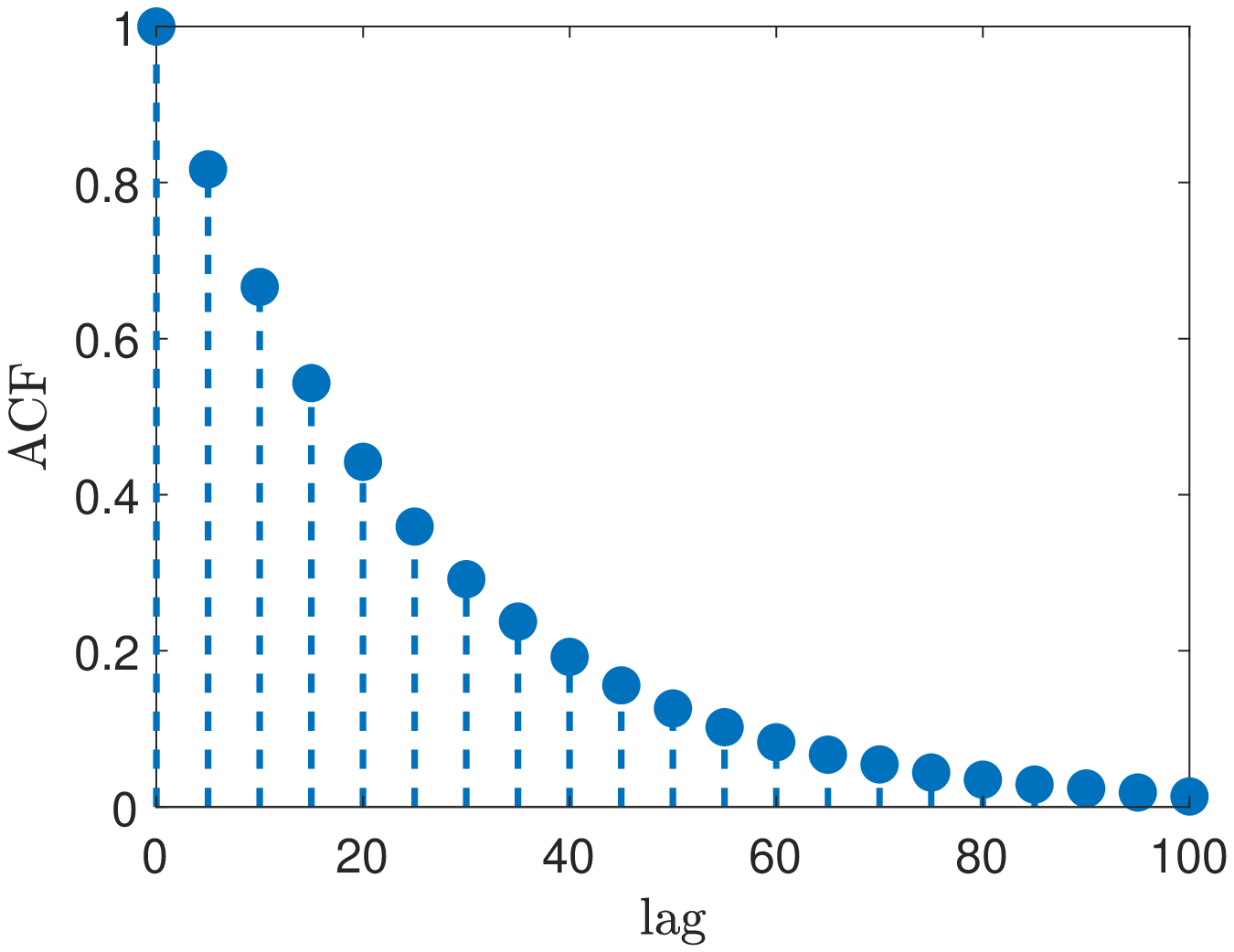}
	}
	\subfloat[ACF, $x_2$]{
		\label{subfig:2dGauss_ULA_wellCond_acf_2ndcomp}
		\includegraphics[scale=.269]{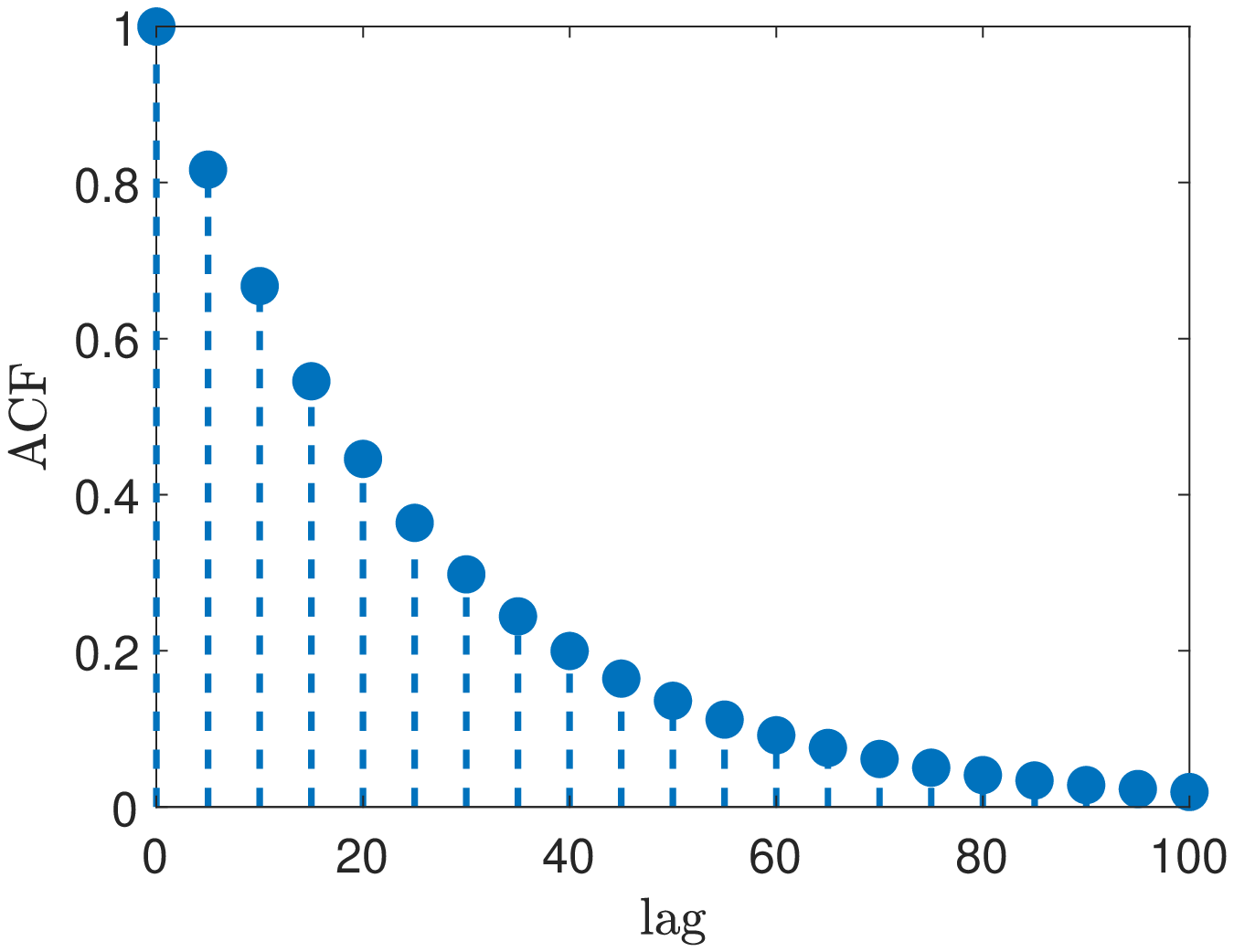}
	} \\
	\subfloat[MYULA, $\mathcal{N}(\mu_2,\Sigma_2)$]{
		\label{subfig:2dGauss_ULA_badCond_pdf}
		\includegraphics[scale=.269]{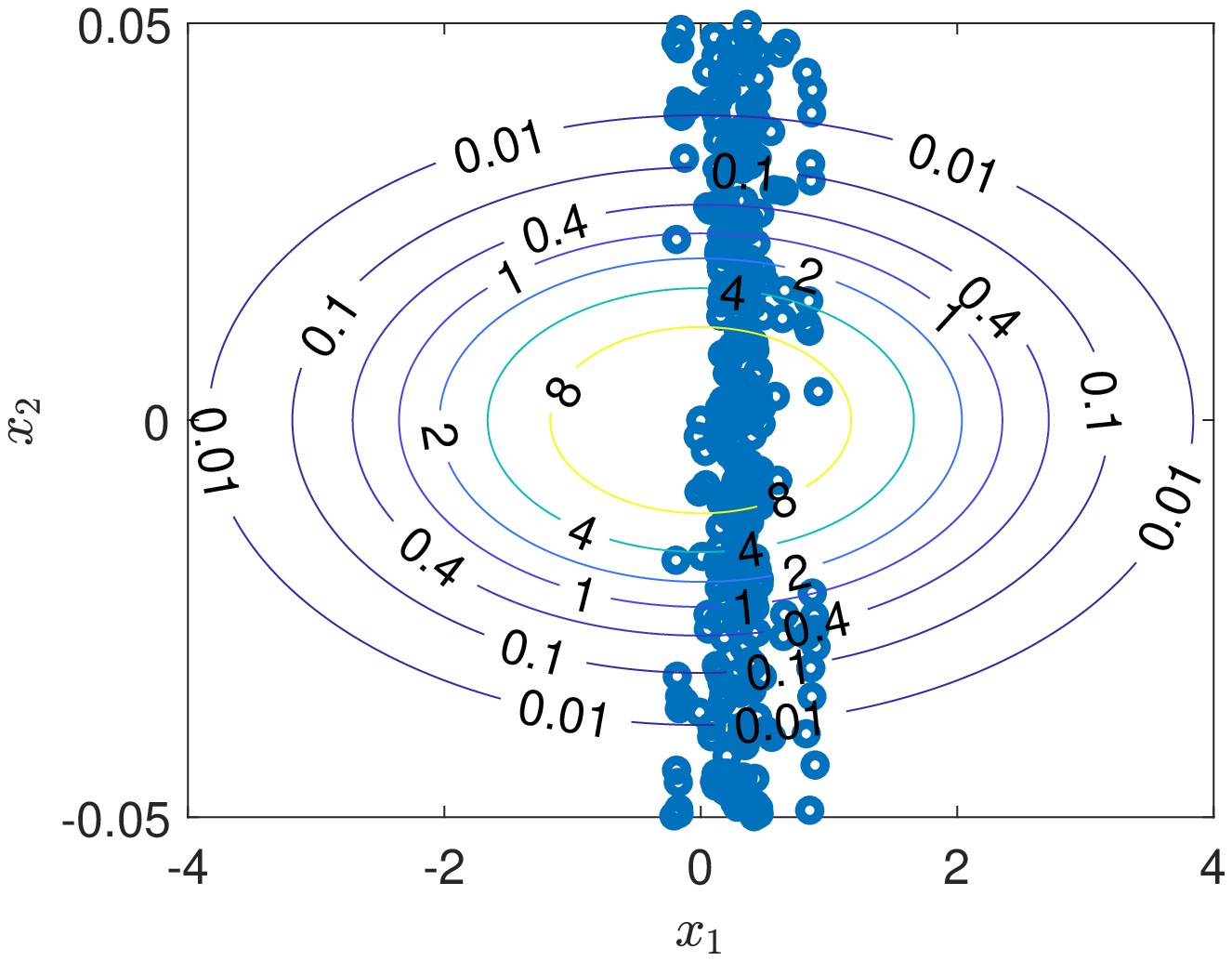}
	}
	\subfloat[ACF, $x_1$]{
		\label{subfig:2dGauss_ULA_badCond_acf_1stcomp}
		\includegraphics[scale=.269]{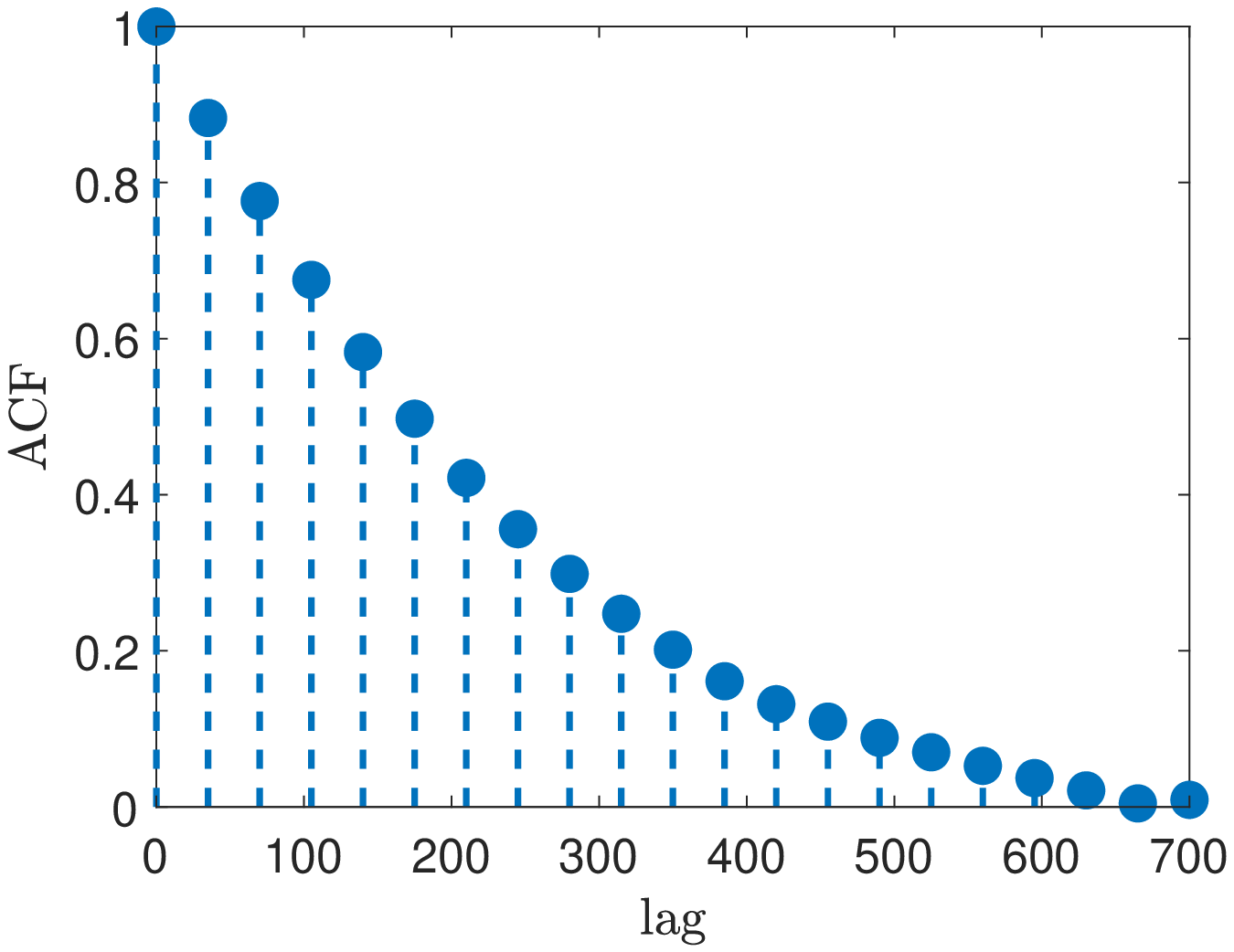}
	}
	\subfloat[ACF, $x_2$]{
		\label{subfig:2dGauss_ULA_badCond_acf_2ndcomp}
		\includegraphics[scale=.269]{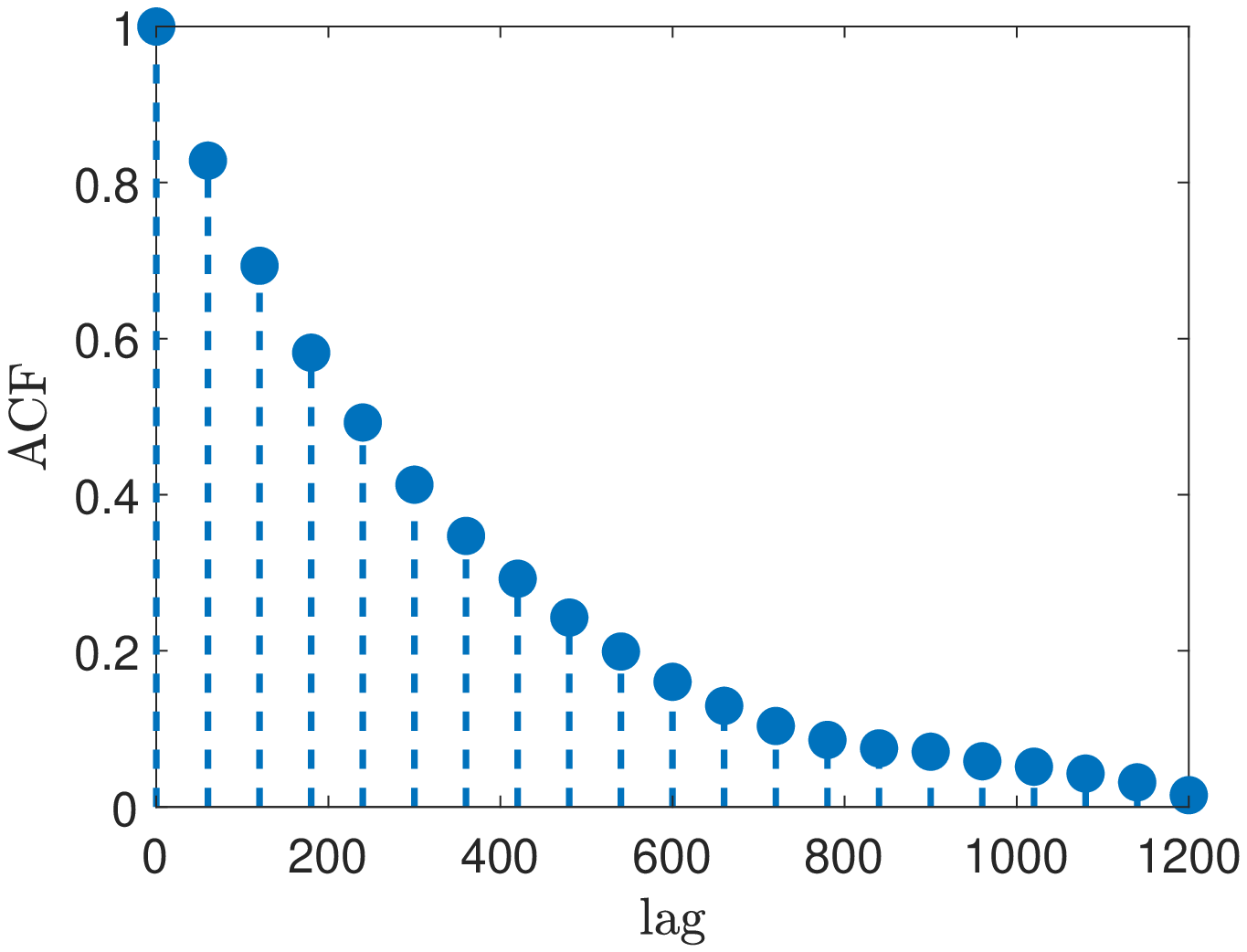}
	}
	\caption{Two-dimensional Gaussian distribution: {\normalfont(a)} $10^3$ samples generated by MYULA using the target distributions $\mathcal{N}(\mu_1,\Sigma_1)$ with $\delta = 2 /( L + \ell )= 1.98 \times 10^{-2}$ where $L = 1/ \sigma_{11}^2 = 100$ and $\ell = 1/ \sigma_{22}^2 = 1$; and {\normalfont(d)} $5 \times 10^3$ samples generated by MYULA using the target distributions $\mathcal{N}(\mu_2,\Sigma_2)$ with $\delta = 2 / (L + \ell) = 1.99 \times 10^{-4}$ where $L = 1/ \sigma_{11}^2 = 10^4$ and $\ell = 1/ \sigma_{22}^2 = 1$. Autocorrelation functions of the {\normalfont(b)-(e)} first and {\normalfont(c)-(f)} second component (i.e., $x_1$ and $x_2$) of the samples generated by the ULA algorithm, having $\mathcal{N}(\mu_1,\Sigma_1)$ and $\mathcal{N}(\mu_2,\Sigma_2)$ as target distributions, respectively.}
	\label{fig:2dGauss_ULA}
\end{figure}

To highlight this issue, we report below two simple illustrative experiments where MYULA is applied to a two-dimensional Gaussian distribution. In this case there is no non-smooth term $g$ and the time-step restriction is dictated by the Lipschitz constant of $f$, but the same phenomenon arises in more general models. In the first experiment we consider $\mu_1 = (0,0)$ and $\Sigma_1 = \textrm{diag}(1,10^{-2})$ (i.e., $L_f=10^2$); whereas in the second experiment we use  $\mu_2 = (0,0)$ and $\Sigma_2 = \textrm{diag}(1,10^{-4})$ (i.e., $L_f=10^4$). The results are presented in Figure \ref{fig:2dGauss_ULA}. Notice that in the first case MYULA explores the distribution very well, showing a good rate of decay in the autocorrelation functions of both components. However, in the second case, MYULA exhibits poor convergence properties as it struggles to explore the first component.

This limitation of the EM approximation could be partially mitigated by preconditioning the gradient $\nabla \log \pi_\lambda$ {by considering a Langevin SDE on an appropriate Riemannian manifold, as recommended in \cite{Girolami2011}, and in a spirit akin to natural gradient {descent} and Newton optimisation methods.  The preconditioning procedure proposed in \cite{Girolami2011} is very effective but too expensive for imaging models because it requires evaluating quantities related to second and third order derivatives of $\log \pi_\lambda$ and performing expensive matrix operations. Conversely, simple procedures such as preconditioning with a pseudo-inverse of the Hessian matrix of the log-likelihood function are computationally efficient but do not typically lead to significant improvements in performance because they do not take into account the geometry of the log-prior. The development of computationally efficient yet effective preconditioning strategies for imaging models is an active research topic, see, e.g., \cite{Marnissi2016,Marnissi2017}.}
	
{Moreover, a different strategy to improve the convergence rate of the EM approximation is to substitute both $f$ and $g$ by their regularised envelopes $f^\lambda$ and $g^\lambda$ so that the Lipschitz constant of $\nabla \log \pi_\lambda$ is bounded by $2/\lambda$, at the expense of additional bias. One can then use a single MYULA kernel targeting $\pi_\lambda$, or alternatively a splitting scheme involving two MYULA kernels with $\lambda = \delta$ to separately address $f^\lambda$ and $g^\lambda$ as recommended recently in \cite{Vono_TSP_2019}. That splitting leads to a Markov chain that is by construction numerically stable and potentially much faster than MYULA at the expense of some further estimation bias. Note that splitting schemes that combine Gibbs sampling with relaxations are a highly promising direction of research as they could potentially lead to algorithms with dimension-free convergence rates \cite{Vono2020}.}
	
{It is worth mentioning at this point that one can also consider other dynamics to derive Markov chains with potentially better convergence properties, namely the Hamiltonian dynamic which leads to the Hamiltonian Monte Carlo (HMC) algorithm \cite{mcmcAlgImaging, Chaari2016}. However, HMC uses a Verlet integrator that, despite being superior in other ways, has the same stepsize restrictions as the Euler method and hence also struggles to address problems that are poorly conditioned. Also, HMC uses a Metropolis correction that {can be} dramatically inefficient in large problems such as imaging problems.}

In this paper we propose to fundamentally improve proximal MCMC methods for imaging by using state-of-the-art numerical SDE approximation strategies that significantly outperform the conventional EM scheme. More precisely, we focus on a class of explicit stabilised methods that are specifically designed to deal with the time-step restriction, called stochastic orthogonal Runge-Kutta-Chebyshev methods (SK-ROCK) \cite{abdulleSKROCK}. The idea, in a nutshell, is to cleverly combine several evaluations of the gradient ${\nabla \log \pi^{\lambda} (x)}$ in a way that allows for taking larger time-steps, and thus breaking the stability barrier of MYULA. As mentioned previously, the same strategy can then be  applied to other MCMC methods that internally use MYULA (e.g., \cite{Vono_TSP_2019}), or variants of MYULA with other approximations of $\pi$ (e.g., \cite{proxMCMC,Atchade2018}), although this is beyond the scope of this paper and will be investigated in future works.

\section{Proposed Bayesian computation method}\label{sec:samplingAlgorithms}
\subsection{Stochastic orthogonal Runge-Kutta-Chebyshev methods}
We propose to significantly accelerate Bayesian computation for imaging problems by using the state-of-the-art explicitly stabilised SK-ROCK scheme \cite{abdulleSKROCK} to approximate the Langevin SDE \eqref{eq:MYLeq} associated with $\pi_\lambda$, instead of the basic EM discretisation scheme that underpins MYULA and other proximal MCMC methods. From a numerical analysis viewpoint, this is a highly advanced Runge-Kutta stochastic integration scheme that extends the deterministic Chebyshev method \cite{abdulleDeterministicChebyshev} to SDEs, and uses a damping strategy to stabilise the stochastic term. Crucially, its implementation is straightforward as it only requires knowledge of the gradient operator $\nabla \log \pi^\lambda(x)$ given by \eqref{eqn:gradientMoreauYoshida}, which is also used in MYULA. However, unlike MYULA that uses a single evaluation of $\nabla \log \pi^\lambda(x)$ per iteration, the considered Runge-Kutta scheme performs $s \in \mathbb{N}^*$ evaluations of $\nabla \log \pi^\lambda(x)$ at carefully chosen extrapolated points determined by Chebyshev polynomials. In this regard, the stochastic integration scheme is morally similar to accelerated optimisation methods that also use several gradient evaluations and extrapolation techniques to significantly improve their convergence properties. In fact, {the deterministic Runge-Kutta-Chebyshev method was recently shown to have {similar} theoretical convergence properties to Nesterov's accelerated optimisation algorithms in the case of strongly convex functions \cite{EVV18}}.


The proposed proximal SK-ROCK method is presented in Algorithm \ref{alg:SKROCK} below, where $T_s$ denotes the Chebyshev polynomial of order $s$ of the first kind, defined recursively by $T_{k+1} = 2xT_k(x) - T_{k-1}(x)$ with $T_0(x) = 1$ and $T_1(x) = x$.  {The two main parameters of the algorithm are the number of stages $s \in \mathbb{N}^*$ and the step-size $\delta \in (0,\delta_s^{max}]$. Notice that the range of admissible values for $\delta$ is controlled by $s$: for any $s \in \mathbb{N}^*$, the maximum allowed step-size is given by $\delta_s^{max}=l_s / ( L_f + 1/\lambda)$ with $l_s=[(s-0.5)^2 (2 - 4/3 \eta) -1.5]$ and $\eta = 0.05$ \cite{abdulleSKROCK}. Violating this upper bound leads to a potentially explosive Markov chain. Also note that in the case of $s=1$ the method reduces to MYULA.}
	
{The values of $\delta$ and $s$ are subject to standard bias-variance trade-offs. On the one hand, to optimise the mixing properties of the algorithm one would like to choose $\delta$ as large as possible. MYULA, based on the EM method, requires setting $\delta < \delta_1^{max}= 1 / ( L_f + 1/\lambda)$ for stability, but in SK-ROCK one can in principle take $\delta$ arbitrarily large by increasing the value of $s$. However, this would also increase the asymptotic bias and the computational cost per iteration. In our numerical experiments we found that a good trade-off in terms of bias, variance, and computational cost per iteration is achieved by setting $3<s<15$ and using a value of $\delta$ that is close to the maximum allowed step-size $\delta_s^{max}$. As a general rule for imaging problems, we recommend using $s = 15$ in problems that are strongly log-concave, and $s = 10$ otherwise.} Lastly, it is worth mentioning at this point that we also considered other alternatives to the EM scheme, namely the Runge-Kutta scheme of \cite{abdulleSROCK}, but found that SK-ROCK delivers the best performance for imaging models (the results with alternative schemes are not reported in the paper because of lack of space).



\begin{algorithm}
\caption{SK-ROCK algorithm}
\label{alg:SKROCK}
\begin{algorithmic}
\STATE \textbf{Set} $X_0 \in \mathbb{R}^d$, $\lambda > 0$, $n \in \mathbb{N}$, $s \in \{ 3,\ldots ,15 \}$, $\eta = 0.05$
\STATE \textbf{Compute} $l_s=(s-0.5)^2 (2 - 4/3 \eta) -1.5$ 
\STATE \textbf{Compute}
\[
\omega_0=1+\frac{\eta}{s^2}, \; \; \omega_1=\frac{T_s(\omega_0)}{T_s^{'} (\omega_0)}, \; \; \mu_1=\frac{\omega_1}{\omega_0}, \; \; \nu_1=s \omega_1 /2, \; \; k_1 = s \omega_1 / \omega_0
\]
\STATE  {\textbf{Choose} $\delta  \in (0,\delta_s^{max} ]$, {where $\delta_s^{max}=l_s / ( L_f + 1/\lambda)$}}
\FOR {$i = 0:n$}
\STATE $Z_{i+1} \sim \mathcal{N}(0,\mathbb{I}_d)$
\STATE $K_0  =  X_i$
\STATE $K_1  =  X_i + \mu_1 \delta \nabla \log \pi^{\lambda}(X_i +\nu_1 \sqrt{2 \delta} Z_{i+1}) + k_1 \sqrt{2 \delta} Z_{i+1}$
\FOR {$j = 2:s$}
\STATE \textbf{Compute}
\[
\mu_j=\frac{2\omega_1 T_{j-1}(\omega_0)}{T_j(\omega_0)}, \; \; \nu_j=\frac{2\omega_0 T_{j-1}(\omega_0)}{T_j(\omega_0)}, \; \; k_j=-\frac{T_{j-2}(\omega_0)}{T_j(\omega_0)}=1-\nu_j
\]
\STATE $K_j  =  \mu_j \delta \nabla \log \pi^{\lambda}(K_{j-1}) + \nu_j K_{j-1} + k_j K_{j-2}$
\ENDFOR
\STATE $X_{i+1} = K_s$
\ENDFOR
\end{algorithmic}
\end{algorithm}

\begin{figure}
\centering
\subfloat[SK-ROCK, $\mathcal{N}(\mu_1,\Sigma_1)$]{
\label{subfig:2dGauss_ROCK_wellCond_pdf}
\includegraphics[scale=.269]{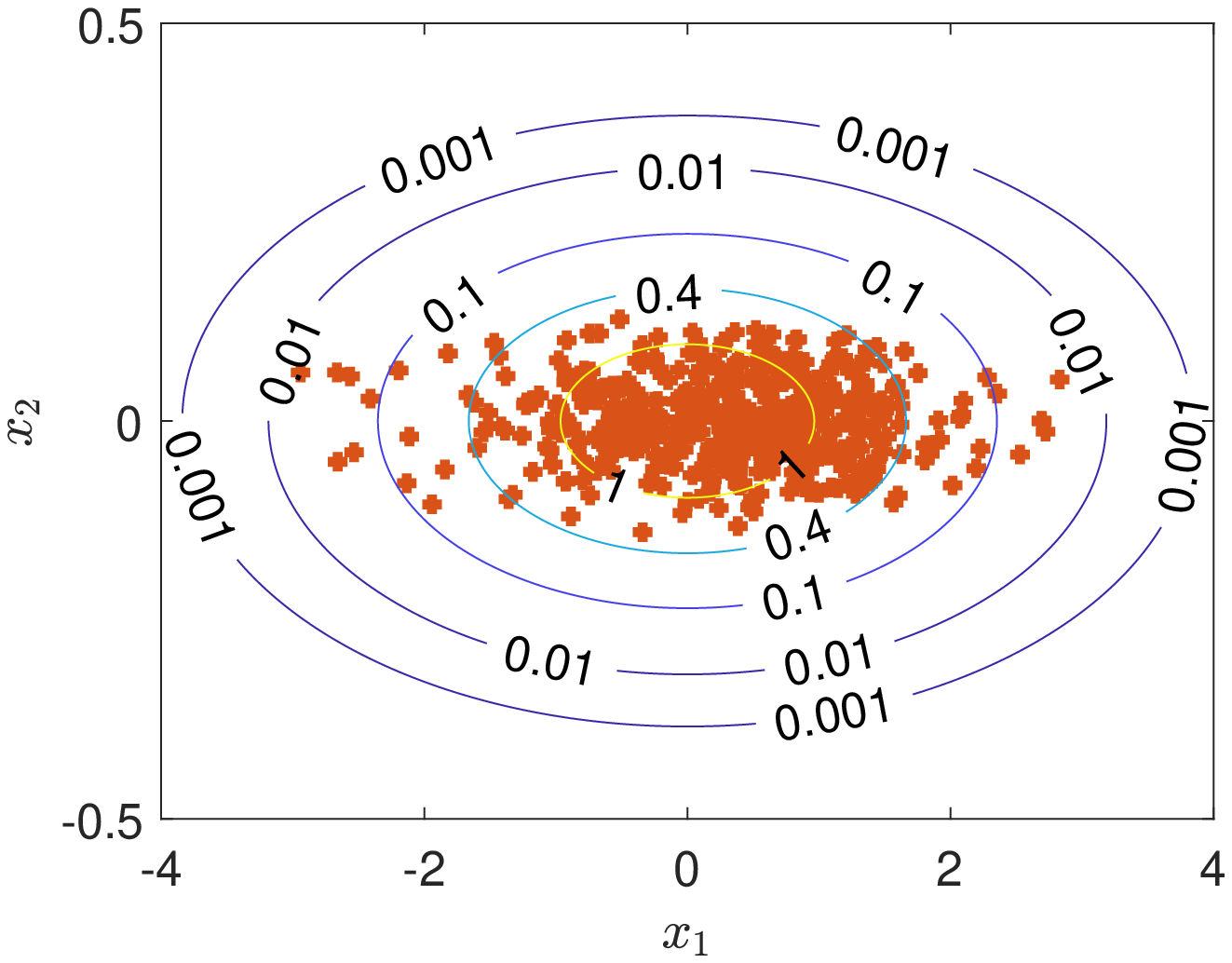}
}
\subfloat[ACF, $x_1$]{
\label{subfig:2dGauss_ROCK_wellCond_acf_1stcomp}
\includegraphics[scale=.269]{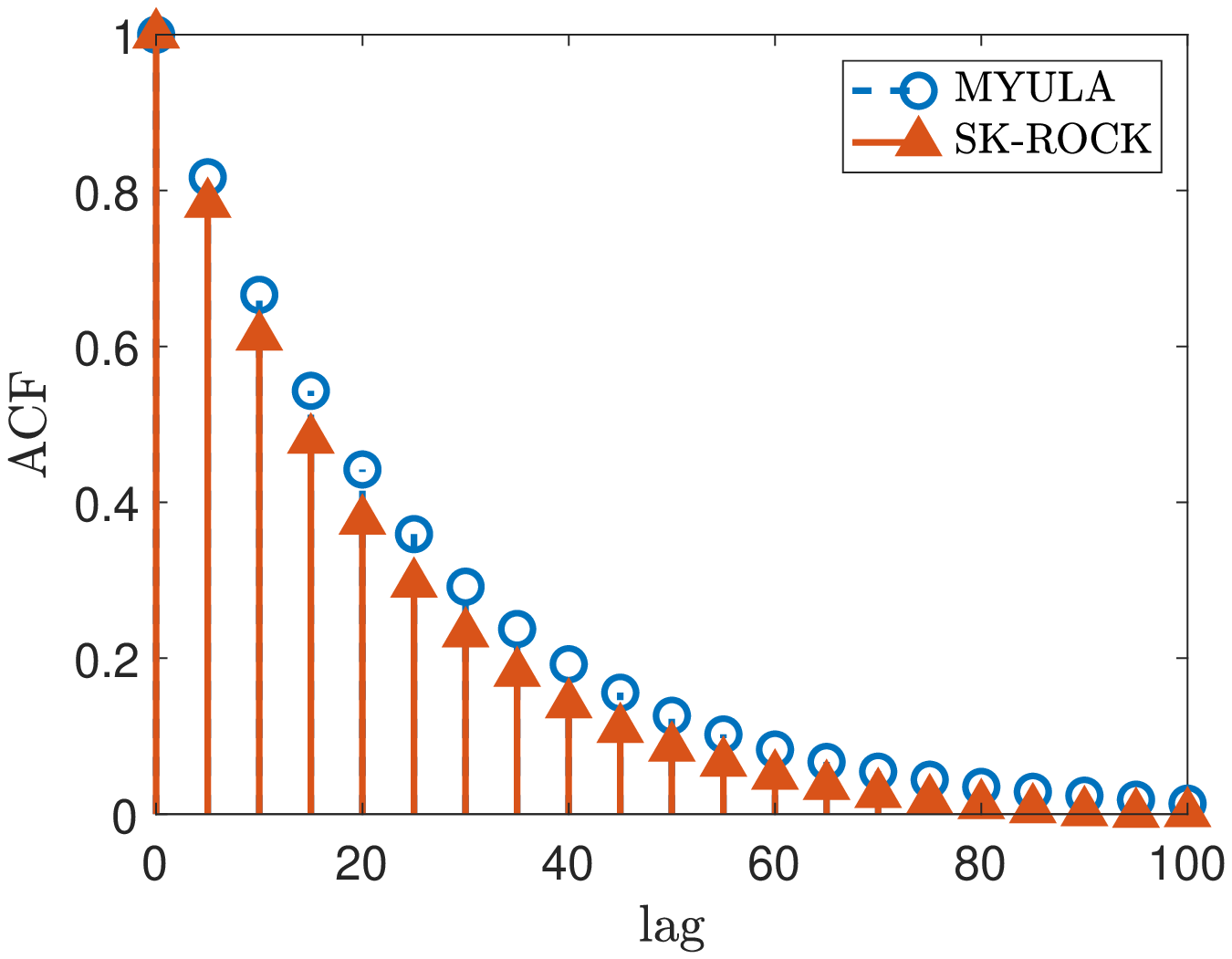}
}
\subfloat[ACF, $x_2$]{
\label{subfig:2dGauss_ROCK_wellCond_acf_2ndcomp}
\includegraphics[scale=.269]{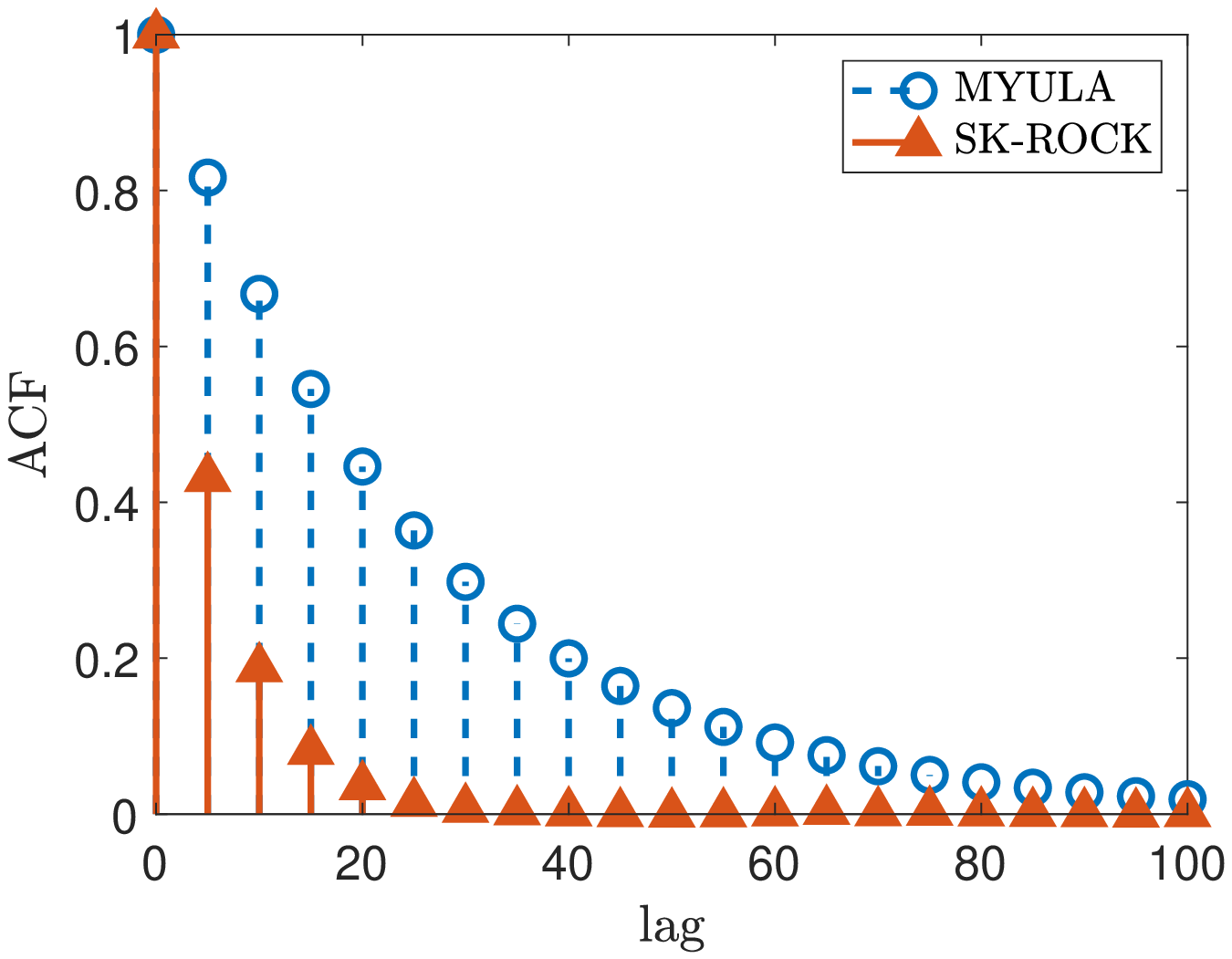}
} \\
\subfloat[SK-ROCK, $\mathcal{N}(\mu_2,\Sigma_2)$]{
\label{subfig:2dGauss_ROCK_badCond_pdf}
\includegraphics[scale=.269]{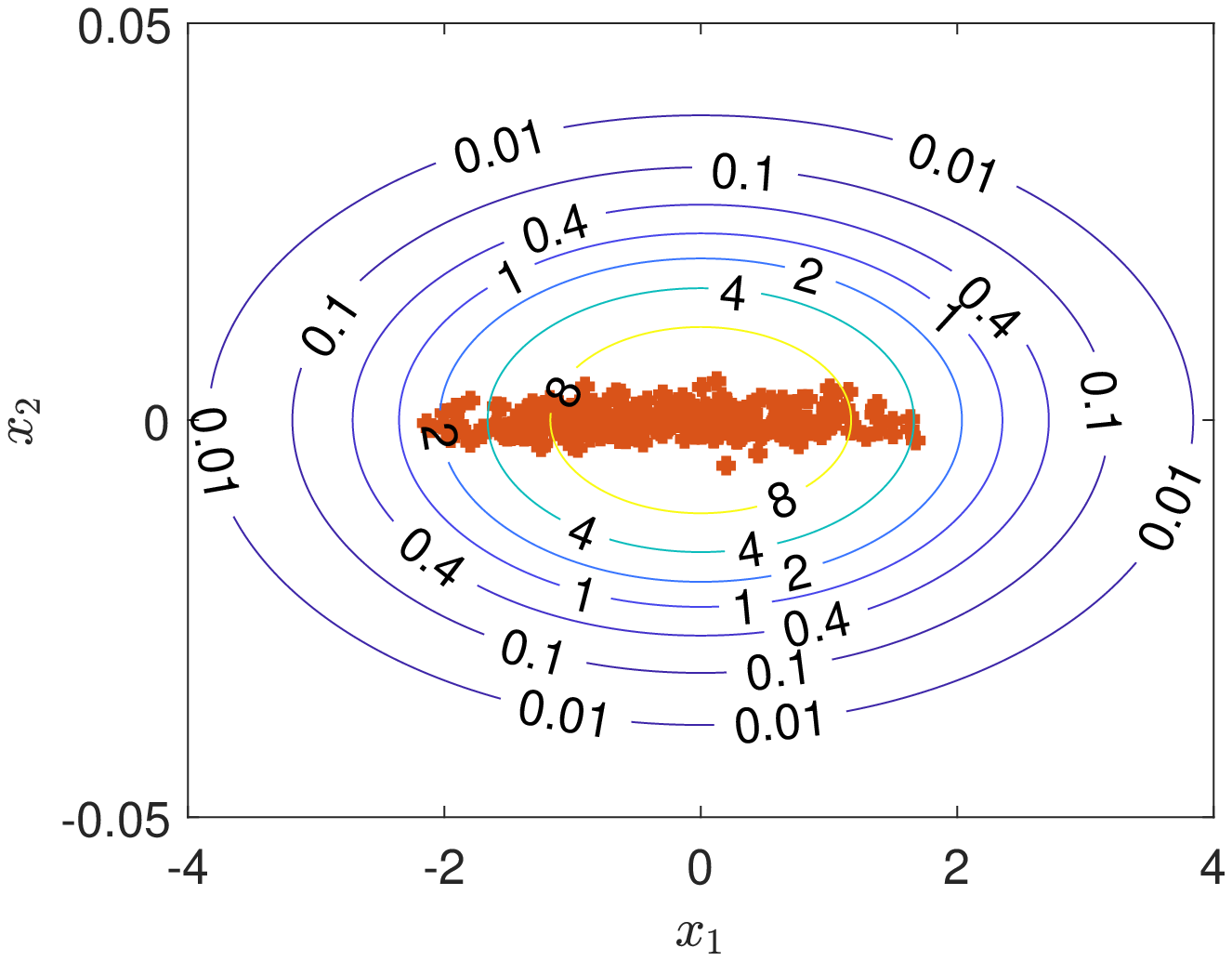}
}
\subfloat[ACF, $x_1$]{
\label{subfig:2dGauss_ROCK_badCond_acf_1stcomp}
\includegraphics[scale=.269]{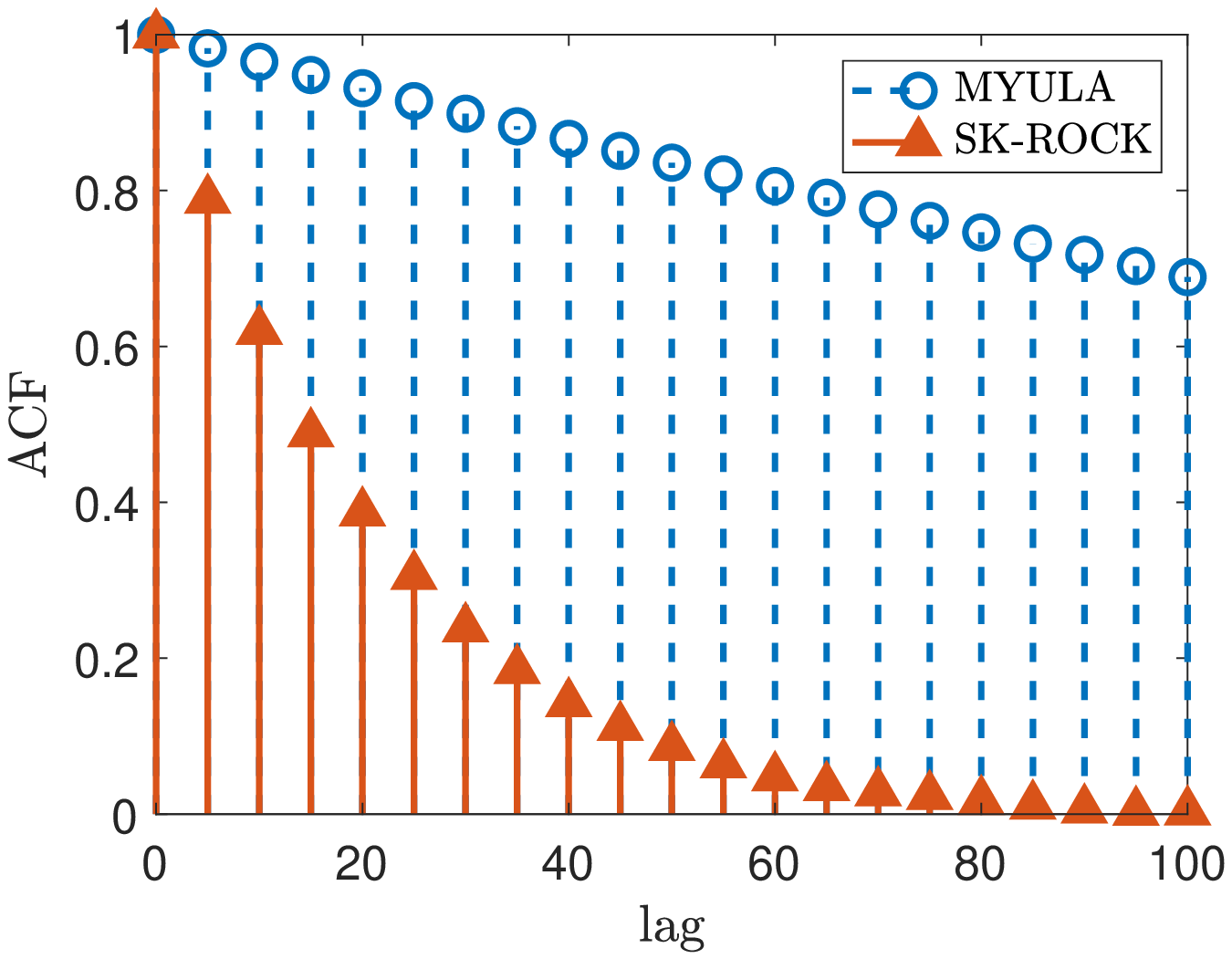}
}
\subfloat[ACF, $x_2$]{
\label{subfig:2dGauss_ROCK_badCond_acf_2ndcomp}
\includegraphics[scale=.269]{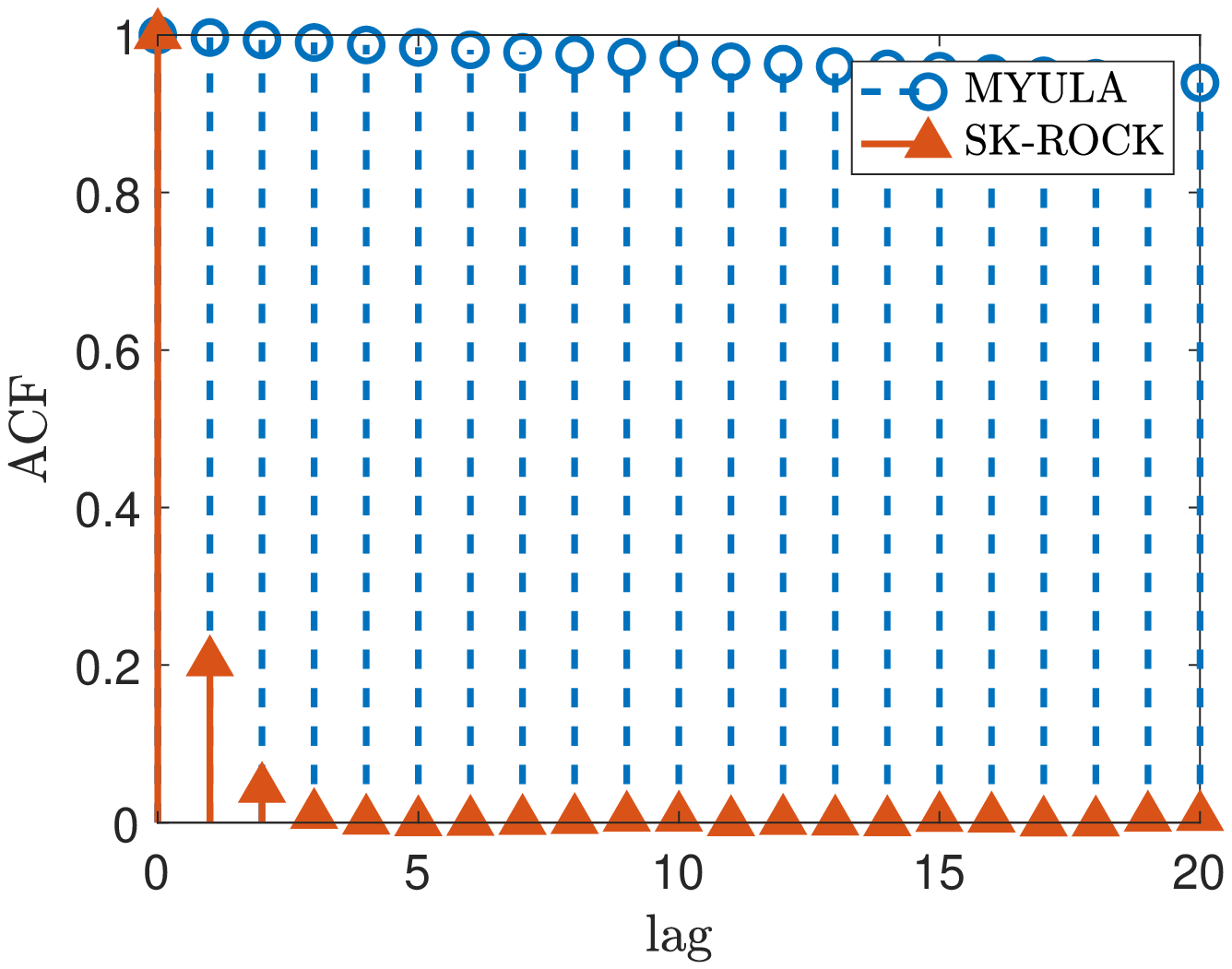}
}
\caption{Two-dimensional Gaussian distribution: {\normalfont(a) $10^3 / s$ samples generated by the SK-ROCK algorithm ($s=2$) using the target distribution  $\mathcal{N}(\mu_1,\Sigma_1)$ with $\delta = 4.82 \times 10^{-2}$ and  {\normalfont(d)} $5 \times 10^3 / s$ samples ($s=16$) using the target distribution $\mathcal{N}(\mu_2,\Sigma_2)$ with $\delta = 4.84 \times 10^{-2}$}. Autocorrelation functions of the {\normalfont(b)-(e)} first and {\normalfont(c)-(f)} second component (i.e., $x_1$ and $x_2$) of the samples generated by the SK-ROCK algorithm, having $\mathcal{N}(\mu_1,\Sigma_1)$ and $\mathcal{N}(\mu_2,\Sigma_2)$ as target distributions, respectively.}
\label{fig:2dGauss_ROCK}
\end{figure}

To illustrate the benefits of using the proximal SK-ROCK method instead of MYULA, we repeat the two Gaussian experiments reported in Figure \ref{fig:2dGauss_ULA} with Algorithm \ref{alg:SKROCK}. The results are shown in Figure \ref{fig:2dGauss_ROCK}, and where we have set the number of $s$ optimally by using \eqref{eqn:bestNStagesSKROCK}. Observe that because the SK-ROCK method is allowed to use a larger stepsize $\delta$ {in a stable manner}, it produces, {for the same computational cost (i.e., number of gradient evaluations)}, samples that are significantly less correlated than MYULA  with respect to the slow component. We also observe in Figure \ref{fig:2dGauss_ROCK} that this allows SK-ROCK to explore the target distribution more accurately.


\subsubsection{Computational Complexity}
{To the best of our knowledge, it is not possible to establish general complexity results for Runge-Kutta-Chebyshev methods by using existing analysis techniques, and we are currently investigating new bespoke techniques to study SK-ROCK. This is an important difference w.r.t. the EM scheme used in MYULA, for which there are detailed non-asymptotic convergence results available that can be used to characterise its computational complexity \cite{convergenceULA}. Nevertheless, it is possible to get an intuition for the computational complexity of SK-ROCK by theoretically analysing its convergence properties for a $d$-dimensional Gaussian target distribution with density $\pi(x) \propto \exp{(-0.5 x^\top \Sigma^{-1} x)}$, and $\Sigma = \text{diag}(\sigma_1^2,...,\sigma_d^2)$. More precisely, we study the convergence of SK-ROCK in the 2-Wasserstein distance, as a function of the number of gradient evaluations and the condition number $\kappa = {\sigma^{2}_{\text{max}}}/{\sigma^{2}_{\text{min}}}$, and compare it with MYULA. This is achieved by analysing in full generality the numerical solution of the Langevin SDE associated with $\pi$, given by}
\begin{equation}
\label{eqn:SDE_Gaussian_1d}
dX_t = - \Sigma^{-1} X_t dt + \sqrt{2} dW_t\, ,
\end{equation}
by a one step numerical integrator, which yields (in general) a recurrence of the form 
\begin{equation}\label{eqn:generalNumMethod}
X_{n+1}^i = R_1(z_i)X_n^i + \sqrt{2 \delta} R_2(z_i) \xi_{n+1}^i, \quad \xi_{n+1}^i \sim N(0,1),
\end{equation}
where $z_i=-\delta / \sigma_i^2$ and $X_0=(x_0^1,...,x_0^d)^T$ is a deterministic initial condition. For the EM scheme used in MYULA we have $R_{1}(z)=1+z$ and $R_{2}(z)=1$, and for the SK-ROCK we have that \cite{abdulleSKROCK}
\begin{equation}\label{eqn:R1R2SKROCK}
R_1(z)=\frac{T_s (\omega_0 + \omega_1 z)}{T_s (\omega_0)}, \; \; R_2(z)=\frac{U_{s-1} (\omega_0 + \omega_1 z)}{U_{s-1} (\omega_0)} \left( 1 + \frac{\omega_1}{2} z \right)\, ,
\end{equation}
where $T_{s},U_{s}$ are Chebyshev polynomials of first and second kind respectively and 
\[
\omega_0=1+\frac{\eta}{s^2}, \; \; \omega_1=\frac{T_s(\omega_0)}{T_s^{'} (\omega_0)}.
\] 
By using the fact that Gaussian distributions are closed under linear transformations, {and assuming that the initial condition  $X_{0}$ is deterministic}, we derive the distribution of $X_{n}$ for any $\delta > 0$ and obtain the following general result that holds for the EM (MYULA) method and for SK-ROCK. The proof is reported in Appedix \ref{app:w2analysis}.
\begin{proposition} \label{prop:first}
	Let $\pi (x) \propto \exp{(-0.5 x^T \Sigma^{-1} x)}$ with $\Sigma = \text{diag}(\sigma_1^2,...,\sigma_d^2)$, and let $Q_{n}$ be the probability measure associated with $n$ iterations of the generic Markov kernel \eqref{eqn:generalNumMethod}. Then the 2-Wasserstein distance between $\pi$ and $Q_{n}$ is given by
	\begin{equation}\label{eqn:wassersteinDistanceFinal}
	W_2(\pi;Q_n )^2 = \sum_{i=1}^d \left(D_n(z_i,x_0^i) + B_n(z_i,\sigma_i) \right)
	\end{equation}
	where
	\[
	D_n(x,u)  = (R_1(x))^{2n} u^2, \quad 
	B_n(x,u) = \left[u - \sqrt{2\delta} R_2(x) \left( \frac{1 - (R_1(x))^{2n}}{1 - (R_1(x))^2} \right)^{1/2} \right]^2. 
	\]
	{In addition the following bound holds
	\begin{equation}\label{eq:bbound}
	W_{2}(\pi;Q_{n+1} )^2  \leq  W_{2}(\pi;\tilde{\pi} )^2+C W_{2}(\tilde{\pi},Q_{n} )^2
	\end{equation}
	where 
	\[
	\tilde{\pi} = \mathcal{N}\left(0,2\delta (R_{2}(z))\left[\frac{1}{1-R^{2}_{1}(z)}\right]   \right),
	\]
	is the numerical invariant measure and
	\begin{equation} \label{eq:contb}
	C=\max_{1\leq i \leq d}R_{1}(z_{i})^{2}.
	\end{equation}
	}
\end{proposition}
{The bound \eqref{eq:bbound} can now be used to compare the EM and the SKROCK method in terms of how many gradient evaluations are required to achieve $W_2(\pi;Q_n ) < \varepsilon$ for some desired accuracy level $\varepsilon > 0$. We see that the $W^2_2$ distance between $\pi$ and $Q_n$ involves two terms. The first term $W_{2}(\pi;\tilde{\pi} )^2$ relates directly to the asymptotic bias of the method \cite{AVZ14} (recall that without a Metropolis correction step, any generic approximation of \eqref{eqn:SDE_Gaussian_1d} will have some asymptotic bias because it will not exactly converge to $\pi$). The second term $C W_{2}(\tilde{\pi},Q_{n} )^2$ related to the convergence of the chain to the stationary distribution $\tilde{\pi}$, with the $C$ controlling the convergence rate. In imaging problems, the computational complexity is usually largely dominated by the second term in \eqref{eq:bbound} because of the dimensionality involved.}

\begin{figure}
	\centering
	\subfloat[$\varepsilon^2 = 10^{-1}$]{
		\label{subfig:rateDecayW2_epsilon_1eminus1}
		\includegraphics[scale=.41]{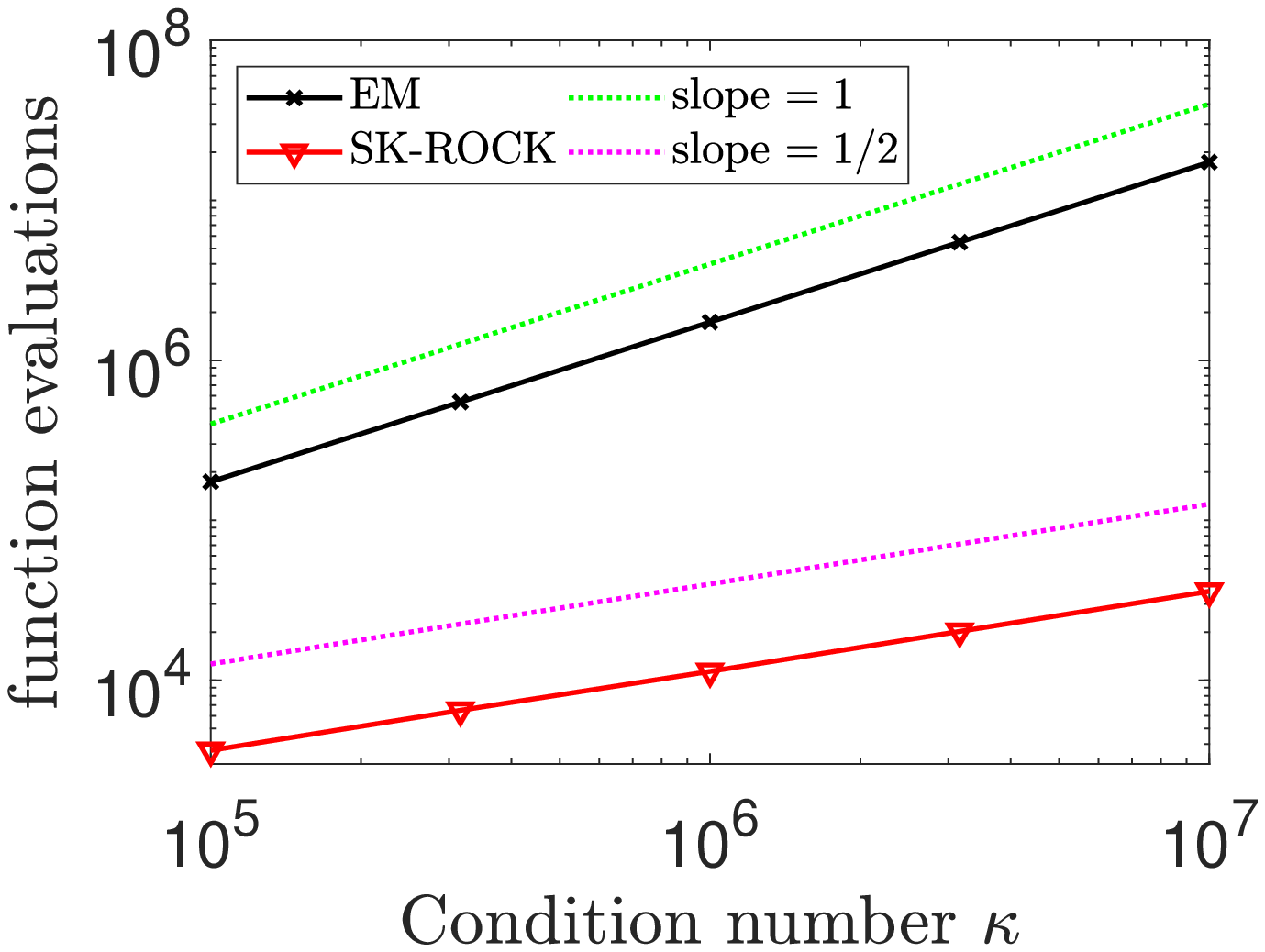}
	}
	\subfloat[$\varepsilon^2 = 10^{-3}$]{
		\label{subfig:rateDecayW2_epsilon_1eminus3}
		\includegraphics[scale=.41]{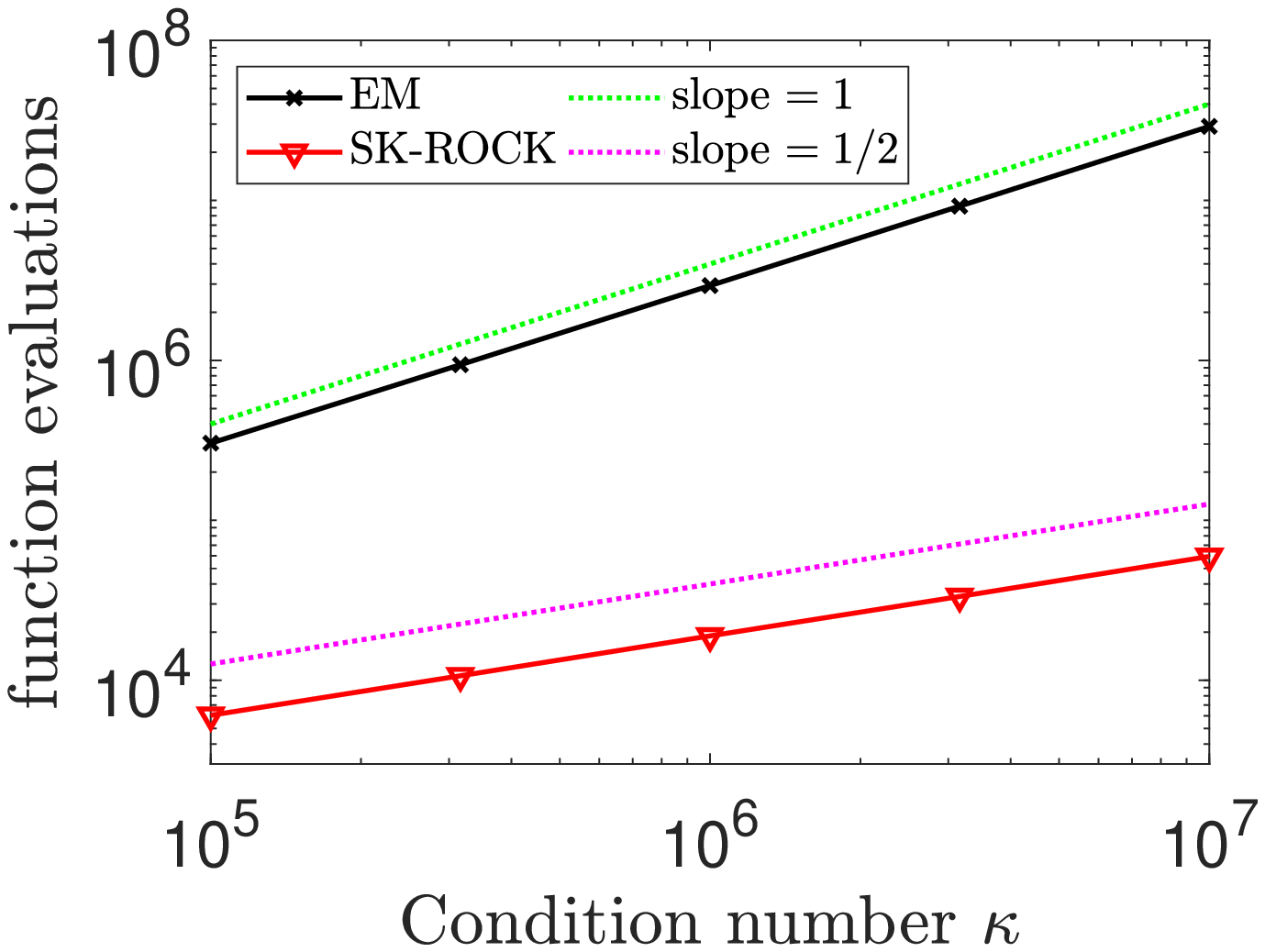}
	}
	\caption{Wasserstein distance bounds, Gaussian analysis: Minimum number of gradient evaluations of the EM and SK-ROCK methods in order to have {$W_2(P ; Q_n )^2 < \varepsilon^2 W_2(P ; Q_0)^2$}, given different condition numbers $\kappa$.}
	\label{fig:rateDecayW2}
\end{figure}

{For the case of the EM (MYULA) method it is known \cite{JMLR:v20:18-173} that, with a suitable choice of $\delta$, the number of gradient evaluations that one needs to take in order to achieve $W_2(\pi;Q_n ) < \varepsilon$ is of order $\mathcal{O}(\kappa)$, where we recall that $\kappa = {\sigma^{2}_{\text{max}}}/{\sigma^{2}_{\text{min}}}$ is the condition number of $\Sigma$. {For SK-ROCK, the number of gradient evaluations depends on the choice of $s$ and $\delta$. Our focus is on problems where $\kappa$ is large, where the optimal performance is achieved by minimising $C$ by setting the number of internals stages $s$ of each step to be
\begin{equation}
\label{eqn:bestNStagesSKROCK}
s = \left[ \sqrt{\frac{\eta}{2}(\kappa -1)} \; \right],
\end{equation}
with $\eta=0.05$, and 
\begin{equation} \label{eq:safety1}
\delta =\frac{\omega_0 - 1}{\ell_s \omega_1}, \quad  \ell_s=\frac{1}{\sigma^{2}_{\text{max}}},
\end{equation}
so that $C \approx (\sqrt{\kappa}-1)^{2}/(\sqrt{\kappa}+1)^{2}$ (see \cite{EVV18} and Appendix B for details). In that case, and under the assumption that $W_2(\pi,\tilde{\pi})\ll \epsilon$ so that $W_2(\pi;Q_n)$ is dominated by the term $C W_{2}(\tilde{\pi},Q_{n} )^2$ related to convergence to $\tilde{\pi}$, we observe that the number of gradient evaluations required to achieve $W_2(\pi;Q_n) < \varepsilon$ is of the order of $\mathcal{O}(\sqrt{\kappa})$ instead of $\mathcal{O}(\kappa)$, similarly to the behaviour of accelerated algorithms in optimization  \cite{CarmonYair2018AMfN}}. These convergence results are illustrated in Figure \ref{fig:rateDecayW2}, where we plot the number of gradient evaluations required to achieve $W_2(\pi;Q_n) < \varepsilon$ as a function of the conditioning number $\kappa$ for the EM method and for SK-ROCK, where $\pi$ is a $100$-dimensional Gaussian distribution with mean zero and covariance $\Sigma = \text{diag}(\sigma_1^2,...,\sigma_d^2)$, with decreasing diagonal elements uniformly spread between $\sigma_{1}=1$ and $\sigma_d = 1/\kappa$.}
	
{One can also simplify the non-asymptotic $W^2_2$ results of Appendix A to obtain non-asymptotic results for the estimation bias of the EM and SK-ROCK methods for the mean of Gaussian target densities (this is a weaker analysis than convergence in $W_2^2$). As in the case of the $W_2^2$ analysis, the number of gradient evaluations to attain a prescribed non-asymptotic bias for the mean is of order $\mathcal{O}(\sqrt{\kappa})$ for SK-ROCK, whereas it is of order $\mathcal{O}(\kappa)$ for the EM method. Both methods are asymptotically unbiased for the mean for Gaussian models.}
	
{We emphasise at this point that there are situations where one would not observe any acceleration by using SK-ROCK, namely situations in which a very accurate solution is required and the bound \eqref{eq:bbound} is dominated by the asymptotic bias term $W_{2}(\pi,\tilde{\pi})$. In that case, instead of using MYULA or SK-ROCK with a very small $\delta$, we would recommend using the P-MALA method described in \cite{proxMCMC}, which combines an EM approximation with a MH correction.}


\subsection{Mean-square stability analysis}\label{sec:meanSqStab_SROCK}
\begin{figure}
\centering
\subfloat[EM (dark grey) and SK-ROCK (light grey) mean-square stability regions.]{
\label{subfig:meanSquareSKROCK_s_10_EM}
\includegraphics[scale=.6]{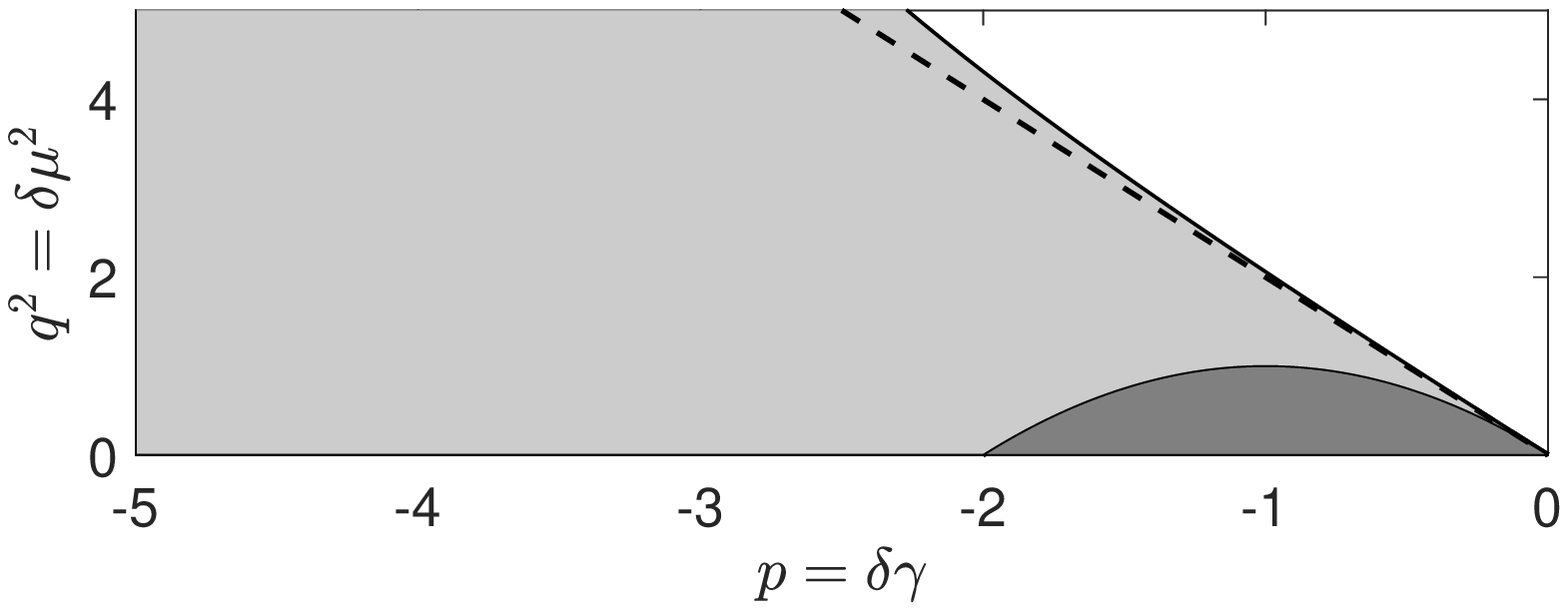}
} \\
\subfloat[SK-ROCK mean-square stability region ($s=10$, $\eta=0.05$)]{
\label{subfig:meanSquareSKROCK_s_10}
\includegraphics[scale=.6]{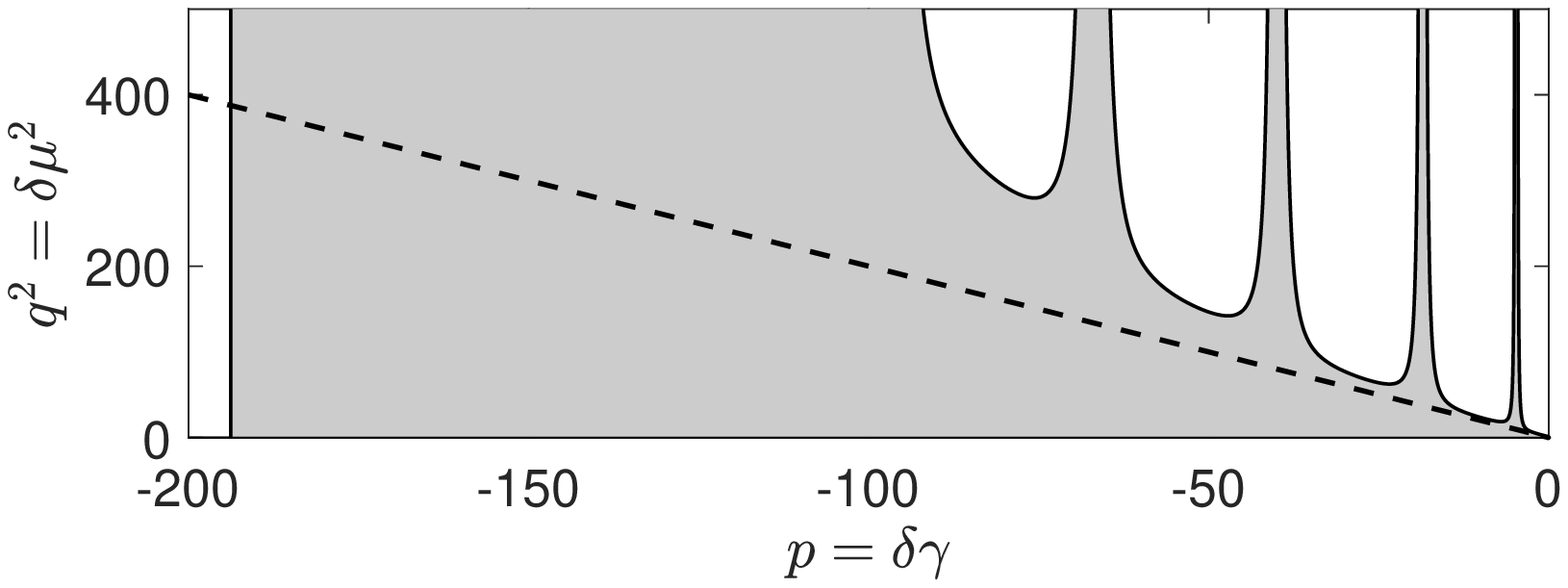}
} \\
\caption{Mean-square stability domains for {\normalfont(a)} EM and and {\normalfont(b)} SK-ROCK (with $s=10$) in the $p-q^2$ plane. The dashed line represents the upper boundary of the true mean-square stability domain.}
\label{fig:meanSquare_EM_SROCK_SKROCK}
\end{figure}

We conclude this section by {discussing the mean-square stability properties of SK-ROCK and the EM method}. In particular, we consider the following test equation that is widely used in the numerical analysis literature {\cite{DH00,DH00a}} to benchmark SDE solvers
\begin{equation}\label{eqn:testEqSDE}
dX(t)=\gamma X(t) dt + \mu X(t) dW(t), \; \; X(0)=1,
\end{equation}
where $\gamma$, $\mu\in\mathbb{R}$, which has the solution  $X(t)=\exp  [(\gamma - 1/2 \mu^2)t +\mu W(t)]$. It is easy to show using Ito calculus that when $2\gamma + \mu^2 < 0$
\[
\lim_{t\rightarrow\infty} \mathbb{E}( \left| X(t) \right|^2 )=0.
\]
We want to understand for what range of  time-steps $\delta$ would a numerical discretisation $X_{n}$ of \eqref{eqn:testEqSDE} behave in a similar manner as $n \rightarrow \infty$, \emph{i.e.} $\mathbb{E}(|X_{n}|^{2}) \rightarrow 0$. In the case of EM one has that 
\[
X_{n+1}=X_{n}+\delta \gamma X_{n}+\sqrt{\delta} \mu  X_{n} Z_{n+1},
 \]
 and hence
 \[
 \mathbb{E}(|X_{n+1}|^{2})=R(p,q) \mathbb{E}(|X_{n}|^{2}), \quad R(p,q)=(1+p)^{2}+q^{2}, \quad p=\delta \gamma, q=\sqrt{\delta} \mu. 
 \]
{In order to have} $\mathbb{E}(|X_{n}|^{2}) \rightarrow 0$ one needs that 
$R(p,q)<1$. We visualise the values of admissible $p,q$ for the EM method in Figure 
\ref{fig:meanSquare_EM_SROCK_SKROCK}(a), where we can see that there is only a very small portion of the true mean-square stability domain  ({$2p+q^{2}<0$}) covered by it (anything on the left hand side of the dotted line in Figure  \ref{fig:meanSquare_EM_SROCK_SKROCK}(a)-(b) belongs to the true stability domain). {This} implies that when one or both  of the parameters $\gamma, \mu$ are large one needs to choose a very small $\delta$ in order to be stable (for example when $\mu=0$ one recovers the stability condition $\delta  < -2\gamma^{-1}$ for the Langevin SDE). In the case of SK-ROCK one has that 
\[
R(p,q)=R_1(p)^2 + R_2(p)^2 q^2 ,
\]
where $R_1$ and $R_2$ are given by \eqref{eqn:R1R2SKROCK}.

{Similarly to the case of the EM method, we now plot} the mean-square stability 
domain of SK-ROCK in Figure 
\ref{fig:meanSquare_EM_SROCK_SKROCK}(b). As we can see, a 
significantly larger portion of the true mean-square stability domain is now 
covered when compared to the EM method. {One can show, using the properties of Chebyshev polynomials \cite{abdulleSKROCK},   that for SK-ROCK  the coverage of the mean-square stability domain increases quadratically in $s$; i.e., that if $(p,q) \in \{ 2p+q^{2}<0 \  \cap  \ p < C(\eta) s^{2}\} $ then $R(p,q)<1$ for the SK-ROCK method.  In contrast, if for comparison one would consider $s$-steps of the EM method, the corresponding coverage of the mean-square stability domain would be  linear in $s$.} This means that for the same number of gradient evaluations $s$, one can choose a much larger time-step $\delta$ for SK-ROCK and still integrate equation \eqref{eqn:testEqSDE} in a stable manner. {The spikes observed in \ref{fig:meanSquare_EM_SROCK_SKROCK}(b) at specific values of $p$ correspond to roots of the polynomial $R_2(p)$ defined in \eqref{eqn:R1R2SKROCK}; these are determined by the values of $s$ and $\eta$, and by the roots of the Chebyshev polynomial of second kind $U_{s-1}$.}


\section{Numerical experiments}
\label{sec:experiments}
In this section we demonstrate the proposed SK-ROCK proximal MCMC methodology with a range of numerical experiments related to image deconvolution, tomographic reconstruction, and hyper-spectral unmixing. We have selected these experiments to represent a wide variety of configurations in terms of ill-posedness and ill-conditioning, strict and strong log-concavity, and dimensionality of $y$ and $x$. {Following our previous recommendation, in the experiments related to image deconvolution and hyper-spectral unmixing the model is strongly log-concave so we use $s=15$, whereas for the tomography experiment we use $s=10$.} We report comparisons with the MYULA method \cite{pereyraMYULA} to highlight the benefits of using the SK-ROCK discretization as opposed to the conventional EM discretization used in Langevin and Hamiltonian algorithms \cite{mcmcAlgImaging}, and because MYULA underpins other proximal MCMC algorithms such as the auxiliary Gibbs sampler of \cite{Vono_TSP_2019}. 

{To make the comparisons fair, in all experiments we use the same number of gradient (and proximal operator) evaluations for MYULA and SK-ROCK, and compare their computational efficiency in several ways (the efficiency of an MCMC method is not an absolute quantity as it depends on the estimator considered). Because our aim is to illustrate the performance of SK-ROCK in Bayesian imaging problems, here we use the MYULA and SK-ROCK samples to compute the following quantities: 1) the minimum mean square error solution given by the posterior mean $\textrm{E}(x|y)$, which is a classic image point estimator; 2) the marginal posterior variances or standard deviations for the image pixels, which provide an indication of the performance of the methods in uncertainty quantification tasks; 3) the effective sample size (ESS) of the fastest mixing component of the chain, calculated in stationarity\footnote{Recall that $\textrm{ESS} = n \{1 + 2\sum_{k} \rho(k)\}^{-1}$, where $n$ is the total number of samples and $\sum_{k} \rho(k)$ is the sum of the $K$ monotone sample auto-correlations which we estimated with the initial monotone sequence estimator \cite{geyer92}.}; and 4) the ESS of the slowest mixing component of the chain, also calculated in stationarity. These fast and slow components correspond to the one-dimensional subspaces where the Markov chains achieve their highest and lowest convergence rates respectively, and that we have identified via an estimate of the first and last eigenvectors of the samples posterior covariance. We choose to report ESS values because these are intuitive quantities that are directly related to the variance of the Monte Carlo estimators, and hence provide an indication of the accuracy of the methods, up to estimation bias\footnote{Note that the computation of ESS values is well-posed because $p(x|y)$ is log-concave. If $p(x|y)$ were heavy-tailed or multi-modal then we would need to consider robust efficiency indicators \cite{Vehtari2019}.}.}
	
In addition to reporting estimates, we use autocorrelation plots to visually compare the convergence properties of both methods (again, we report the autocorrelation function for the fastest and the slowest components of the Markov chains). We also show the evolution of the estimation MSE across iterations, and display the estimates of the marginal (pixelwise) standard deviations. {These latter are useful for illustrating the differences in the performance of the methods, as second order moments are more difficult to estimate by Monte Carlo integration than the posterior mean.}
	
{Notice that because the methods are compared at equal computational budget they do not produce the same number of samples, as their complexity per iteration is different. More precisely, if the MYULA chain has $n$-samples, then the SK-ROCK chain has only $n/s$ samples, which is considerably lower. However, experiments show that SK-ROCK usually delivers higher ESS values because of its superior convergence properties. Similarly, to make the comparison of autocorrelation plots fair with regards to computational complexity, in all autocorrelation plots we apply a 1-in-$s$ thinning to the MYULA chain to artificially boost its autocorrelation function decay rate by a factor of $s$.}

\subsection{One dimensional distributions}
\label{sub:OnedimExp}
We start our numerical experiments by studying two simple one dimensional distributions, namely the Laplace distribution and the uniform distribution in $[-1,1]$, for which we can also perform computations exactly. Since both of these distributions are not Lipschitz differentiable we employ the corresponding Moreau-Yosida approximation using $\lambda=10^{-5}$ to bring $\pi_\lambda$ very close to $\pi$ and deliver a good approximation. This implies that the largest stepsize $\delta$ that can be used for MYULA is $2 \times 10^{-5}$, which is dramatically small. We set $\delta=10^{-5}$ for MYULA and run the corresponding chain for $n=15 \times 10^{6}$ iterations to create a situation where MYULA struggles to deliver a good approximation and that highlights the superior performance of SK-ROCK.

For SK-ROCK we use $s=15$ and set $\delta$ {as it is explained in Algorithm \ref{alg:SKROCK}}. Notice that we choose the (regularised) Laplace and the uniform distributions to illustrate the performance of the methods in {two different scenarios}: the regularised Laplace distribution is strongly log-concave near the mode and only strictly log-concave in the tails, which is problematic for the Langevin diffusion because the gradient remains constant as $|x|$ grows, whereas the regularised uniform distribution is flat over $[-1,1]$ and hence has most of its mass in regions where the gradient is zero, and then strongly log-concave in the tails.

\begin{figure}
\centering
\subfloat[MYULA]{
\label{subfig:histMYULALaplace}
\includegraphics[scale=.269]{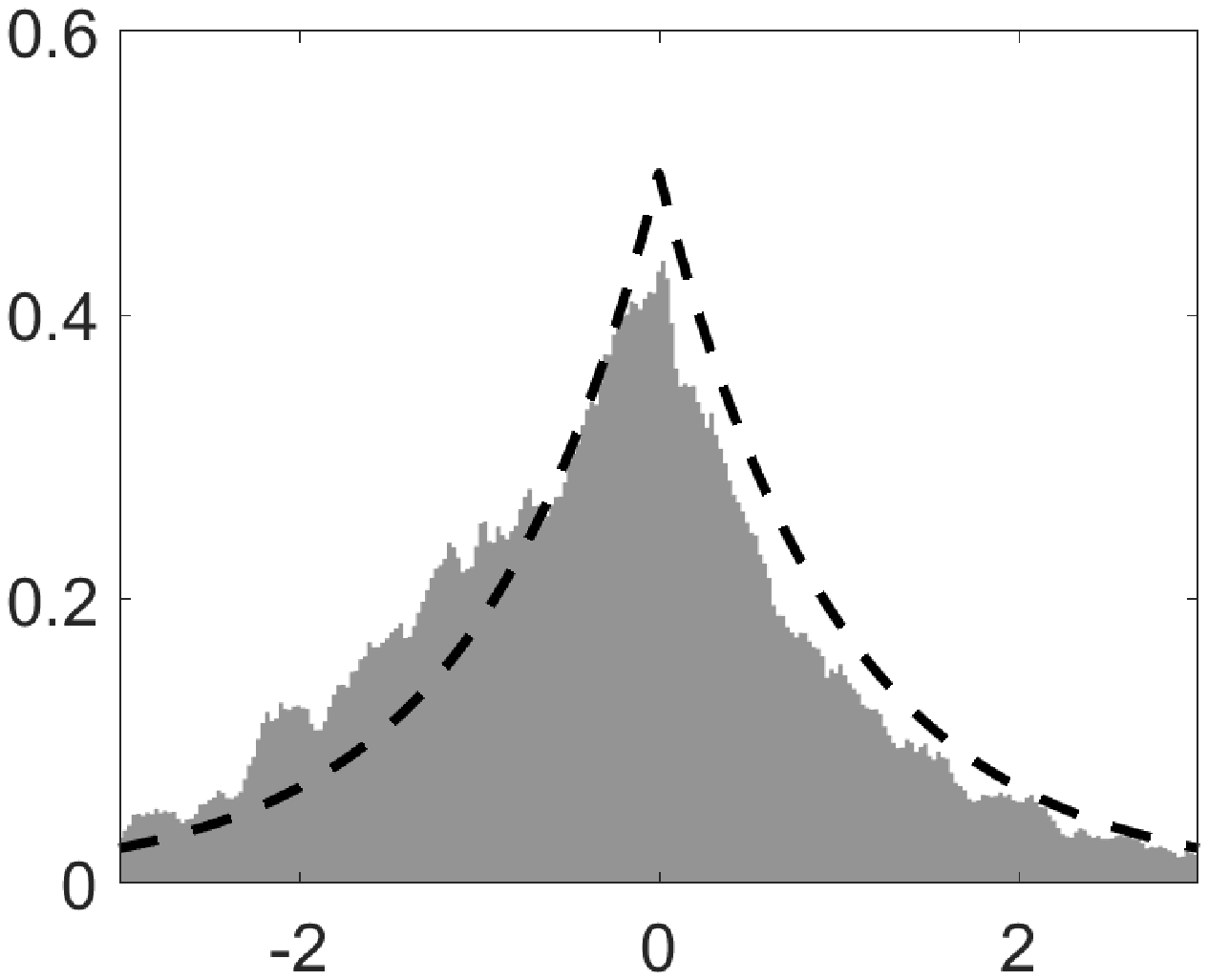}
} 
\subfloat[SK-ROCK, $s=15$]{
\label{subfig:histSKROCKLaplace_s15}
\includegraphics[scale=.269]{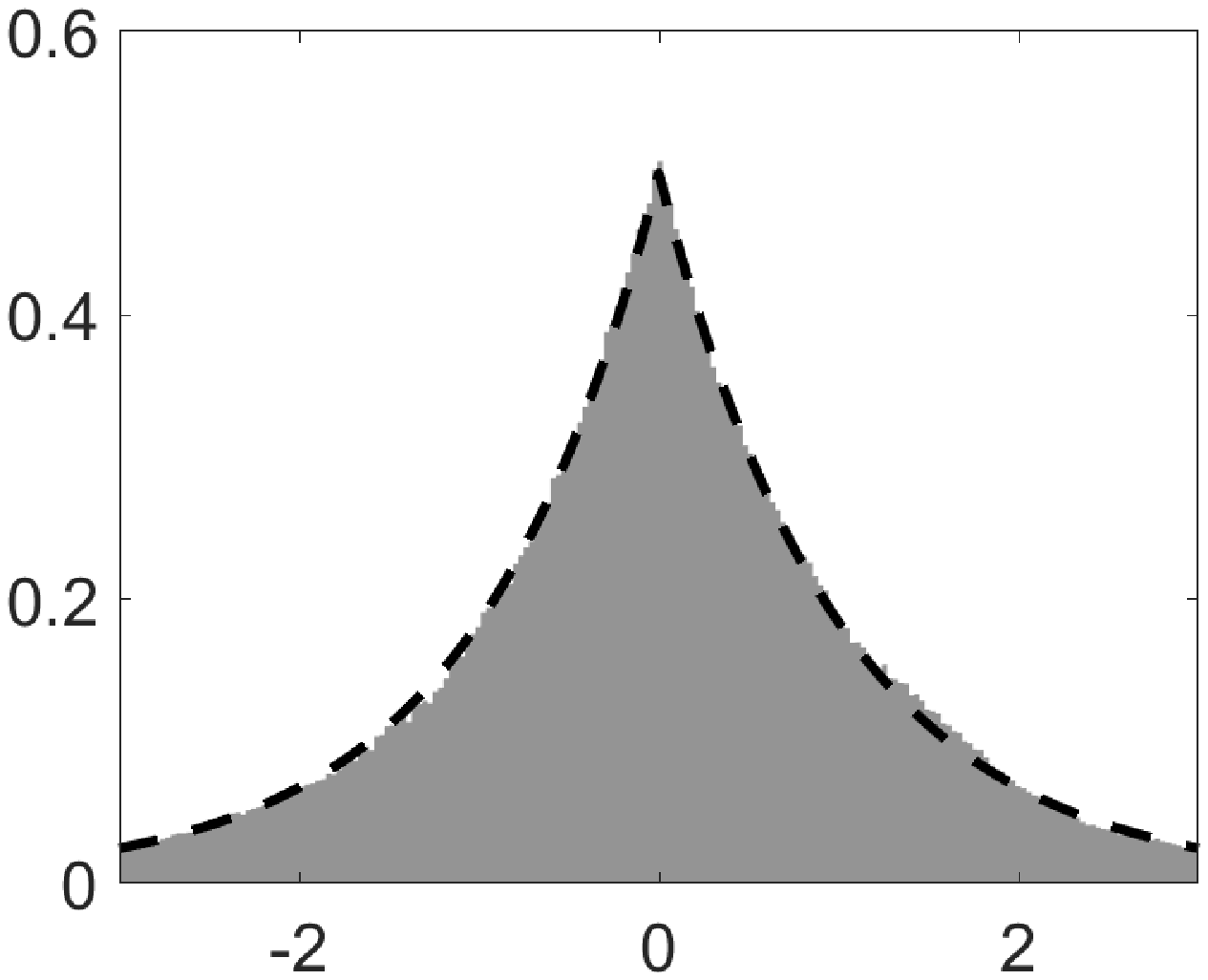}
}
\subfloat[ACF]{
\label{subfig:acfLaplace_s15}
\includegraphics[scale=.269]{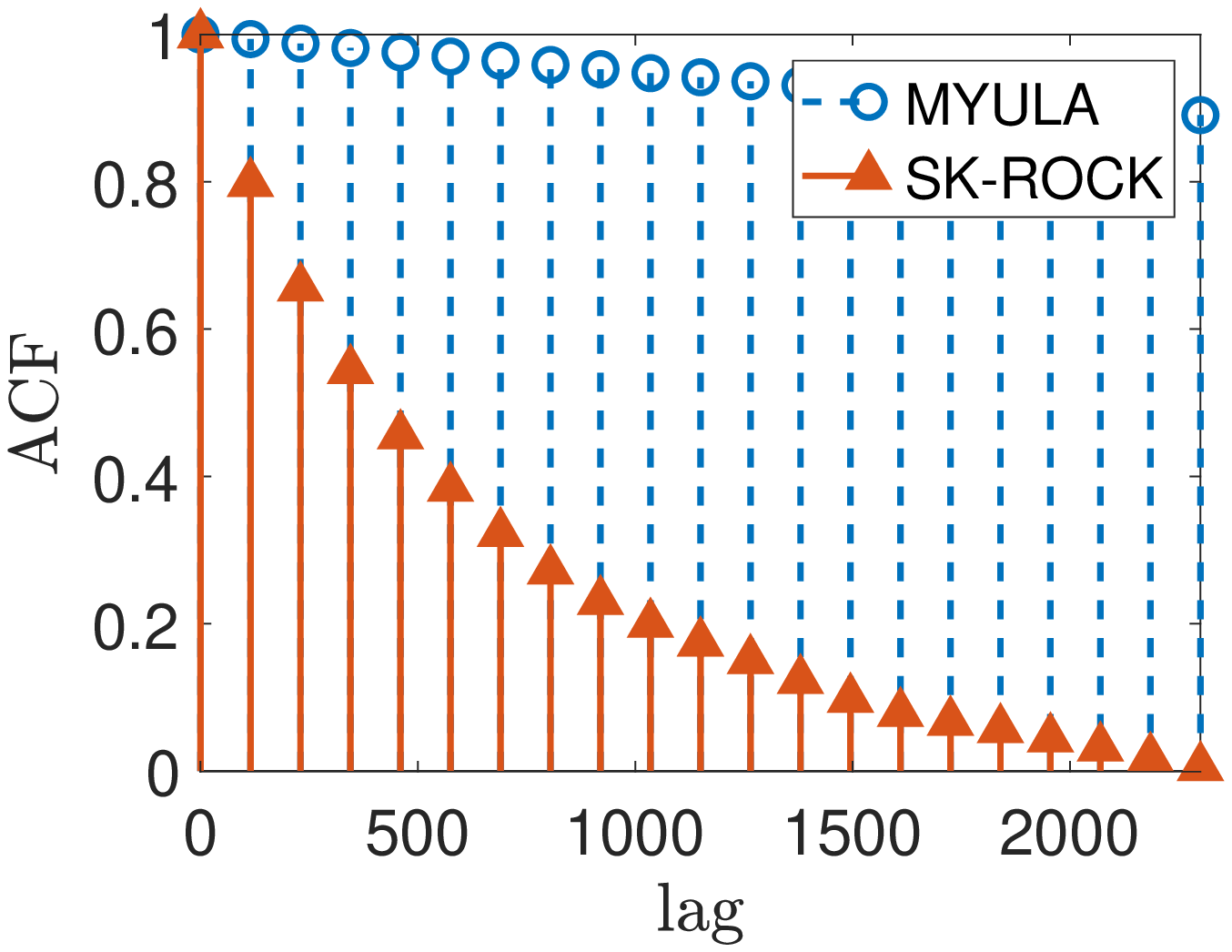}
}
\caption{One-dimensional Laplace distribution: Histograms computed with {\normalfont(a)} $15 \times 10^6$ samples generated by MYULA and {\normalfont(b)} $15 \times 10^6 / s$ samples generated by SK-ROCK from the approximated Laplace distribution, using an approximation parameter $\lambda = 10^{-5}$ and {$s=15$}  for the SK-ROCK method. {\normalfont(c)} Autocorrelation functions of the samples.}
\label{fig:histLaplaceDist}
\end{figure}

\begin{figure}
\centering
\subfloat[MYULA]{
\label{subfig:histMYULAUniform}
\includegraphics[scale=.269]{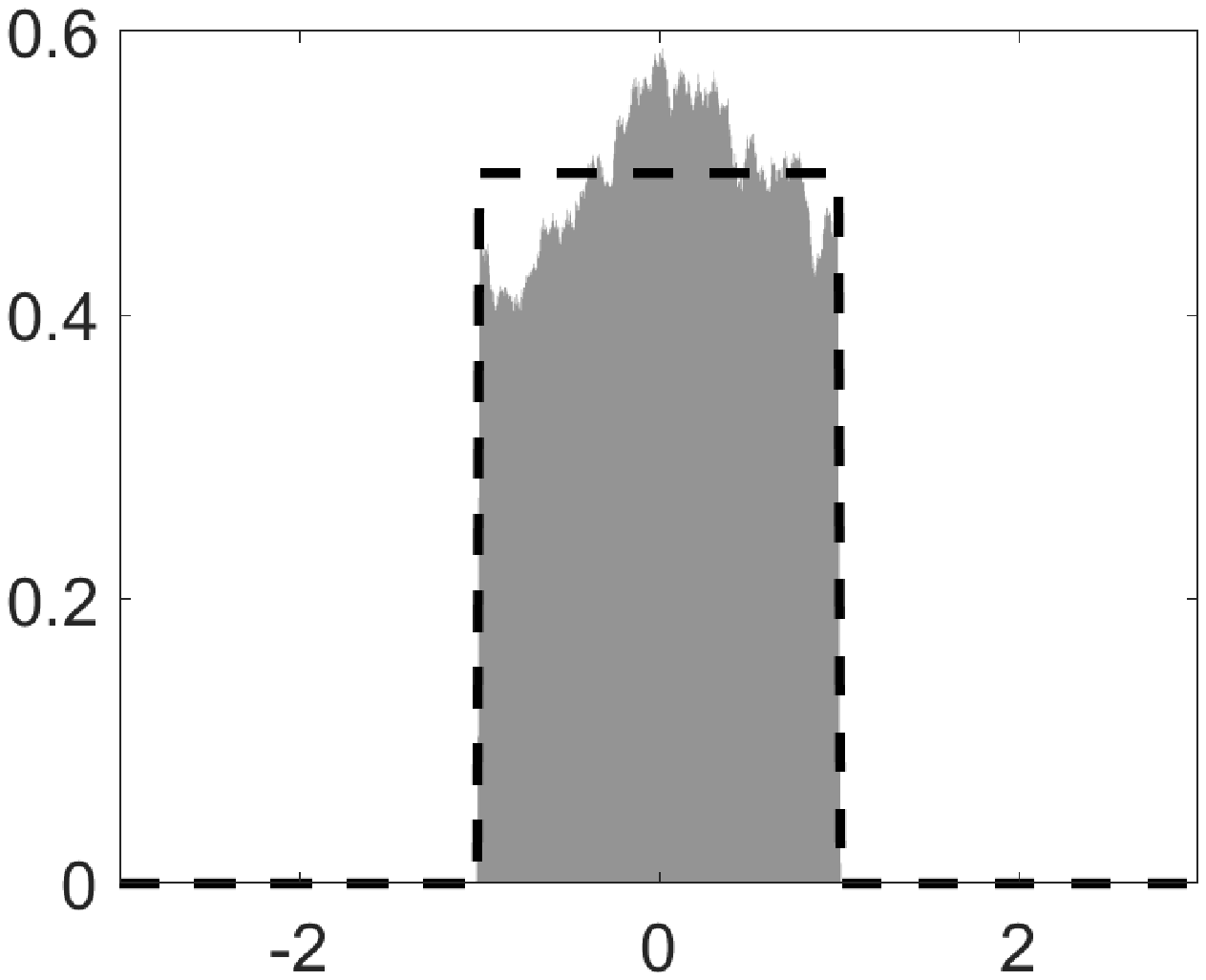}
} 
\subfloat[SK-ROCK, $s=15$]{
\label{subfig:histSKROCKUniform_s15}
\includegraphics[scale=.269]{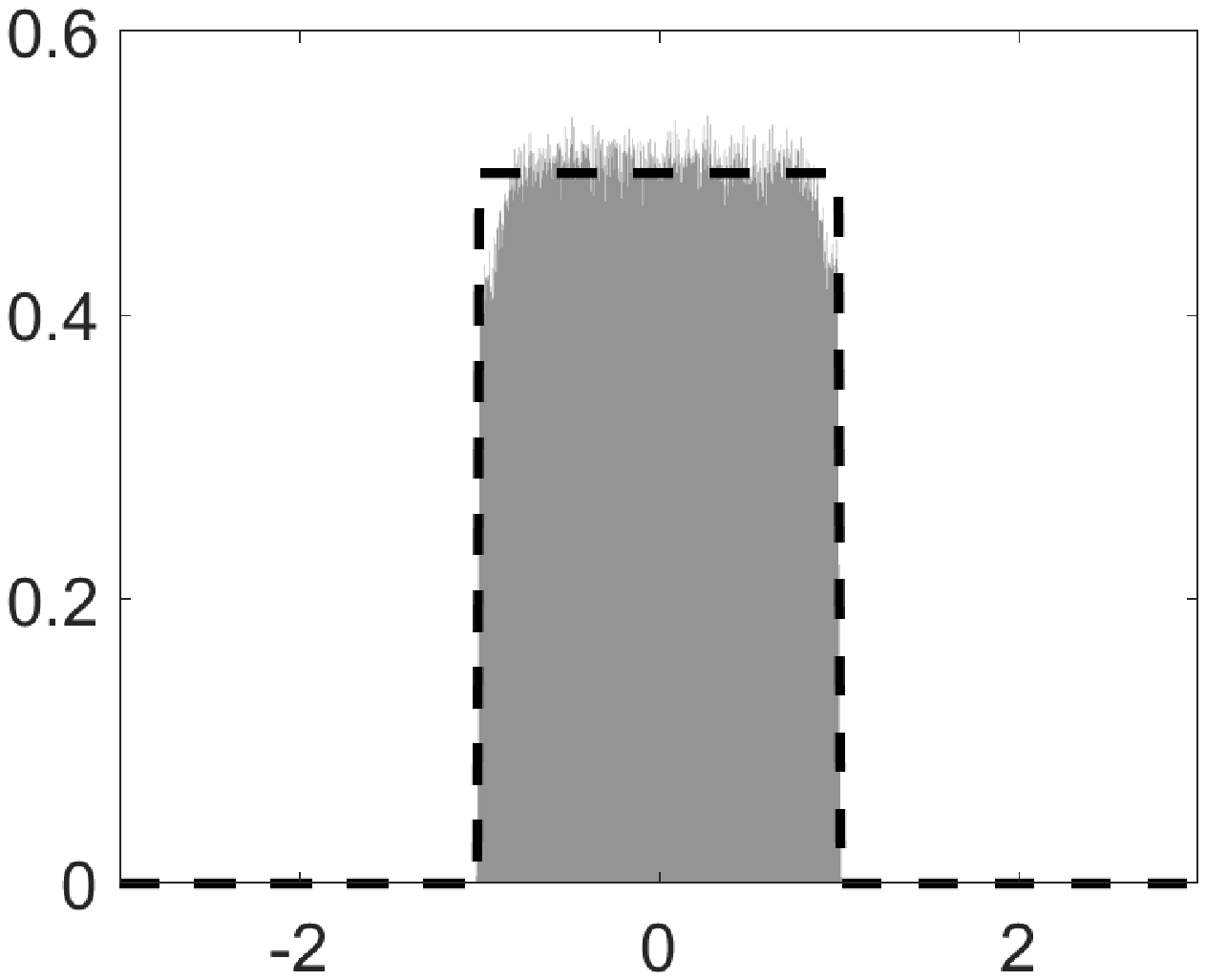}
}
\subfloat[ACF]{
\label{subfig:acfuniform_s15}
\includegraphics[scale=.269]{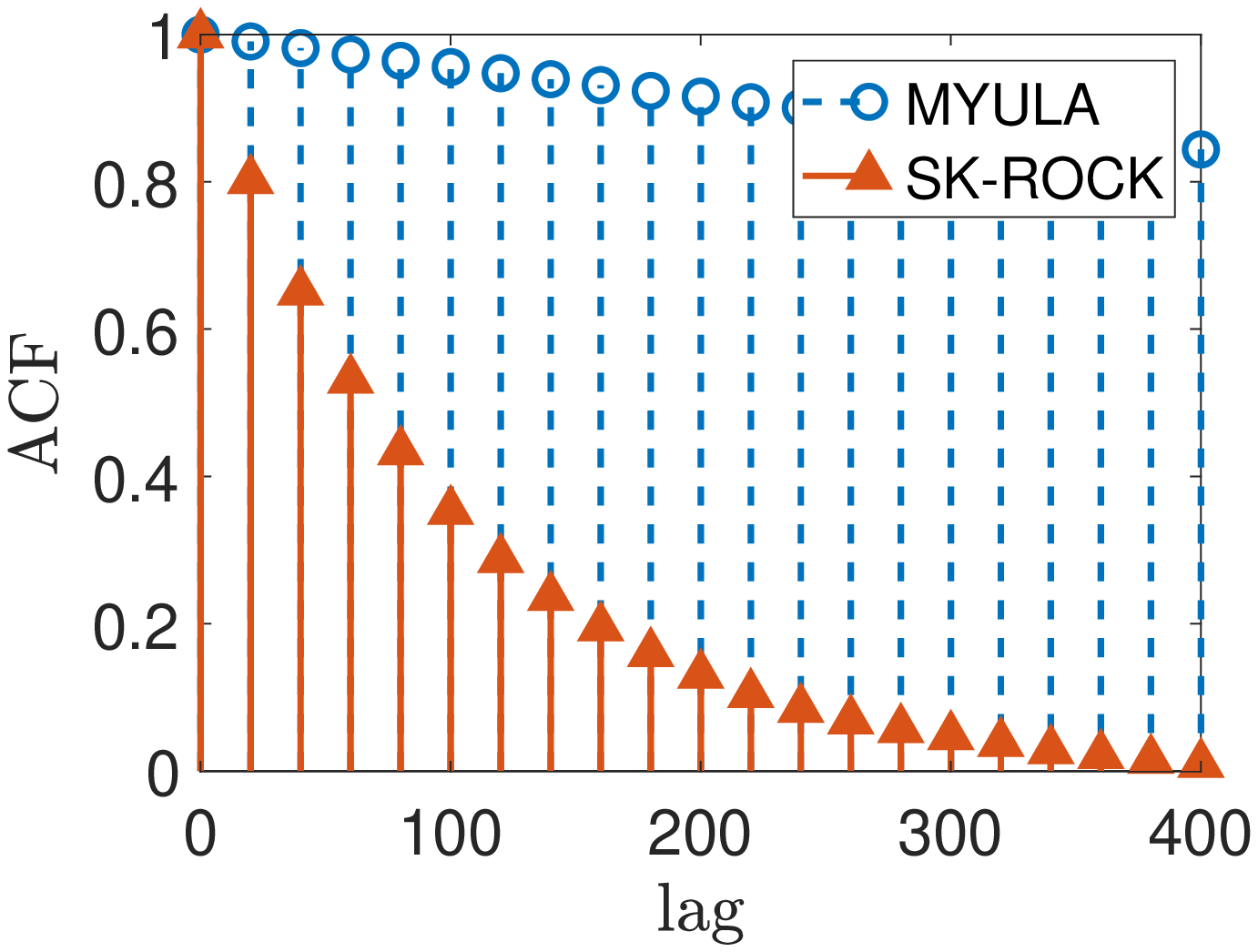}
}
\caption{One-dimensional uniform distribution: Histograms computed with {\normalfont(a)} $15 \times 10^6$ samples generated by MYULA and {\normalfont(b)} $15 \times 10^6 / s$ samples generated by SK-ROCK from the approximated uniform distribution, using an approximation parameter $\lambda = 10^{-5}$ and $s=15$ for the SK-ROCK method. {\normalfont(c)} Autocorrelation functions of the samples.}
\label{fig:histUniformDist}
\end{figure}

Figures \ref{fig:histLaplaceDist} and \ref{fig:histUniformDist} display the histogram approximations of the distributions obtained with the two methods, as well as the autocorrelation functions of the generated Markov chains. Observe that in both cases SK-ROCK significantly outperforms MYULA, which struggles to deliver a good approximation {due to} the stepsize limitation and the limited number of iterations (this phenomenon is particularly clearly captured by the difference in decay speed in the autocorrelation plots). These results are quantitatively summarised in Tables \ref{tab:stepsizeOneDimLaplace} and \ref{tab:stepsizeOneDimUniform} respectively, where we highlight that SK-ROCK delivers an ESS that is over $25$ times larger than MYULA, while also achieving higher accuracy as measured by the Kullback-Leibler (KL) divergence between the empirical distribution and $\pi_\lambda$. For completeness, we also report the results using SK-ROCK with $s=10$.

\begin{table}
{\footnotesize
  \caption{Values of the stepsize $\delta$, effective sample sizes (ESS) and KL-divergence of the EM and SK-ROCK algorithms {for the one dimensional Laplace distribution}.}  \label{tab:stepsizeOneDimLaplace}
\begin{center}
  \begin{tabular}{|c|c|c|c|c|c|} \hline
   \bf Stages $s$ & \bf Method & \bf Stepsize $\delta$ & \bf ESS & \bf KL-Divergence & \bf Speed-up  \\ \hline
    - & MYULA & $1.0 \times 10^{-5}$ & $3.6 \times 10^1$ & $4.8 \times 10^{-2}$ & -\\ 
    $s=10$ 	
    		& SK-ROCK & $1.7 \times 10^{-3}$ & $6.0 \times 10^2$ & $1.4 \times 10^{-2}$ & 16.67 \\ 
    $s=15$ 
    		& SK-ROCK & $4.0 \times 10^{-3}$ & $9.5 \times 10^{2}$ & $1.0 \times 10^{-2}$  & 26.39\\
		\hline
  \end{tabular}
\end{center}
}
\end{table}

It is worth emphasising at this point that we could improve the ESS performance of both methods by increasing the value of $\lambda$, at the expense of some additional bias. In the case of the uniform distribution this would lead to a considerable number of samples outside the true support $[-1,1]$.

\begin{table}
{\footnotesize
  \caption{Values of the stepsize $\delta$, effective sample sizes (ESS) and KL-divergence of the EM and SK-ROCK algorithms {for the one dimensional uniform distribution.}}   \label{tab:stepsizeOneDimUniform}
\begin{center}
  \begin{tabular}{|c|c|c|c|c|c|} \hline
   \bf  Stages $s$ & \bf Method & \bf Stepsize $\delta$ & \bf ESS & \bf KL-Divergence & \bf Speed-up \\ \hline
    - & MYULA & $1.0 \times 10^{-5}$ & $1.7 \times 10^2$ & $1.3 \times 10^{-2}$ & - \\ 
    $s=10$ 	
    		& SK-ROCK & $1.7 \times 10^{-3}$ & $3.4 \times 10^3$ & $3.2 \times 10^{-2}$ & 20 \\ 
    $s=15$ 	
    		& SK-ROCK & $4.0 \times 10^{-3}$ & $4.9 \times 10^3$ & $3.9 \times 10^{-2}$ & 28.82 \\
\hline
  \end{tabular}
\end{center}
}
\end{table}

\subsection{Image deconvolution with total-variation prior}
\label{sub:cameramanExp}
We now consider a non-blind image deconvolution problem, where we seek to recover a high-resolution image $x\in \mathbb{R}^d$ from a blurred and noisy observation $y=Hx + \epsilon$, where $H$ is a known blur operator and $\epsilon\sim\mathcal{N}(0,\sigma^2 \mathbb{I}_d)$. This problem is ill-conditioned i.e., $H$ is nearly singular, thus yielding highly noise-sensitive solutions. To make the estimation problem well posed, we use a total-variation norm prior that promotes solutions with spatial regularity. The resulting posterior distribution is given by
\begin{equation}\label{eqn:posteriorDist_ImagingExp_cam}
p(x|y) \propto \text{exp} \left( -\| y-Hx \|^2 / 2\sigma^2 - \beta TV(x) \right),
\end{equation}
where $TV(x)$ represents the total-variation pseudo-norm \cite{tvprior1,chambolleAlgorithm}, and $\sigma, \beta \in \mathbb{R}^+$ are model hyper-parameters that we assume fixed (in this experiment we use $\beta=0.047$, determined using the method of \cite{Vidal2019}). 

\begin{figure}
\centering
\subfloat[true image $x$]{
\label{subfig:cameraman_exact_Image_snr_42}
\includegraphics[scale=.416]{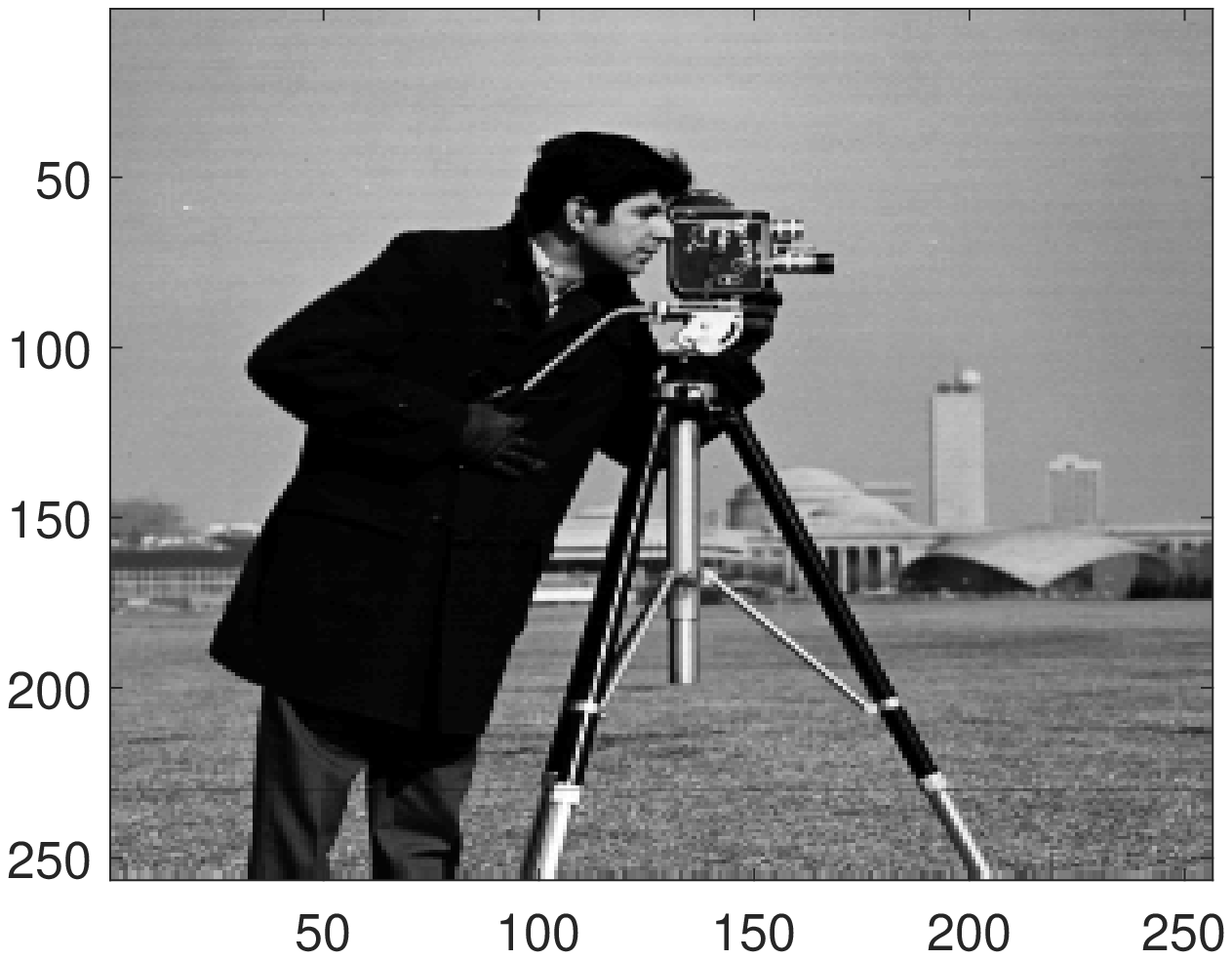}
}
\subfloat[noisy and blurred observation $y$]{
\label{subfig:cameraman_blurred_Image_snr_42}
\includegraphics[scale=.416]{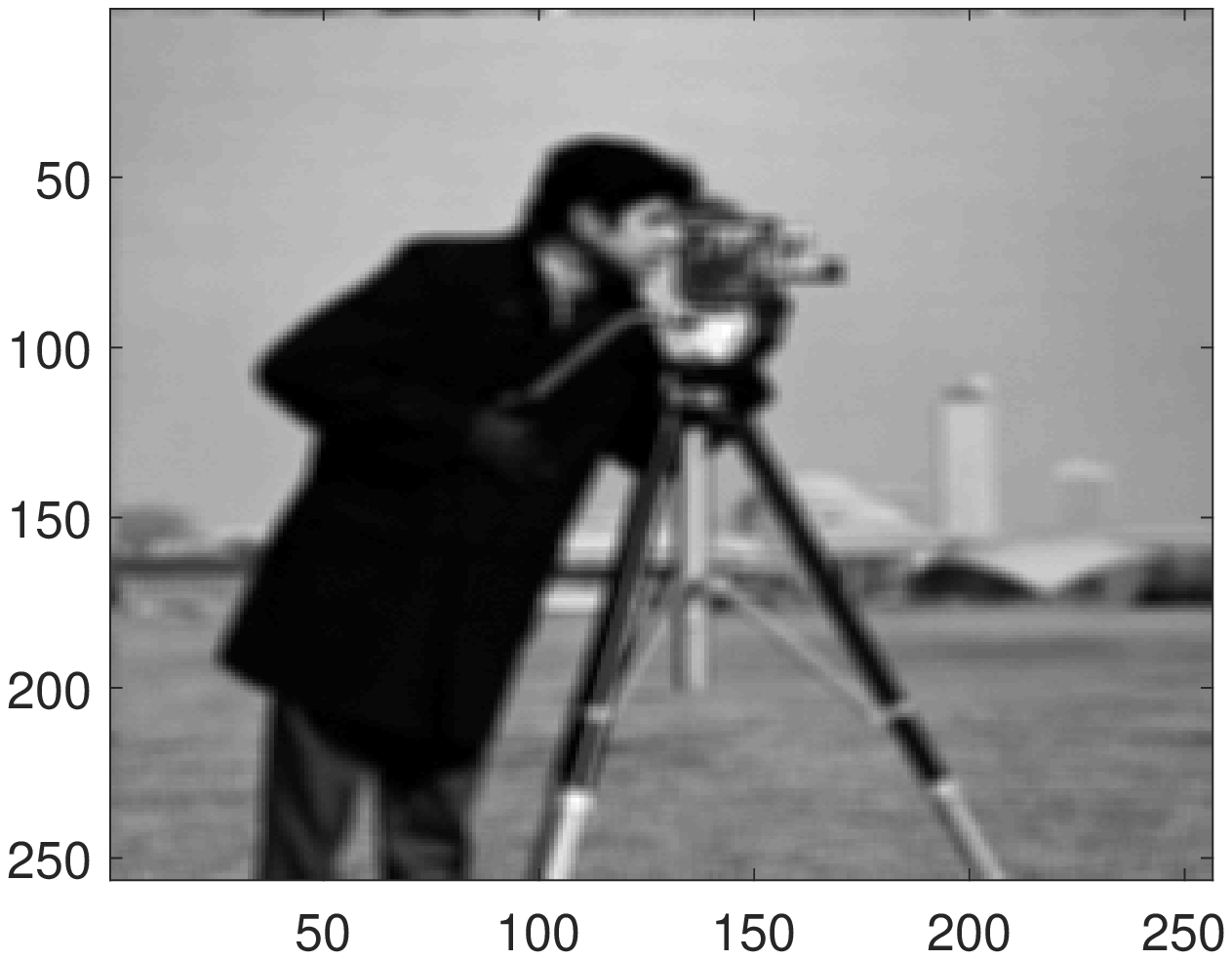}
}\\
\subfloat[MYULA: posterior mean]{
\label{subfig:cameraman_mean_samples_MYULA}
\includegraphics[scale=.416]{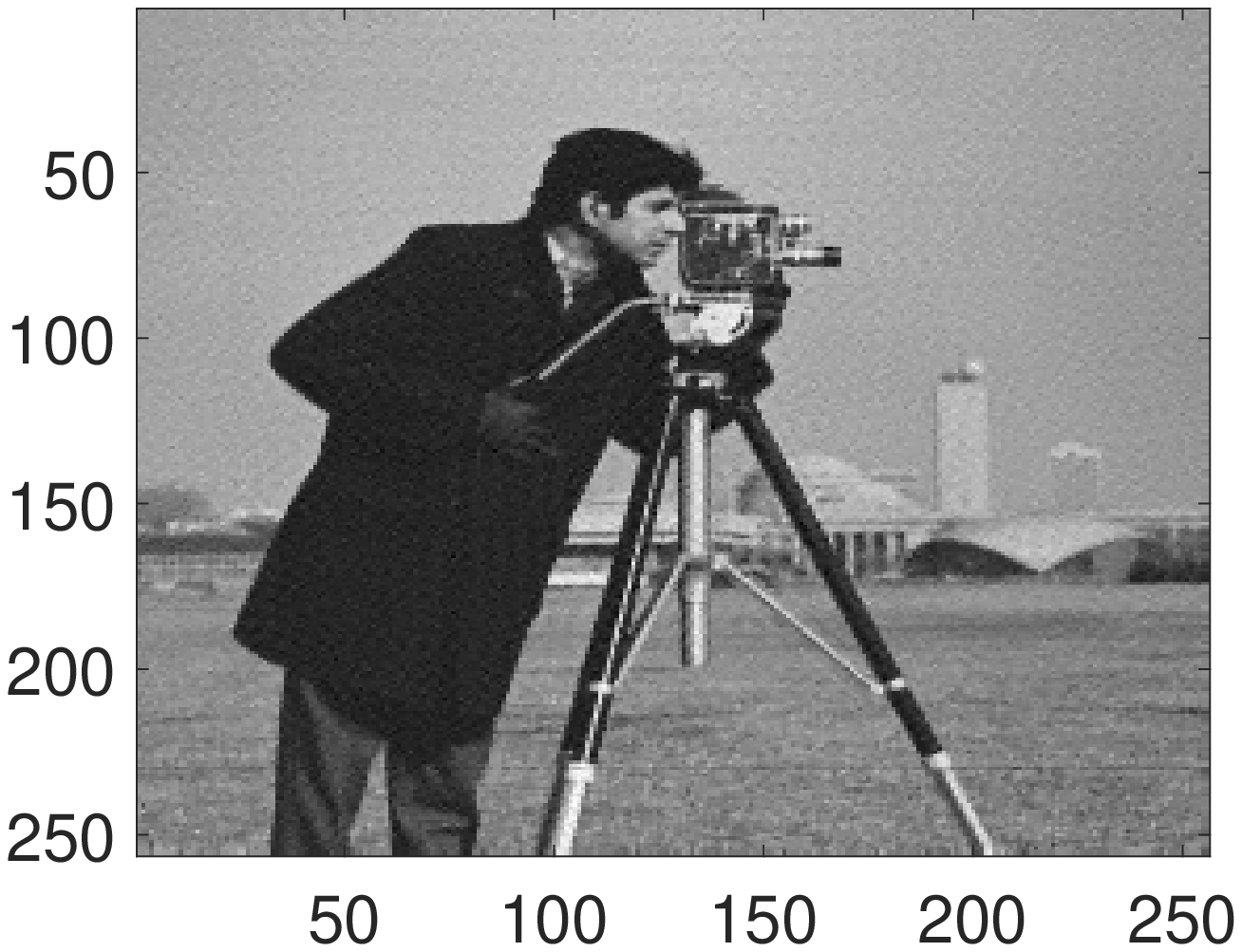}
}
\subfloat[SK-ROCK: posterior mean ($s=15$)]{
\label{subfig:cameraman_mean_samples_SKROCK_s15}
\includegraphics[scale=.416]{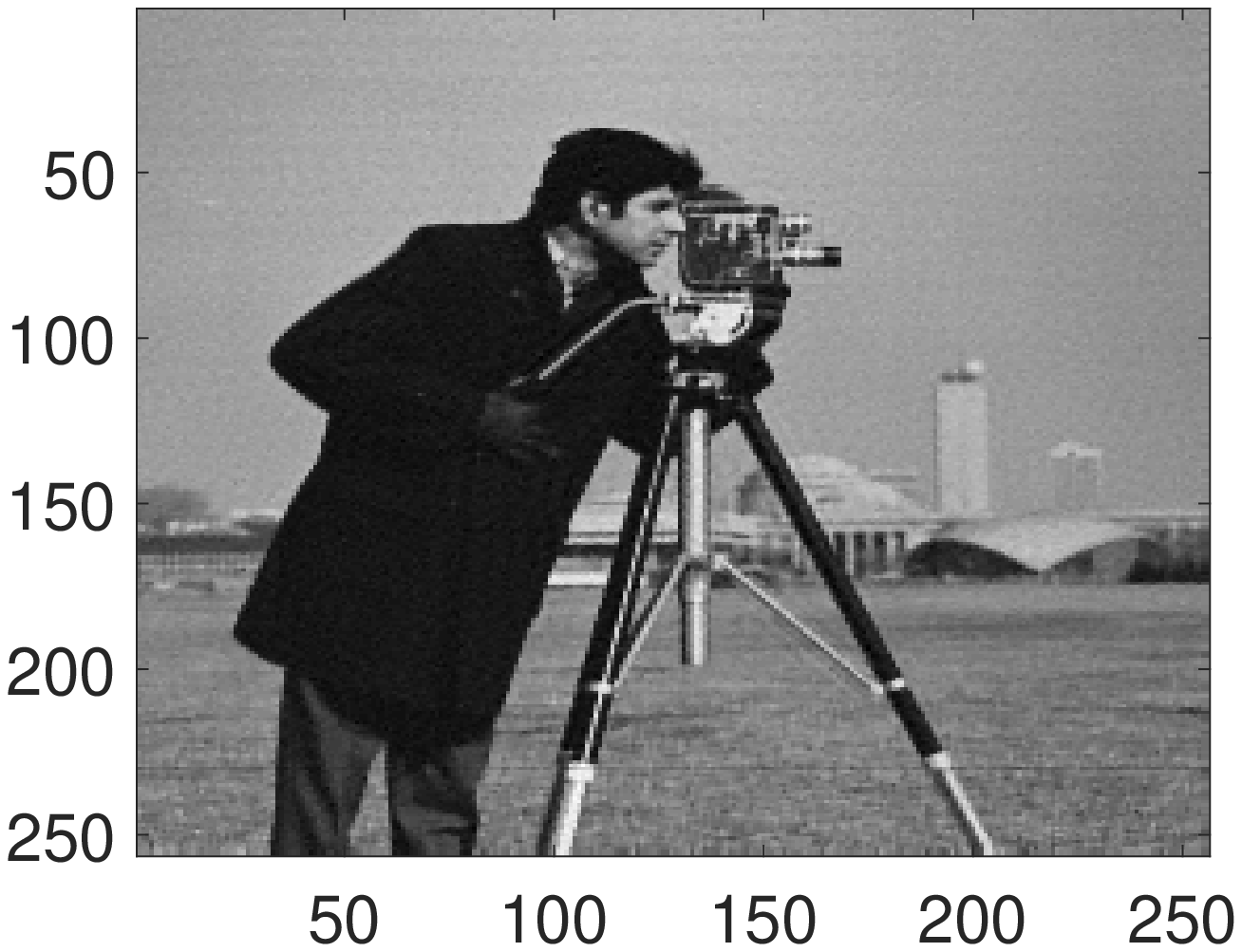}
}\\
\subfloat[MYULA: standard deviation]{
\label{subfig:cameraman_std_samples_MYULA}
\includegraphics[scale=.416]{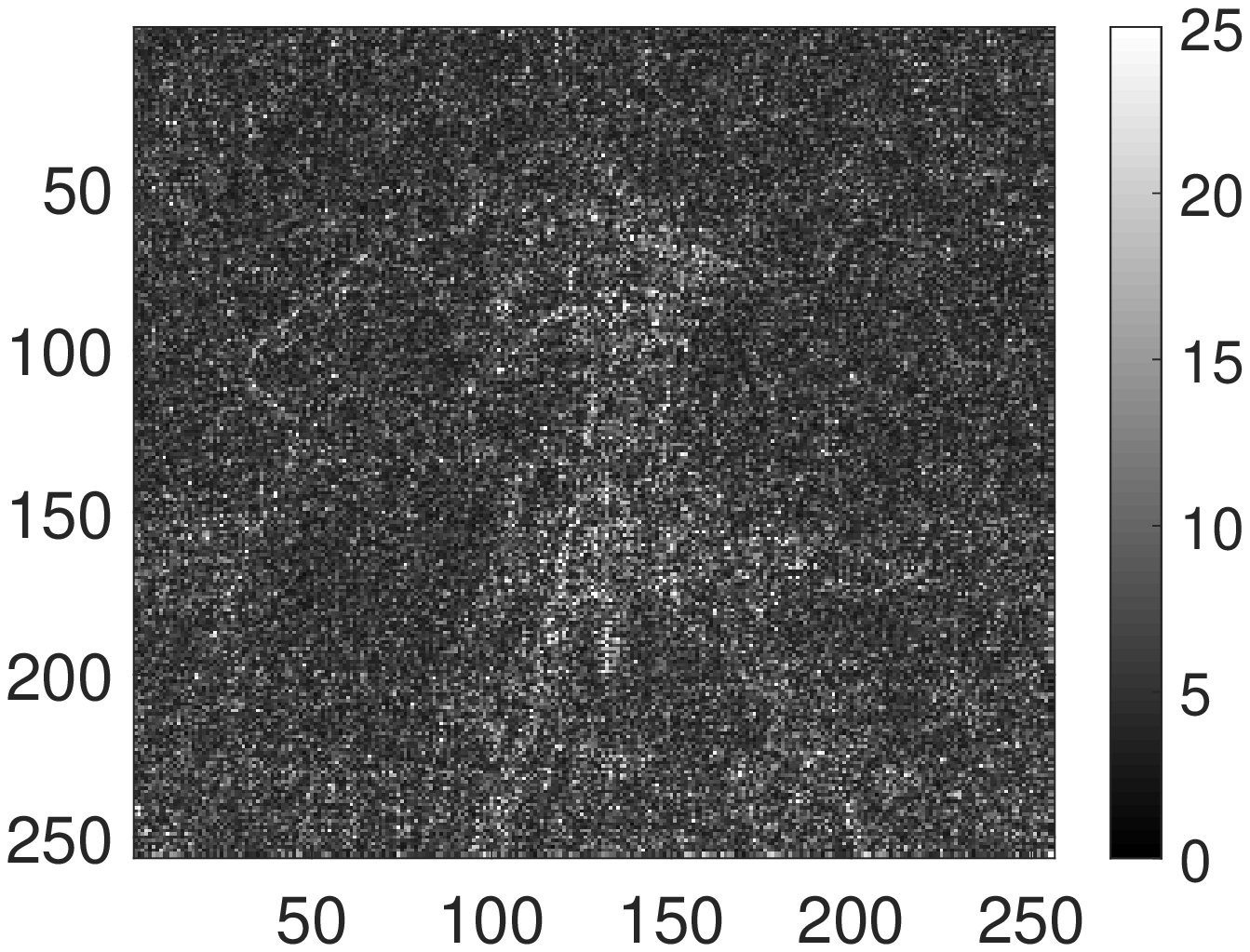}
}
\subfloat[SK-ROCK: standard deviation ($s=15$)]{
\label{subfig:cameraman_std_samples_SKROCK_s15}
\includegraphics[scale=.416]{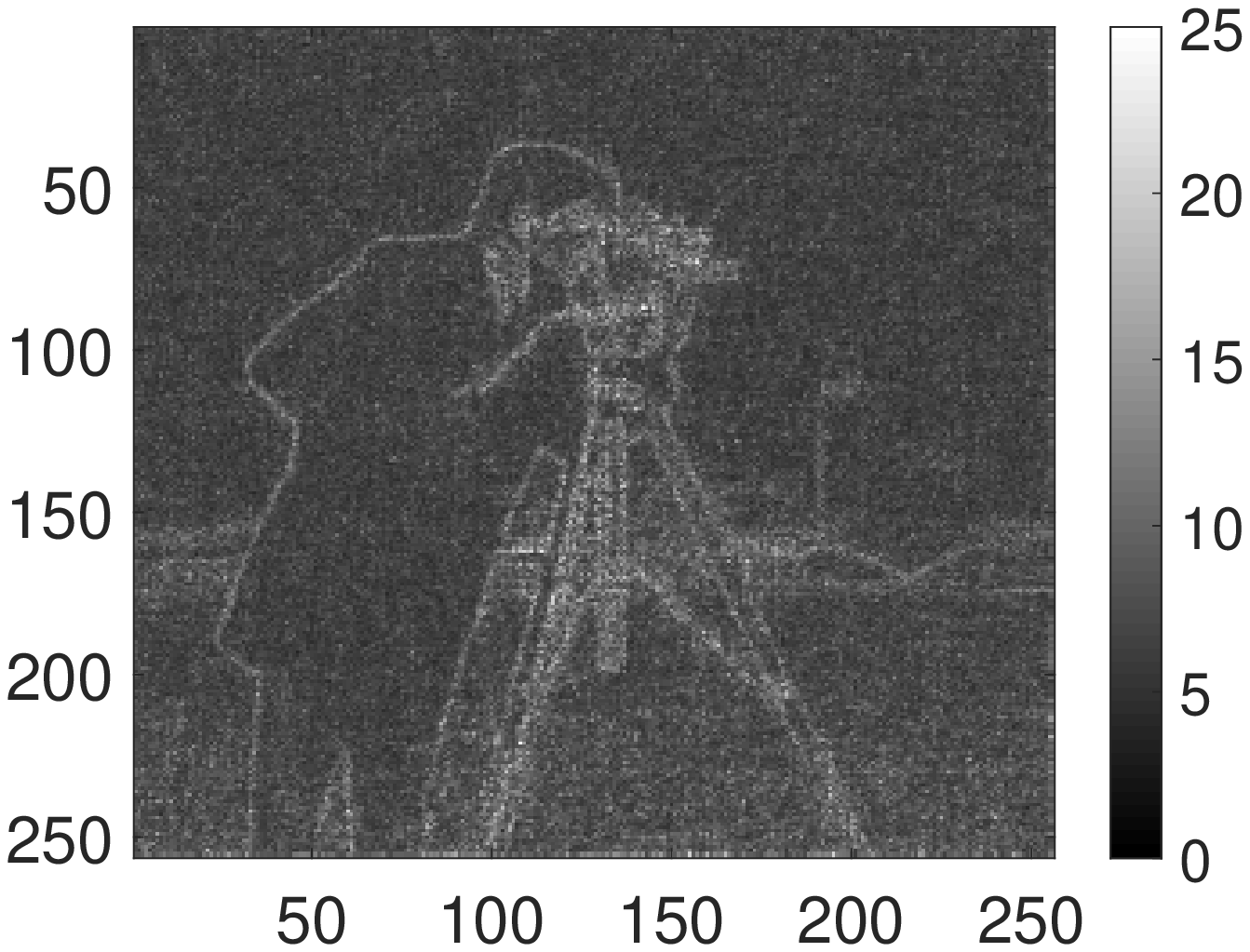}
}
\caption{\texttt{Cameraman} experiment:
{\normalfont(a)} Original image of dimension $256 \times 256$ pixels; {\normalfont(b)} blurred observation with SNR$=40$. {\normalfont(c)} Mean of $10^3$ samples generated by MYULA and {\normalfont(d)} mean of $10^3 / s$ samples generated by SK-ROCK. {\normalfont(e)} Standard deviation of the samples generated by MYULA and {\normalfont(f)} SK-ROCK.}
\label{fig:cameraman_experiment_snr_42}
\end{figure}

Figure \ref{fig:cameraman_experiment_snr_42} presents an experiment with the \texttt{cameraman} test image of size $d=256 \times 256$ pixels, depicted in Figure\ref{fig:cameraman_experiment_snr_42}(a). Figure \ref{fig:cameraman_experiment_snr_42}(b) shows an artificially blurred and noisy observation $y$, generated by using a $5 \times 5$ uniform blur and $\sigma=0.47$, related to a blurred signal-to-noise ratio of $40$dB. We use MYULA and SK-ROCK to draw Monte Carlo samples from \eqref{eqn:posteriorDist_ImagingExp_cam} using $\lambda = L_f^{-1} = 0.21$. To make the comparison fair, we generate $10^3$ samples using MYULA and $10^3 / s$ samples using SK-ROCK {for} $s = 15$. We then use the generated samples to compute two quantities: 1) the minimum mean square error (MMSE) estimator of $x|y$, given by the posterior mean; and 2) the pixel-wise (marginal) posterior standard deviation, which provides an indication of the level of confidence in each pixel value, as measured by the model. This quantity is useful {to highlight} features in the image that are difficult to accurately determine; in the case of in image deconvolution problems these are the exact locations of edges and contours in the image. {Notice that computing standard deviations requires computing second order statistical moments, which is more difficult than estimating the posterior mean, and hence requires a larger number of effective samples to produce stable estimates.}

Observe in Figures \ref{fig:cameraman_experiment_snr_42}(c)-(f) that while the estimates of the posterior mean obtained with MYULA and SK-ROCK are visually similar, the estimates of the pixel-wise standard deviations obtained with SK-ROCK are noticeably more accurate and in agreement with the results obtained by sampling the true posterior with an asymptotically unbiased Metropolised algorithm{,} see \cite[Example 4.1]{proxMCMC}. In particular, the standard deviations estimated with SK-ROCK accurately capture the uncertainty in the location of the contours in the image, whereas MYULA produces very noisy results as it struggles to estimate second order moments because of the stepsize limitation and limited computation budget (with a sufficiently large number of iterations, MYULA would produce similar results to SK-ROCK).

Moreover, to rigorously analyse the convergence properties of the two methods and compute autocorrelation functions, we generated $10^7$ samples with MYULA and $10^7 / s$ samples using SK-ROCK ($s=15$). We then used these samples to determine the fastest and slowest components of each chain\footnote{{The chain's slowest (fastest) component was identified by computing the approximated singular value decomposition of the chain's covariance matrix and choosing on the samples the component with the largest (smallest) singular value.}} and measured their autocorrelation functions. We also computed trace plots for the chains by using $T(x) = \log \pi_\lambda(x|y)$ as scalar statistic, which is particularly interesting because it determines the typical set of $x|y$ \cite{Pereyra2017}. These trace plots clearly illustrate how the methods behave during their transient regime, and then how they behave once the chains have converged to the typical set.

\begin{figure}
\centering
\subfloat[$\log \pi^{\lambda}(x)$]{
\label{subfig:cameraman_logpitrace}
\includegraphics[scale=.269]{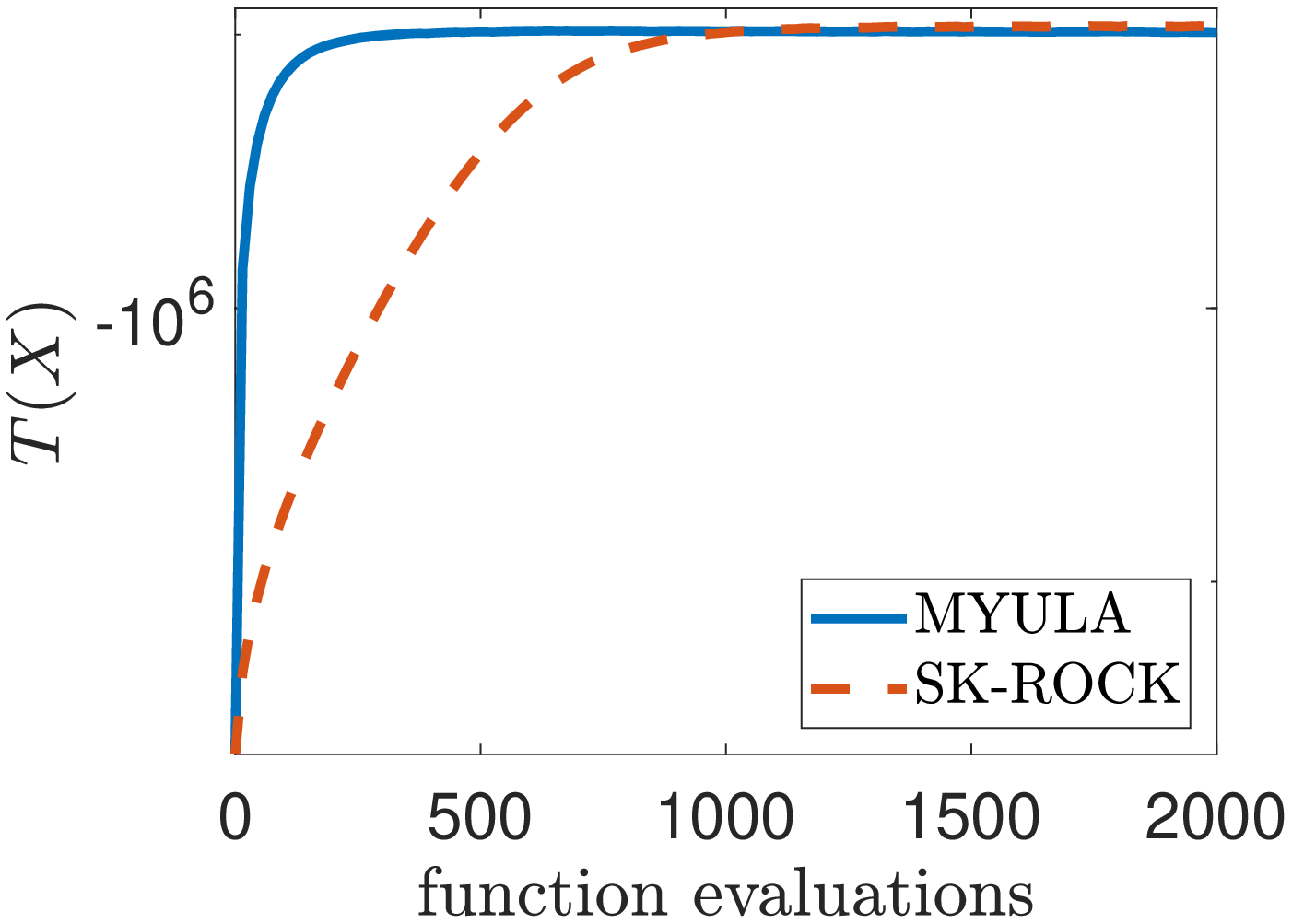}
}
\subfloat[$\log \pi^{\lambda}(x)$]{
\label{subfig:cameraman_chains_s15}
\includegraphics[scale=.269]{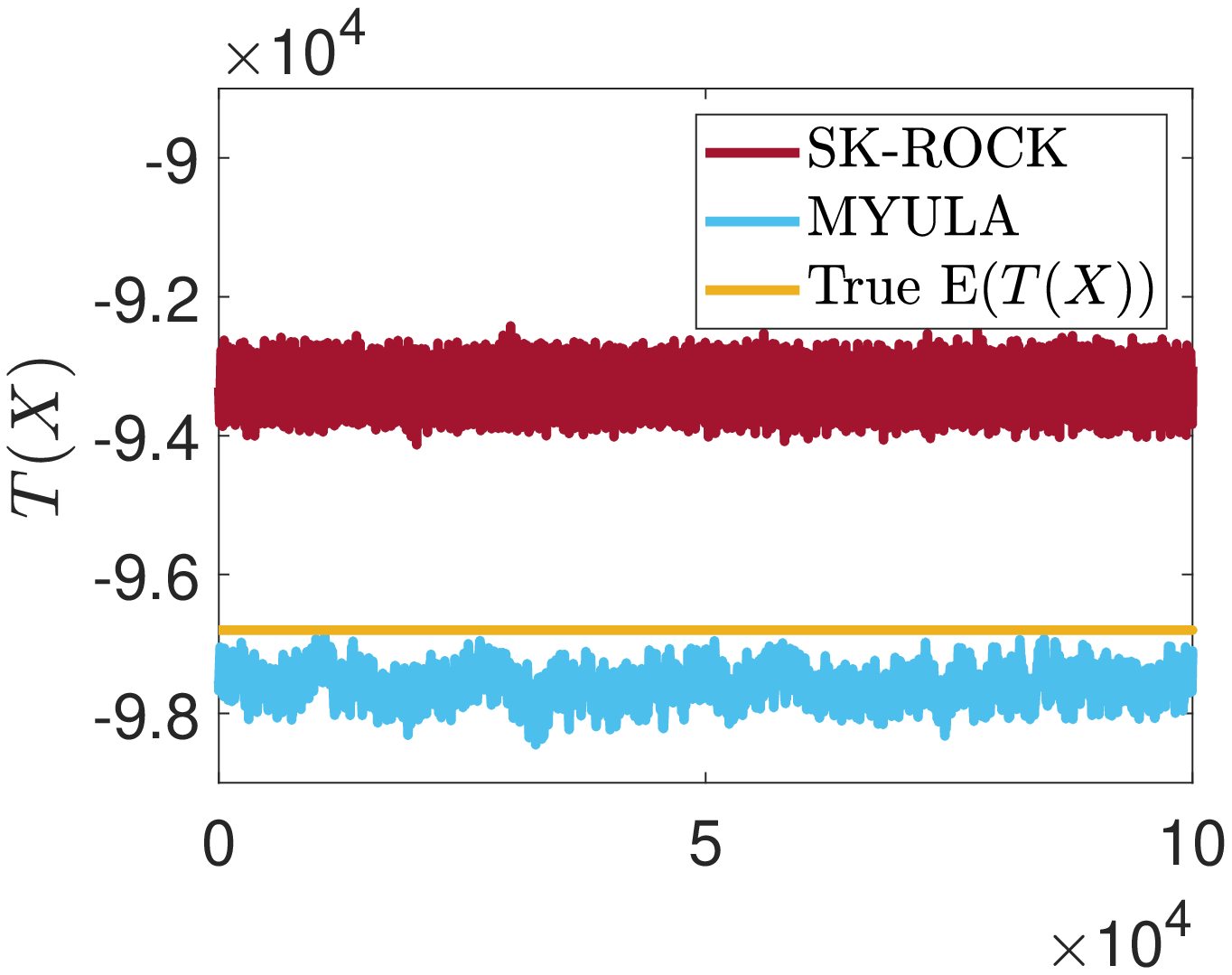}
} 
\subfloat[ACF slow component]{
\label{subfig:cameraman_acf_slow_s15}
\includegraphics[scale=.269]{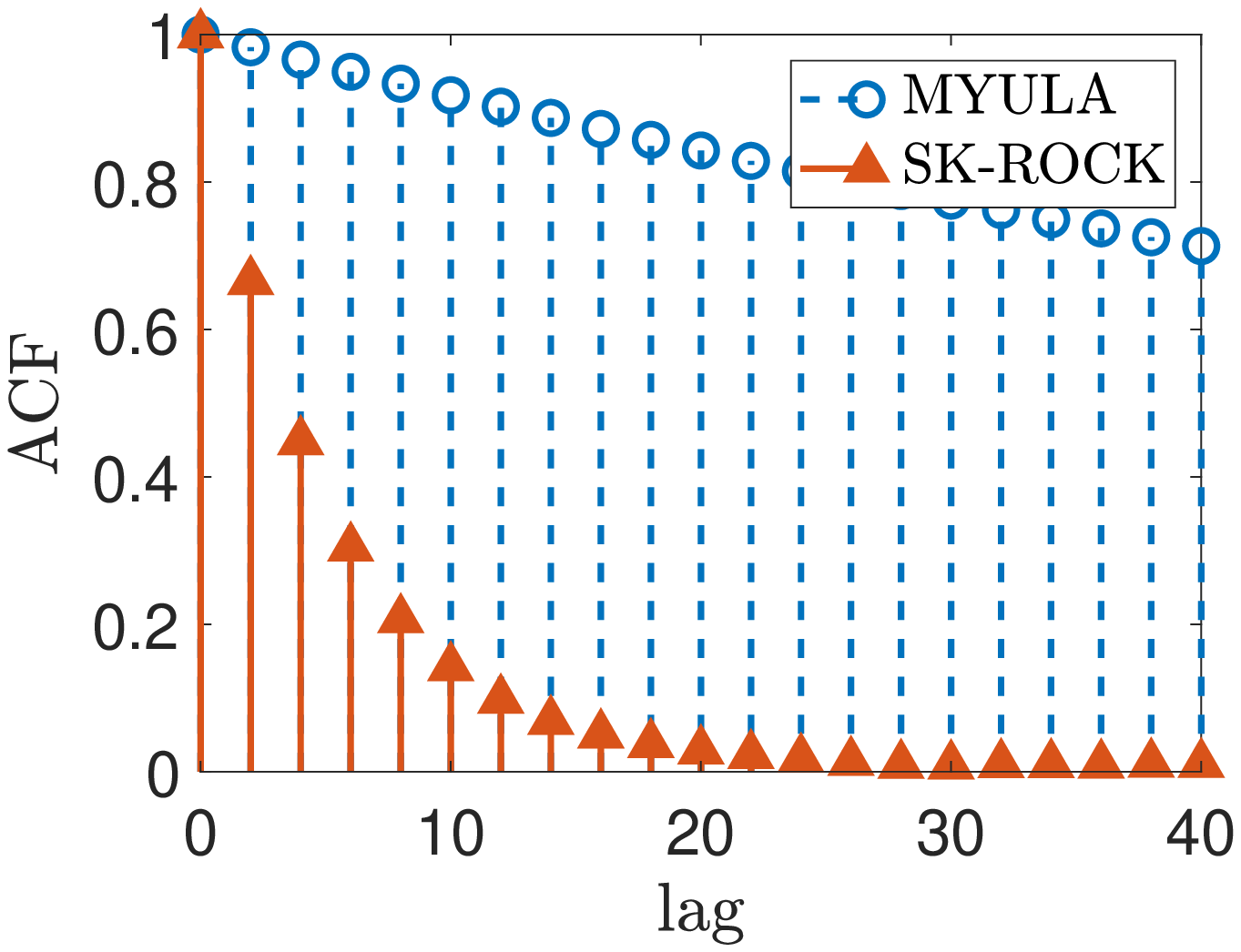}
}
\caption{\texttt{Cameraman} experiment: {\normalfont(a)} Convergence to the {typical set of the } posterior distribution \eqref{eqn:posteriorDist_ImagingExp_cam}  {for the first $2 \times 10^3$ MYULA samples and the first $2 \times 10^3 / s$ SK-ROCK  ($s=15$) samples}. {\normalfont(b)} {Last $10^5$ values of $\log \pi(x)$}. {\normalfont(c)} {Autocorrelation function for the slowest component.}}
\label{fig:cameraman_experiment_logPiTrace_chains_hist}
\end{figure}

Figure \ref{fig:cameraman_experiment_logPiTrace_chains_hist}(a) shows the convergence of the Markov chains to the typical set $\{x : T(x) \approx \textrm{E}[T(x)|y]\}$. 
Moreover, Figure \ref{fig:cameraman_experiment_logPiTrace_chains_hist}(b) shows the last $10^5$ samples of the chains (again with a 1-in-$s$ thinning for MYULA). {Additionally, we  have included the summary statistic $\mathbb{E}(T(X))$ calculated by  a very long run of the P-MALA algorithm \cite{proxMCMC}, which targets (\ref{eqn:posteriorDist_ImagingExp_cam}) exactly, in order to study the bias of the methods\footnote{The statistics $T(x) = \log p(x|y)$ is very useful for analysing the bias of high-dimensional log-concave distributions because these concentrate sharply on the typical set $T(x) \approx E(T(X))$ \cite{Pereyra2017}.}. We can see that, for this experiment, the bias of SK-ROCK is slightly increased in comparison to MYULA, however,} it has significantly better mixing properties that result in a better exploration of the typical set. Lastly, the superior convergence properties of SK-ROCK are also clearly illustrated by the autocorrelation plots of Figure \ref{fig:cameraman_experiment_logPiTrace_chains_hist}(c), which shows the autocorrelation functions for the slowest components of the chains, and where again we observe a dramatic improvement in decay rate ({we have again used  a 1-in-$s$ thinning for MYULA for fair comparison}). Table \ref{tab:resultsCameramanExp} reports the associated ESS values for this experiment, where we note that SK-ROCK with $s = 15$ outperforms MYULA by a factor of 21.77 in terms of computational efficiency for the slowest component (see Table \ref{tab:resultsCameramanExp}).

\begin{table}
{\footnotesize
  \caption{\texttt{Cameraman} experiment: Summary of the results after generating $10^7$ samples with MYULA and $10^7 / s$ samples with SK-ROCK {with $s=15$}. Computing time $35$ hours per method.}  \label{tab:resultsCameramanExp}
\begin{center}
  \begin{tabular}{|c|c|c|c|c|c|c|} \hline
   \bf Method & \bf Stepsize & \bf ESS & \bf ESS & \bf Speed-up & \bf Speed-up  \\
   & $\delta$ & \bf slow comp. & \bf fast comp. & \bf slow comp. & \bf fast comp.\\ \hline
    MYULA & $0.106$ & $2.88 \times 10^{3}$ & $1.00 \times 10^6$ & - &-  \\ 
    SK-ROCK ($s=10$) & $14.65$  & $4.00 \times 10^4$ & $2.63 \times 10^4$ & $13.89$ & $2.63 \times 10^{-2}$ \\ 
    SK-ROCK ($s=15$)& $34.30$ & $6.27 \times 10^{4}$ & $6.92 \times 10^4$ & $21.77$ & $6.92 \times 10^{-2}$ \\
		\hline
  \end{tabular}
\end{center}
}
\end{table}

\begin{figure}
\centering
\includegraphics[scale=.416]{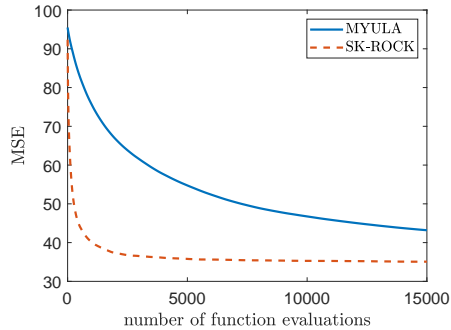}
\caption{\texttt{Cameraman} experiment: Mean squared error (MSE) between the mean of the algorithms and the true image, measured using $15 \times 10^3$ samples from MYULA and $15 \times 10^3 / s$ samples from SK-ROCK ($s=15$), in stationary regime.}
\label{fig:cameraman_experiment_mse}
\end{figure}

We conclude this experiment by comparing the two methods {in terms of estimation of the MSE against the true image}. Figure \ref{fig:cameraman_experiment_mse} shows the evolution of the estimation error for the MMSE solution, as estimated by MYULA and {SK-ROCK}, and as a function of the number of gradient and proximal operator evaluations. Again, observe that the acceleration properties of SK-ROCK lead to dramatic improvement in convergence speed, and consequently to a significantly more accurate computation of the MMSE estimator for given {computational} budget. 

\subsection{Hyperspectral Unmixing}
\label{subsec:expHyperUnmixing}
We now present an application to hyperspectral unmixing \cite{hyperspectral0}. Given a hyperspectral image $y \in \mathbb{R}^{m \times d}$ with $m$ spectral bands and $d$ pixels, the unmixing problem assumes that the observed scene is composed of $k$ materials or \emph{endmembers}, each with a characteristic spectral response $a_j \in \mathbb{R}^{m}$ for $j \in \{1,\ldots, k\}$, and seeks to determine the proportions or abundances $x_{j,i}$ of each material $j \in \{1,\ldots, k\}$ in each image pixel $i \in \{1,\ldots,d\}$. Here we consider the widely used linear mixing model $y = Ax + w$, where $A = \{a_1,\ldots, a_k\}\in \mathbb{R}^{m \times k}$ is a spectral library gathering the spectral responses of the materials, $x \in \mathbb{R}^{k \times d}$ gathers the abundance maps, and $w \sim \mathbb{N}(0, \sigma^2 \mathbb{I}_{m \times d})$ is {additive Gaussian noise}. Moreover, following \cite{hyperspectral1}, we expect $x$ to be sparse since most image pixels contain only a subset of the materials. Also, we expect materials to exhibit some degree of spatial coherence and regularity. In order to promote solutions with these characteristics, we use the $\ell_1$-TV prior proposed in \cite{hyperspectral1} for this type of problem
$$
p(x)\propto\exp\{- \alpha \|x\|_1 - \beta TV(x)\}\boldsymbol{1}_{\mathbb{R}_+^n}(x) ,
$$
where $\alpha >0$ and $\beta >0$ are hyper-parameters that we assume fixed (in this experiment we use $\alpha=25$ and $\beta=185$, determined using the method of \cite{Vidal2019}). The resulting posterior distribution is given by \cite{hyperspectral1}
\begin{equation}
\label{eqn:hyperspectral_posterior_dist}
p(x|y) \propto \exp \left[ -\|y-Ax\|^2 / 2\sigma^2 - \alpha \|x\|_1 - \beta TV(x) \right]\, \boldsymbol{1}_{\mathbb{R}_+^n}(x).
\end{equation}

\begin{figure}
\centering
\subfloat[MSE]{
\label{subfig:hyperspectral_mse}
\includegraphics[scale=.38]{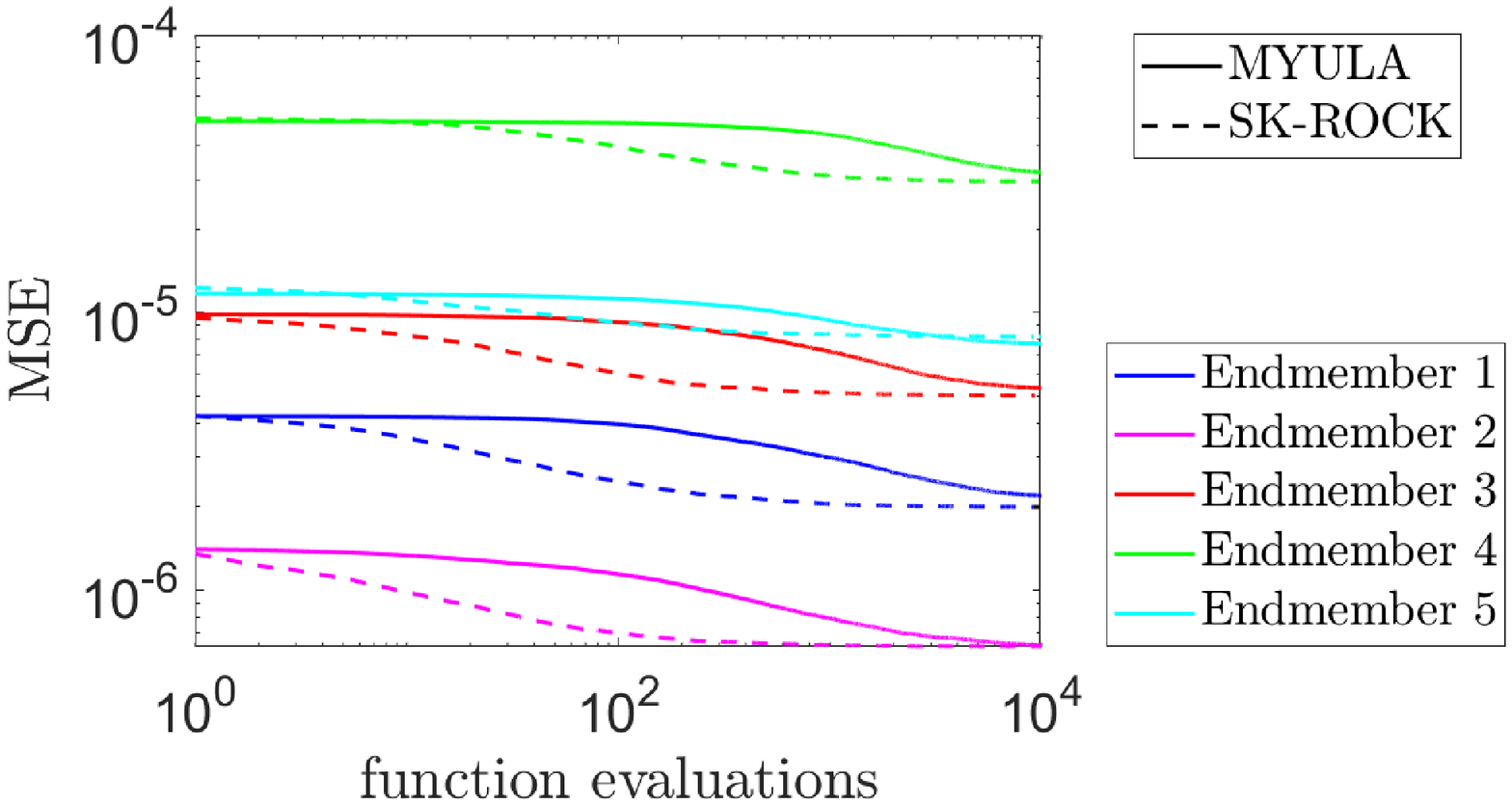}
}
\subfloat[true image $x$]{
\label{subfig:hyperspectral_exact_Image_snr_40}
\includegraphics[scale=.27]{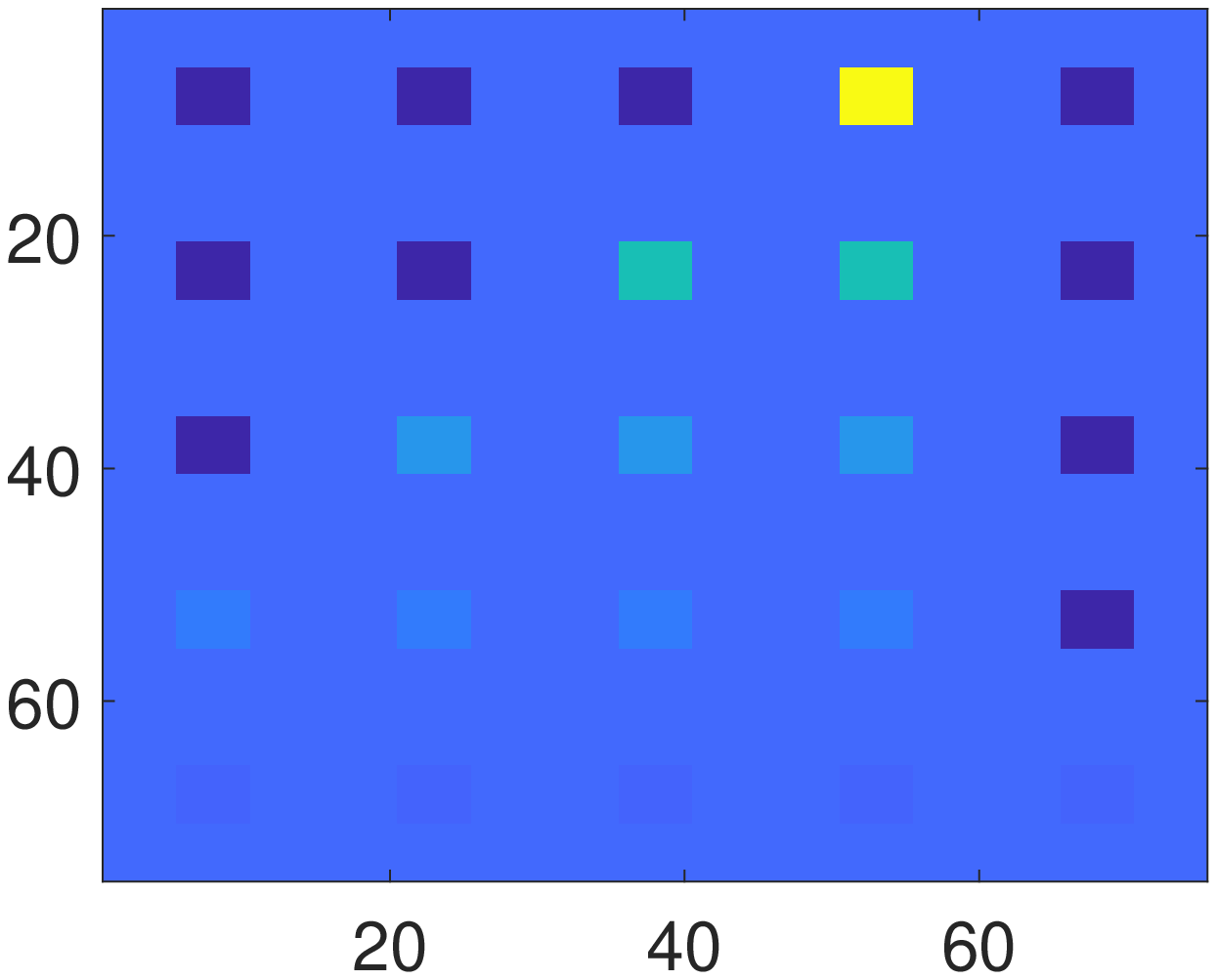}
}\\
\subfloat[MYULA: posterior mean]{
\label{subfig:hyperspectral_mean_samples_MYULA}
\includegraphics[scale=.416]{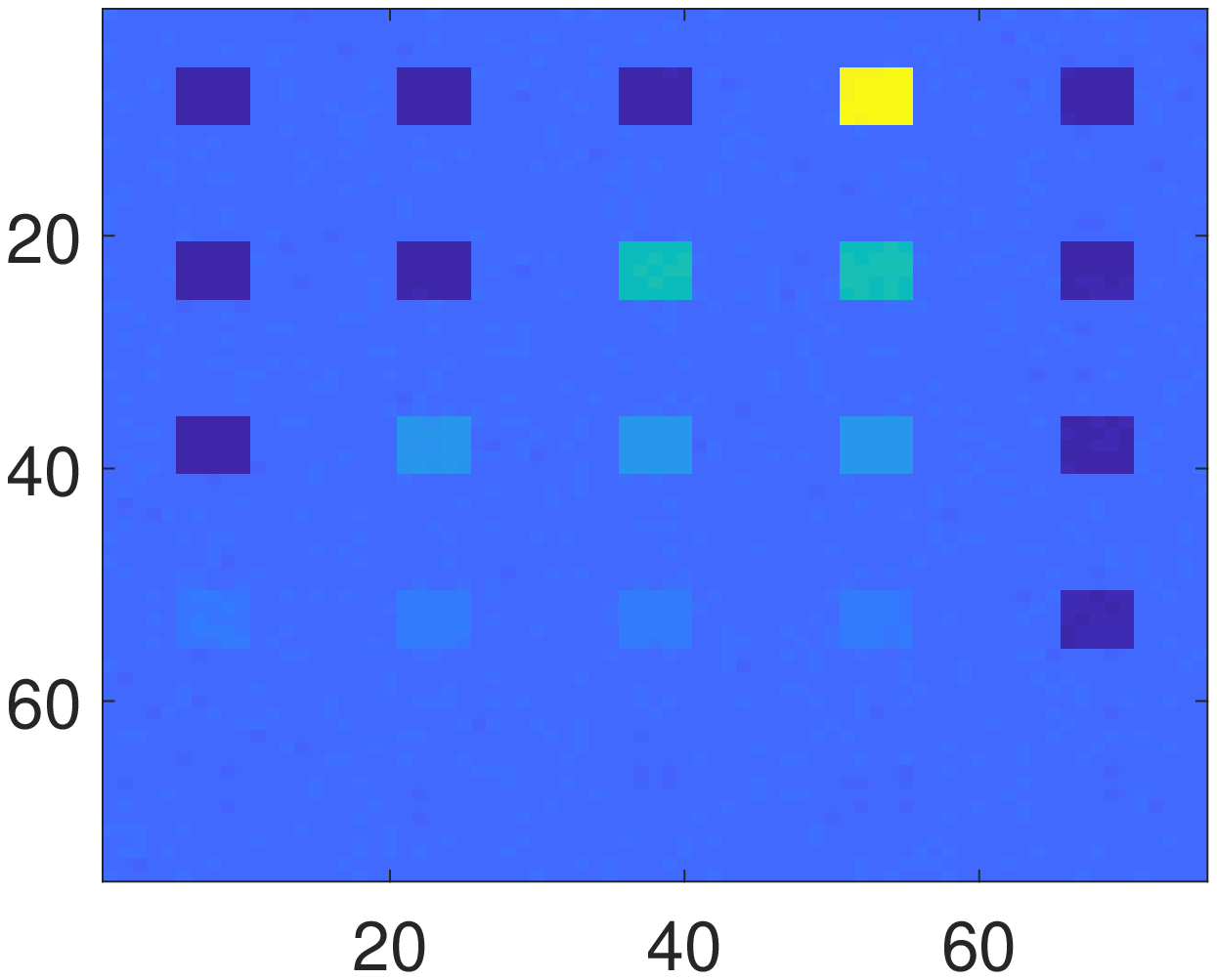}
}
\subfloat[SK-ROCK: posterior mean ($s=15$)]{
\label{subfig:hyperspectral_mean_samples_SKROCK_s10}
\includegraphics[scale=.416]{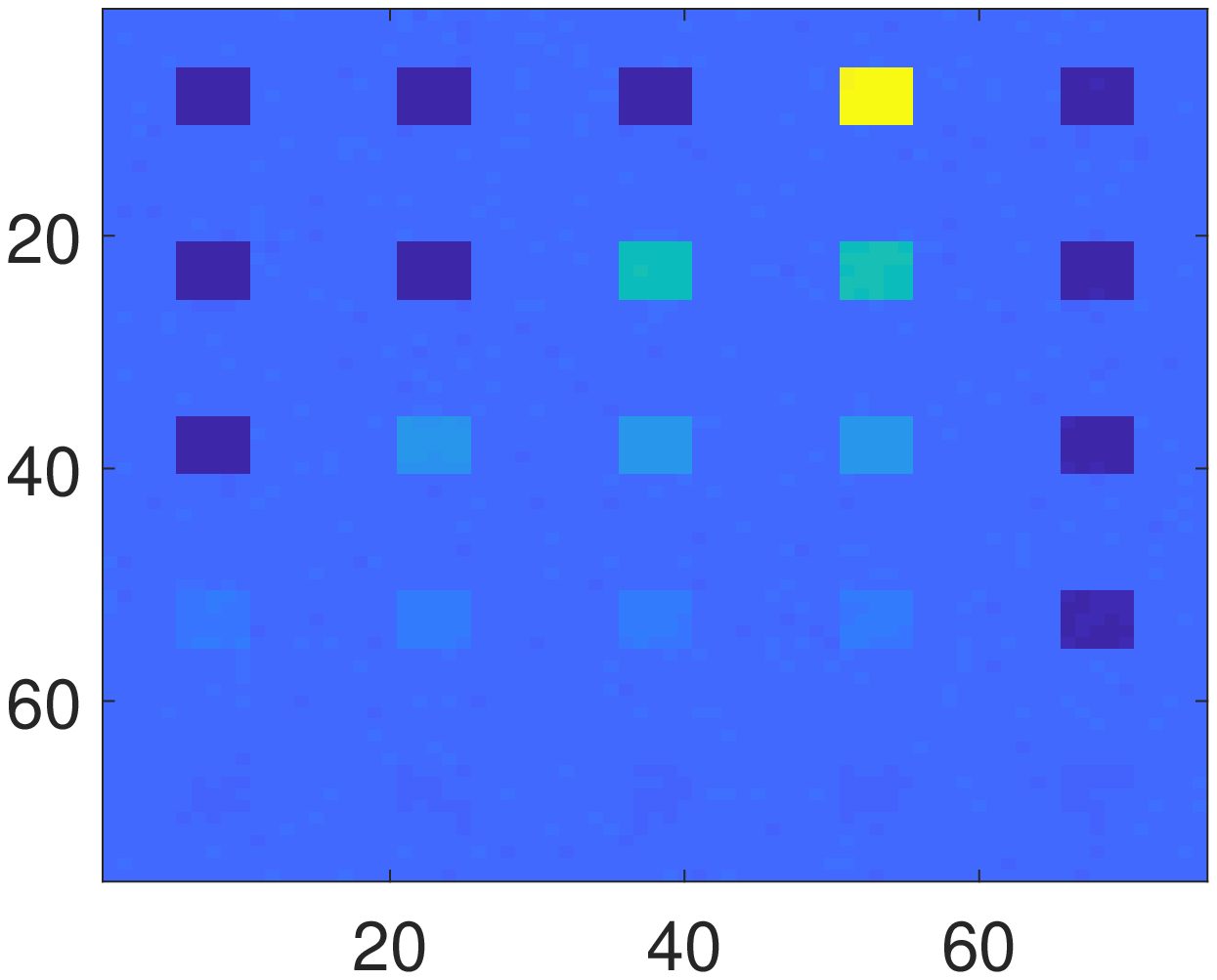}
} \\
\subfloat[MYULA: standard deviation]{
\label{subfig:hyperspectral_std_samples_MYULA}
\includegraphics[scale=.416]{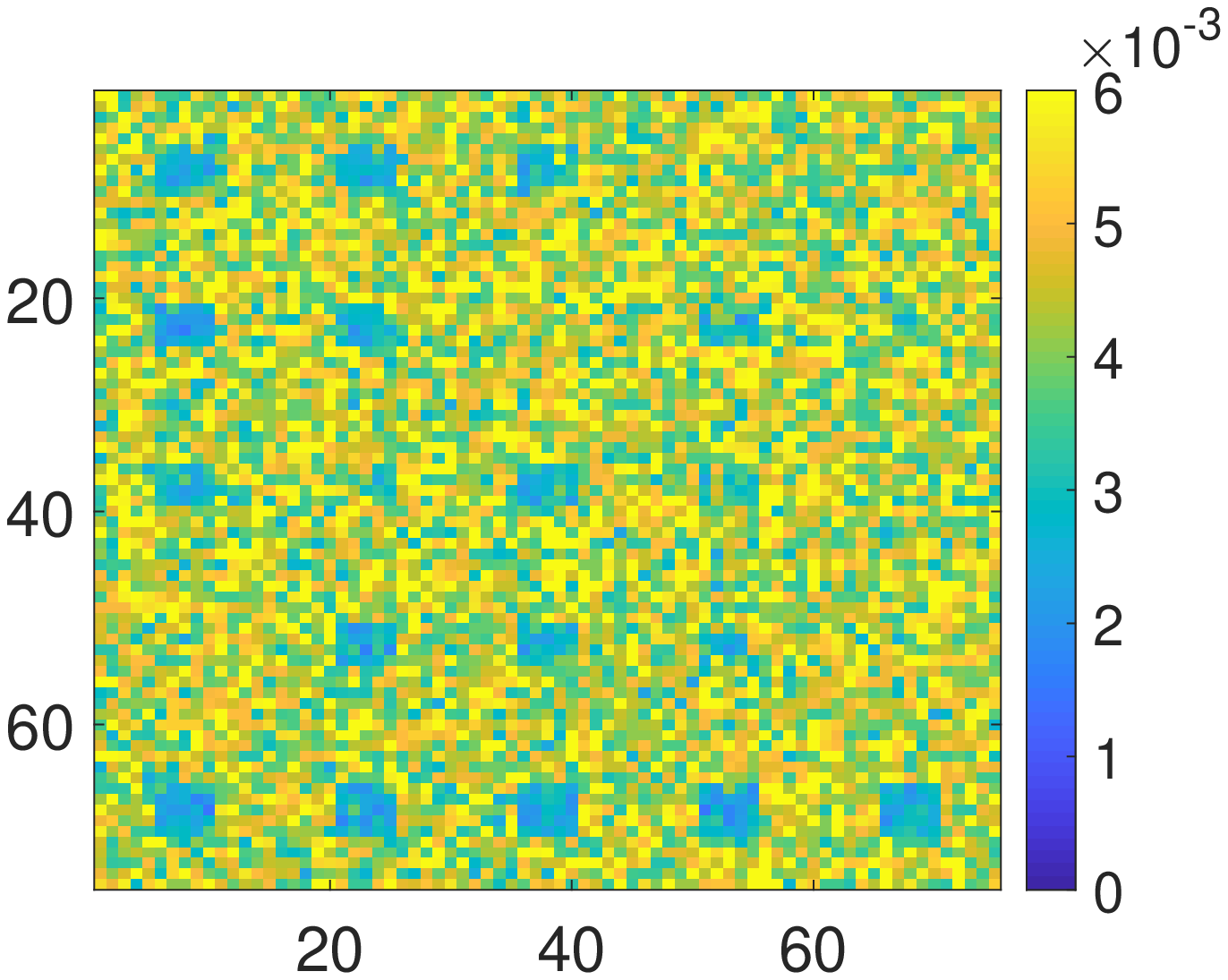}
}
\subfloat[SK-ROCK: standard deviation ($s=15$)]{
\label{subfig:hyperspectral_std_samples_SKROCK_s15}
\includegraphics[scale=.416]{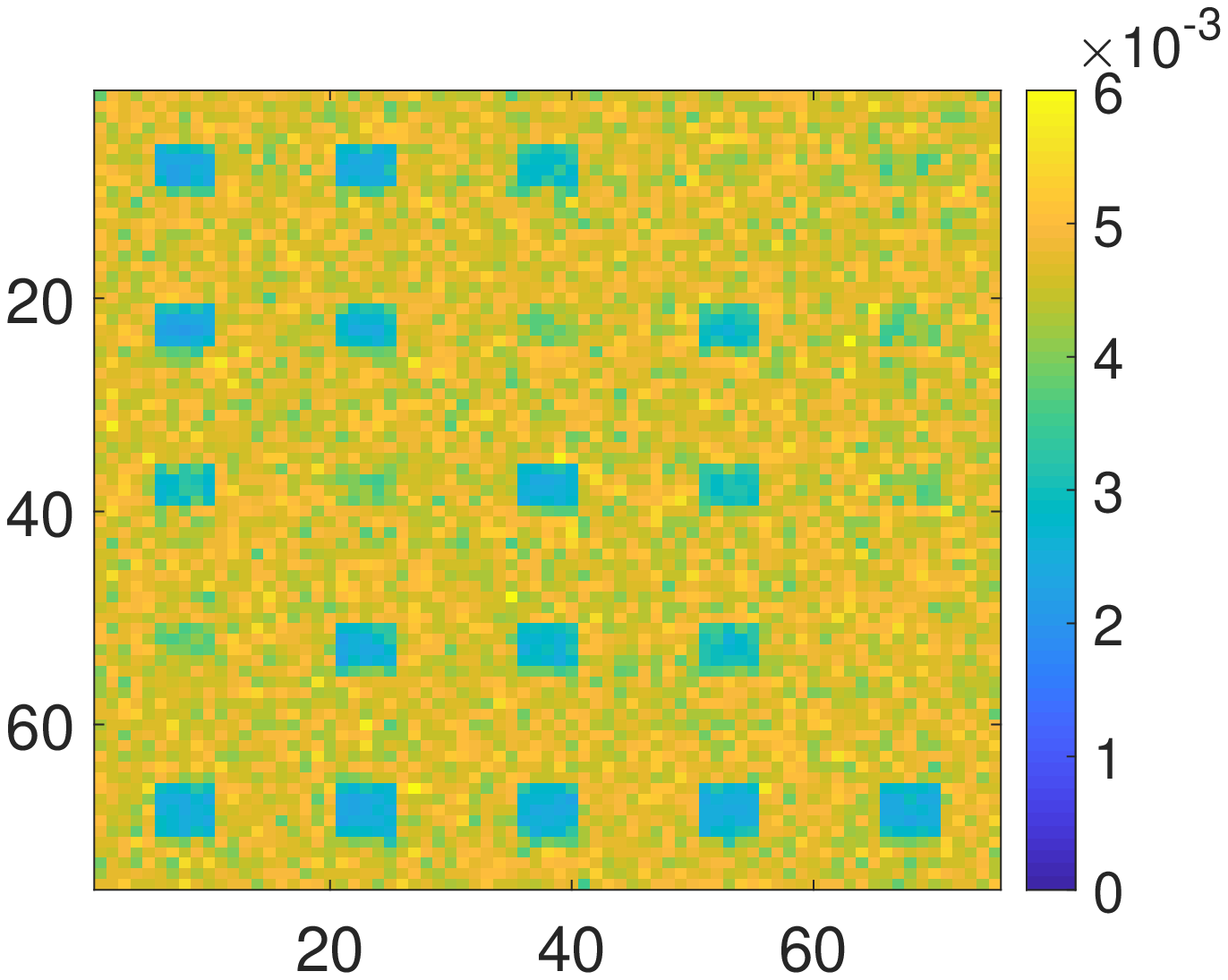}
}
\caption{\texttt{Hyperspectral} experiment: {\normalfont(a)} Mean squared error (MSE) between the mean of the algorithms and the true image (fractional abundances of endmembers 1 to 5) measured using $10^4$ samples from MYULA (solid line) and $10^4 / s$ samples from SK-ROCK (dash-dot line, $s=15$), in logarithmic scale. {\normalfont(b)} True fractional abundances of the endmember 4 ($75 \times 75$ pixels), {\normalfont(c)} posterior mean as estimated with $10^5$ samples generated with MYULA and {\normalfont(d)} $10^5 / s$ samples generated by SK-ROCK. {\normalfont(e)} Standard deviation of the samples generated by MYULA and {\normalfont(f)} SK-ROCK.
}
\label{fig:hyperspectral_experiment_observation_means}
\end{figure}

Figure \ref{fig:hyperspectral_experiment_observation_means} presents an experiment with a synthetic dataset from \cite{hyperspectral1} of size $n = 75 \times 75 = 5625$, with $5$ materials, and noise amplitude $\sigma=8.4 \times 10^{-4}$ related to a signal-to-noise-ratio of $40$dB{,} see \cite{hyperspectral1} for details. Figure \ref{fig:hyperspectral_experiment_observation_means}(a) presents the evolution of the estimation MSE between the true abundance maps and the posterior mean as estimated by MYULA and SK-ROCK (with $s = 15$), and as a function of the number of gradient and proximal operator evaluations (using $\lambda = 7.08 \times 10^{-7}$ {which is in the order of $L_f^{-1}$, as it is recommended in \cite[Section 3.3]{pereyraMYULA}}). As in previous experiments, observe that the posterior means estimated with SK-ROCK converge dramatically faster than the ones calculated with MYULA, clearly exhibiting the benefits of the proposed methodology. Moreover, for illustration, Figures \ref{fig:hyperspectral_experiment_observation_means}(c)-(e) respectively show the estimated abundance maps for the fourth endmember for MYULA ($5 \times 10^5$ samples) and SK-ROCK {($5 \times 10^5/s$ samples, $s = 15$)}, as well as the pixel-wise (marginal) standard deviations for the abundances of this material. Again, as in previous experiments, we notice that the estimates obtained with SK-ROCK are noticeably more precise than the ones of MYULA, which would require a larger number of iterations to accurately estimate these second order statistical moments.

\begin{figure}
\centering
\subfloat[$\log \pi^{\lambda}(x)$]{
\label{subfig:hyperspectral_logpitrace}
\includegraphics[scale=.269]{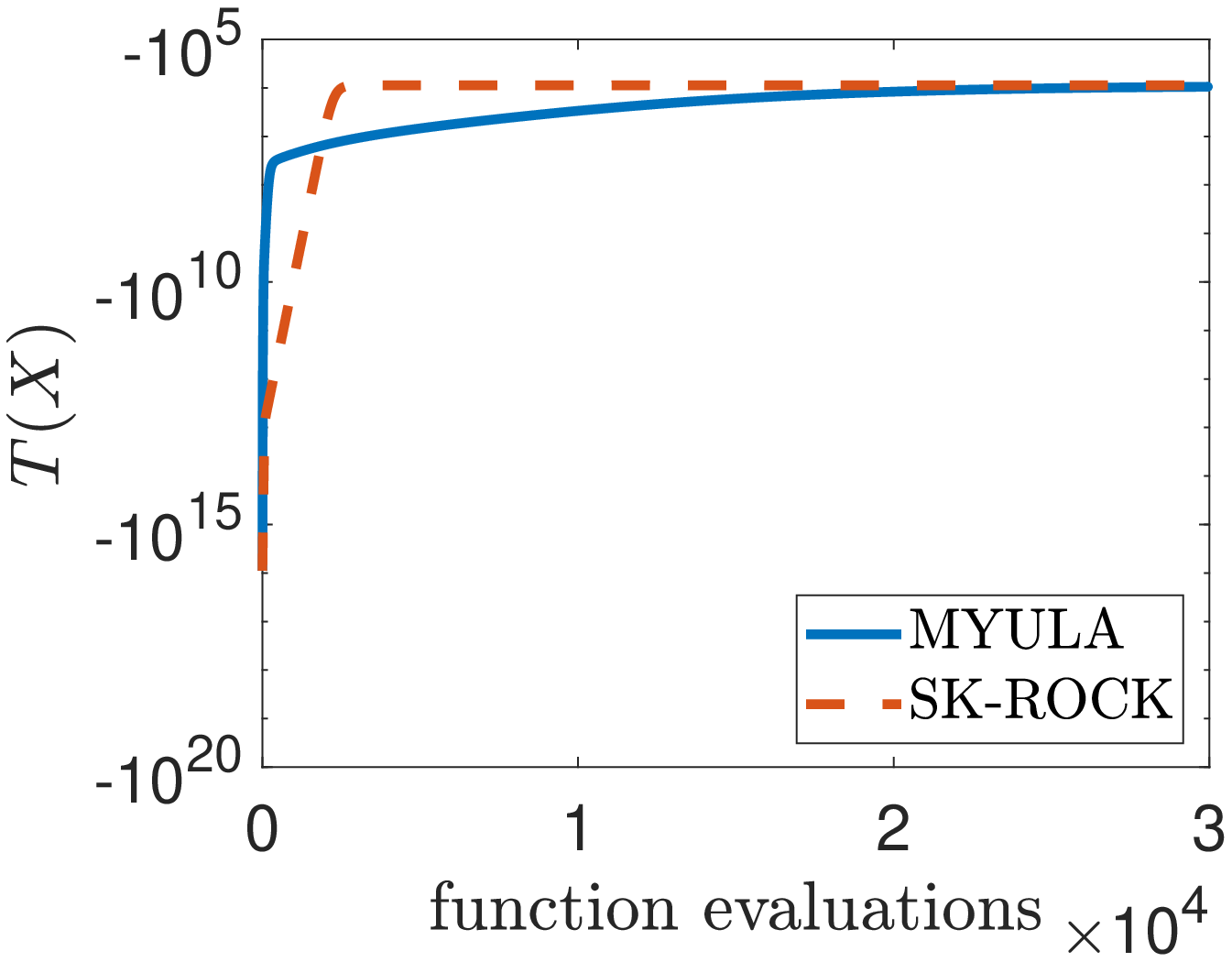}
}
\subfloat[$\log \pi^{\lambda}(x)$]{
\label{subfig:hyperspectral_chains_s15}
\includegraphics[scale=.269]{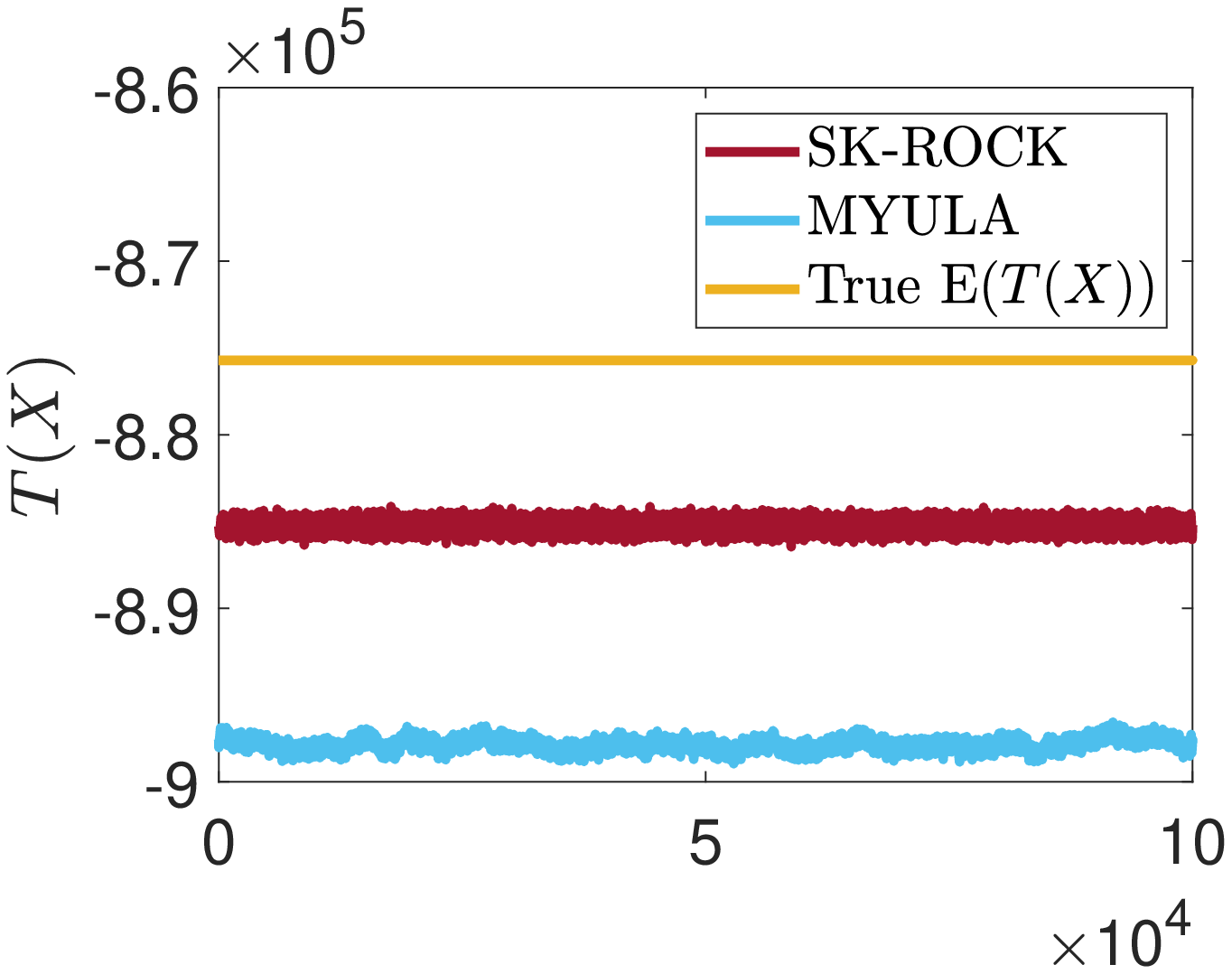}
} 
\subfloat[ACF slow component]{
\label{subfig:hyperspectral_acf_slow_s15}
\includegraphics[scale=.269]{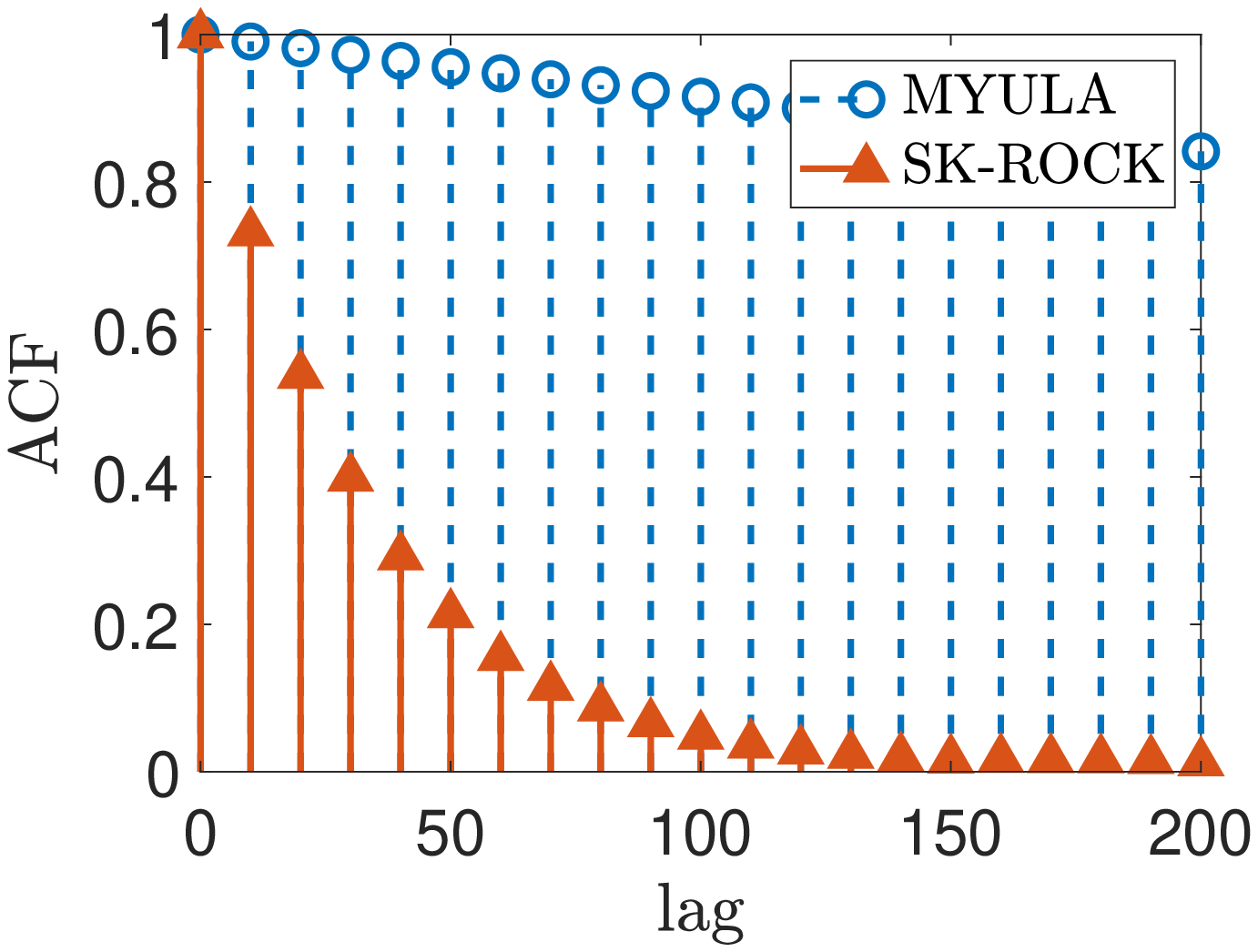}
}
\caption{\texttt{Hyperspectral} experiment:  {\normalfont(a)} Convergence to the {typical set of the} posterior distribution (\ref{eqn:hyperspectral_posterior_dist}) for  the first $3 \times 10^4$ MYULA samples and the first $3 \times 10^4 / s$ SK-ROCK ($s=15$) samples. {\normalfont(b)} Last $10^5$ values of $\log \pi(x)$. {\normalfont(c)} {Autocorrelation function for the slowest component}.}
\label{fig:hyperspectral_experiment_logPiTrace_chains_hist}
\end{figure}

To further compare the convergence properties of the two methods we repeated the experiment and generated $5\times 10^6$ samples with MYULA and {$5 \times10^6/s$} samples with SK-ROCK {for} $s=15$ to make the comparisons fair. Figure \ref{fig:hyperspectral_experiment_logPiTrace_chains_hist}(a) presents trace plots for the two chains during their transient regimes using $T(x) = \log p(x|y)$ as summary statistic, as a function of the number of gradient and proximal operator evaluations; observe that SK-ROCK attains the typical set of $x|y$ significantly faster than MYULA, similarly to the previous experiments. Figure \ref{fig:hyperspectral_experiment_logPiTrace_chains_hist}(b) presents similar trace plots for the two chains in stationarity. {Additionally, as we did in \textit{cameraman} experiment, we  have included the summary statistic $\mathbb{E}(T(X))$ calculated by a very long run of the P-MALA algorithm, which targets (\ref{eqn:hyperspectral_posterior_dist}) exactly, in order to study the bias of the methods. As can be seen clearly, SK-ROCK presents a lower bias than MYULA}, and also exhibits better mixing properties. The good convergence properties of SK-ROCK can be clearly observed in the autocorrelation plots of Figure \ref{fig:hyperspectral_experiment_logPiTrace_chains_hist}(c), which correspond to the slowest components of the chains as determined by their covariance structure, and where we {have again applied the {$1$-in-$15$} thinning to the MYULA chain for fairness of comparison}. Table \ref{tab:resultsHyperspectralExp} reports the ESS values for this experiment. In particular, observe that SK-ROCK outperforms MYULA by a factor of $37.9$ in terms of ESS for the slowest component of the chain, and by a factor of $5.76$ for the {fastest} component.

\begin{table}
{\footnotesize
  \caption{\texttt{Hyperspectral} experiment: Summary of the results after generating $5 \times 10^6$ samples with MYULA and $5 \times 10^6 / s$ samples with SK-ROCK. Computing time $88$ hours per method.}  \label{tab:resultsHyperspectralExp}
\begin{center}
  \begin{tabular}{|c|c|c|c|c|c|c|} \hline
   \bf Method & \bf Stepsize & \bf ESS & \bf ESS & \bf Speed-up & \bf Speed-up  \\
   & $\delta$ & \bf slow comp. & \bf fast comp. & \bf slow comp. & \bf fast comp.\\ \hline
    MYULA & $1.79 \times 10^{-9}$ & $1.50 \times 10^2$ & $0.63 \times 10^4$ & -  & - \\ 
    SK-ROCK ($s=10$) & $3.11 \times 10^{-7}$ & $2.90 \times 10^3$ & $1.70 \times 10^4$ & $19.33$ & $2.69$ \\ 
    SK-ROCK ($s=15$) & $7.28 \times 10^{-7}$ & $5.69 \times 10^3$ & $3.63 \times 10^4$ & $37.93$ & $5.76$ \\ \hline
  \end{tabular}
\end{center}
}
\end{table}

\subsection{Tomographic image reconstruction}
\label{sub:tomographyExp}
We conclude this section with a tomographic image reconstruction experiment. We have selected this problem to illustrate the proposed methodology in a setting where the posterior distribution is strictly log-concave. The lack of strong log-concavity has a clear negative impact on the convergence properties of the continuous-time Langevin SDE \eqref{eqn:LangevinSDE} \cite{convergenceULA}, and also impacts the convergence properties of the MYULA and SK-ROCK approximations.

In tomographic image reconstruction we seek to recover an image $x \in \mathbb{R}^d$ from an observation $y \in \mathbb{C}^p$ related to $x$ by a linear Fourier model $y = A F x + \xi$, where $F$ is the discrete Fourier transform operator on $\mathbb{C}^d$, $A \in \mathbb{C}^{p \times d}$ is a (sparse) tomographic subsampling mask and $\xi \sim N(0,\sigma^2 \mathbb{I}_{2p})$. Typically $d \gg p$, making the estimation problem strongly ill-posed. We address this difficulty by using a total-variation prior to regularise the estimation problem and promote solutions with certain spatial regularity properties. From Bayes' theorem, the posterior $p(x|y)$ is given by:
\begin{equation}
\label{eqn:tomography_posterior_dist}
p(x|y) \propto \exp \left[ - \|y-AFx\|^2 / 2\sigma^2 - \beta TV(x) \right],
\end{equation}
with hyper-parameters $\sigma, \beta \in \mathbb{R}^+$ assumed fixed (in this experiment we use $\beta=10^2$).

\begin{figure}
	\centering
	\subfloat[true image $x$]{
		\label{subfig:tomography_exact_Image_snr_42}
		\includegraphics[scale=.416]{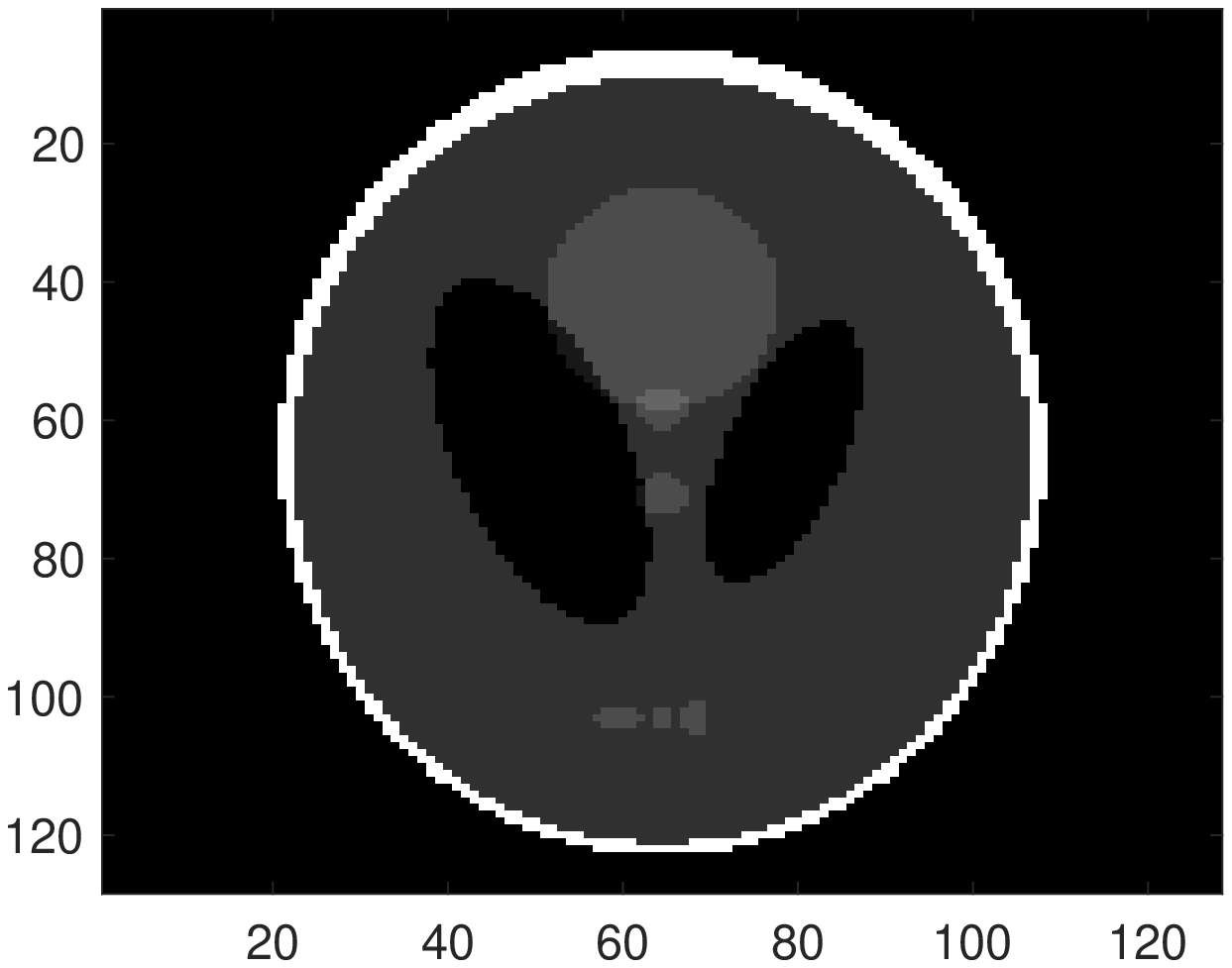}
	}
	\subfloat[observation $y$]{
		\label{subfig:tomography_incomplete_obs}
		\includegraphics[scale=.416]{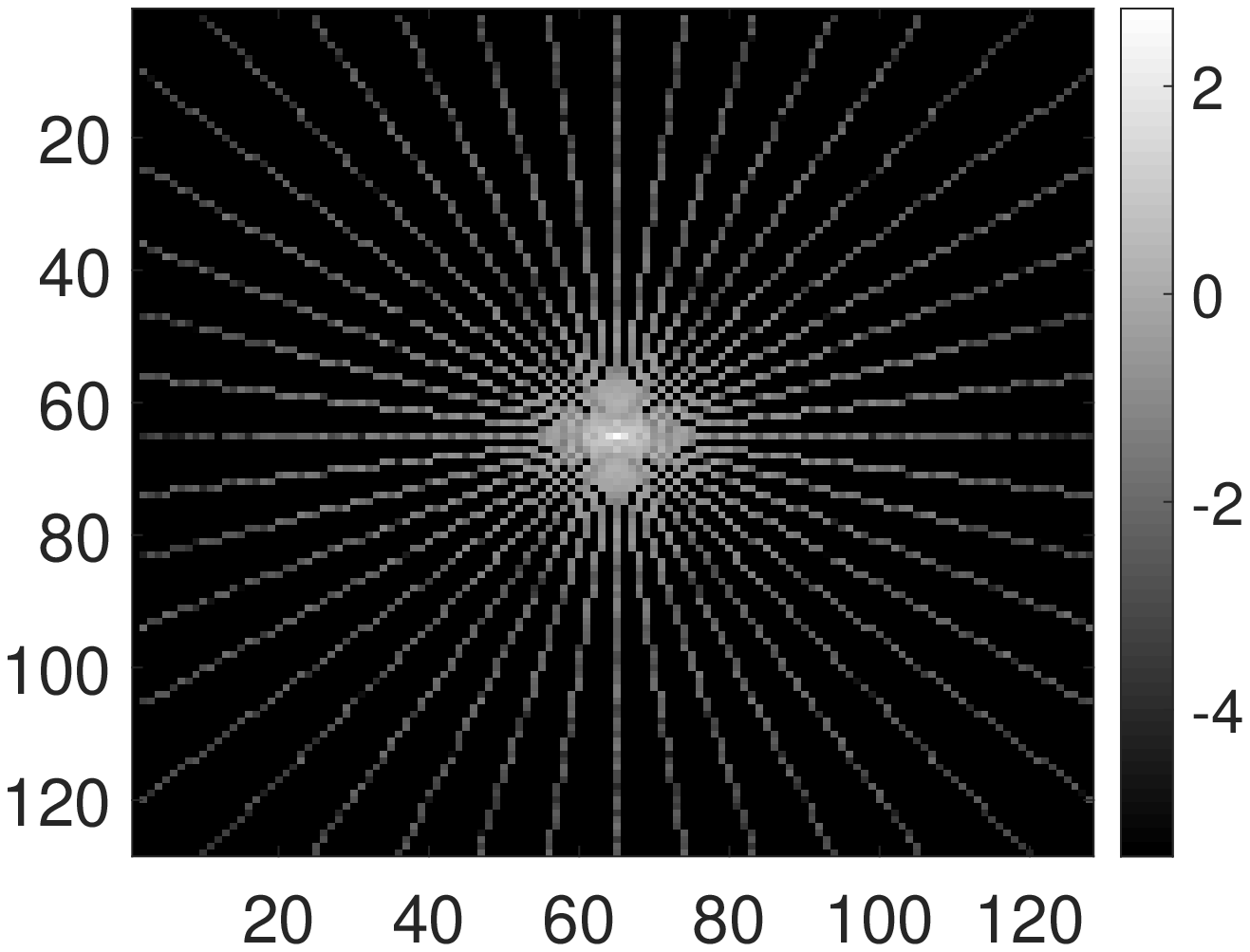}
	} \\
	\subfloat[MYULA: posterior mean]{
		\label{subfig:tomography_mean_samples_MYULA}
		\includegraphics[scale=.416]{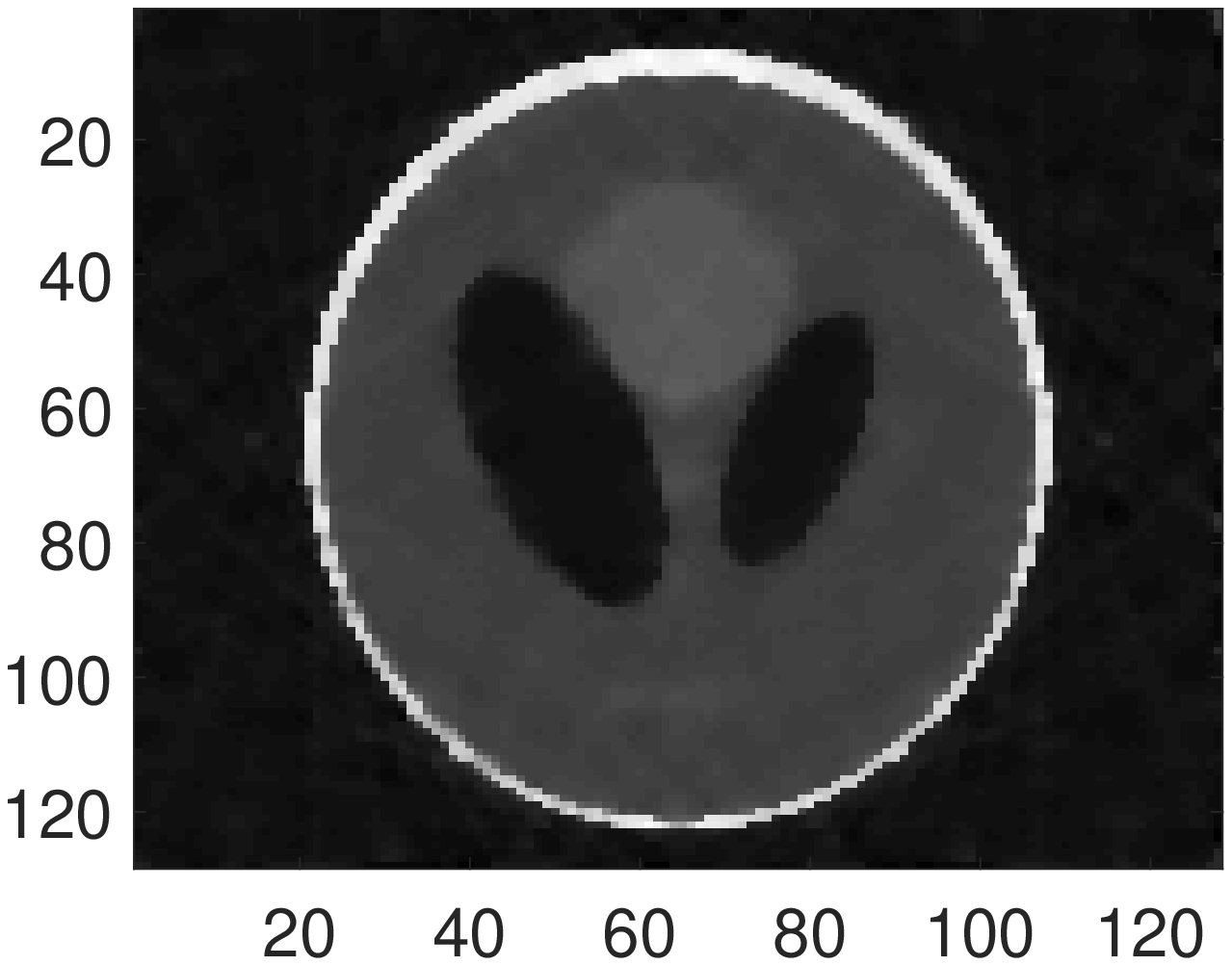}
	}
	\subfloat[SK-ROCK: posterior mean ($s=10$)]{
		\label{subfig:tomography_mean_samples_SKROCK_s10}
		\includegraphics[scale=.416]{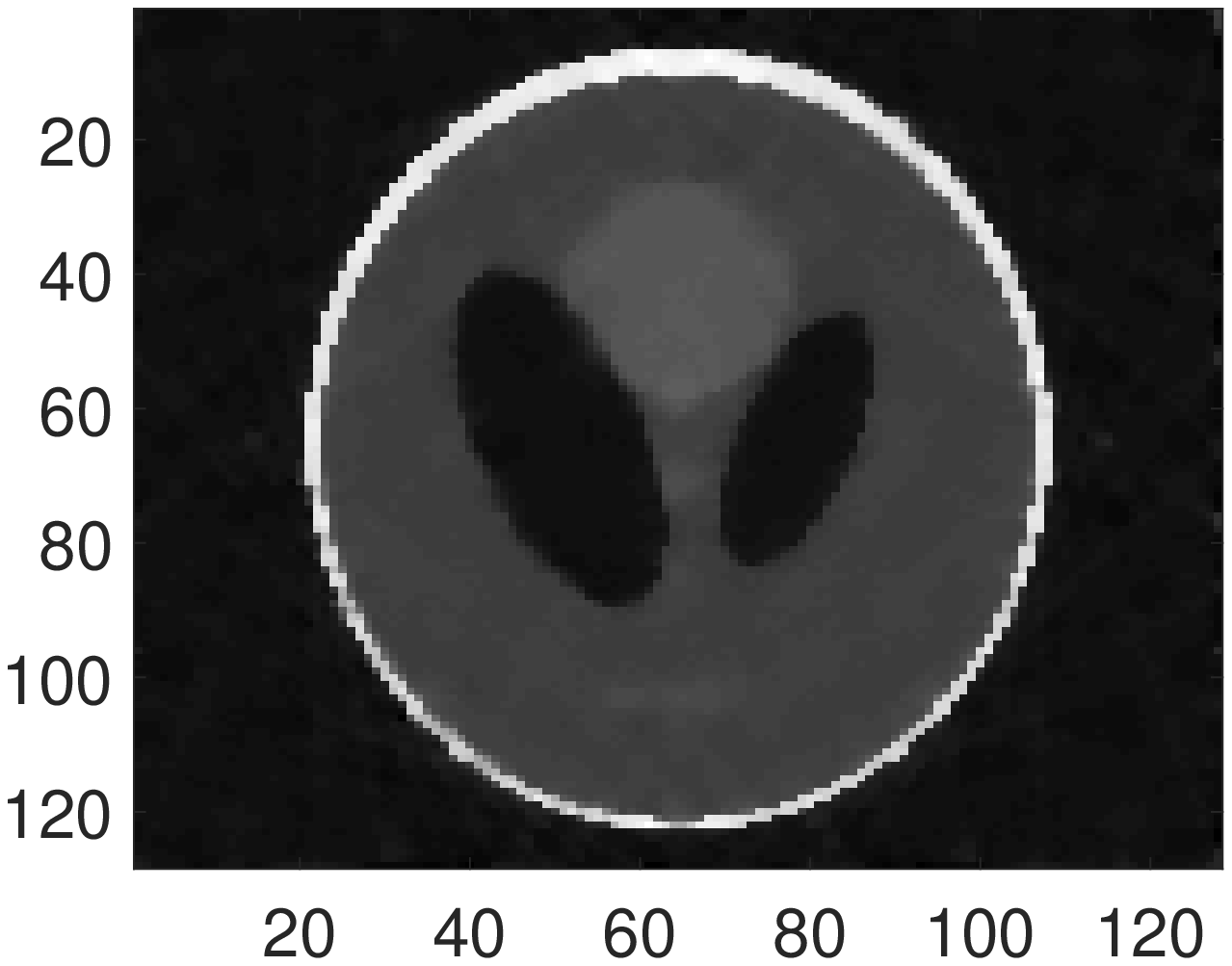}
	}\\
	\subfloat[MYULA: std. dev. - Fourier coefs. ]{
	\includegraphics[scale=.416]{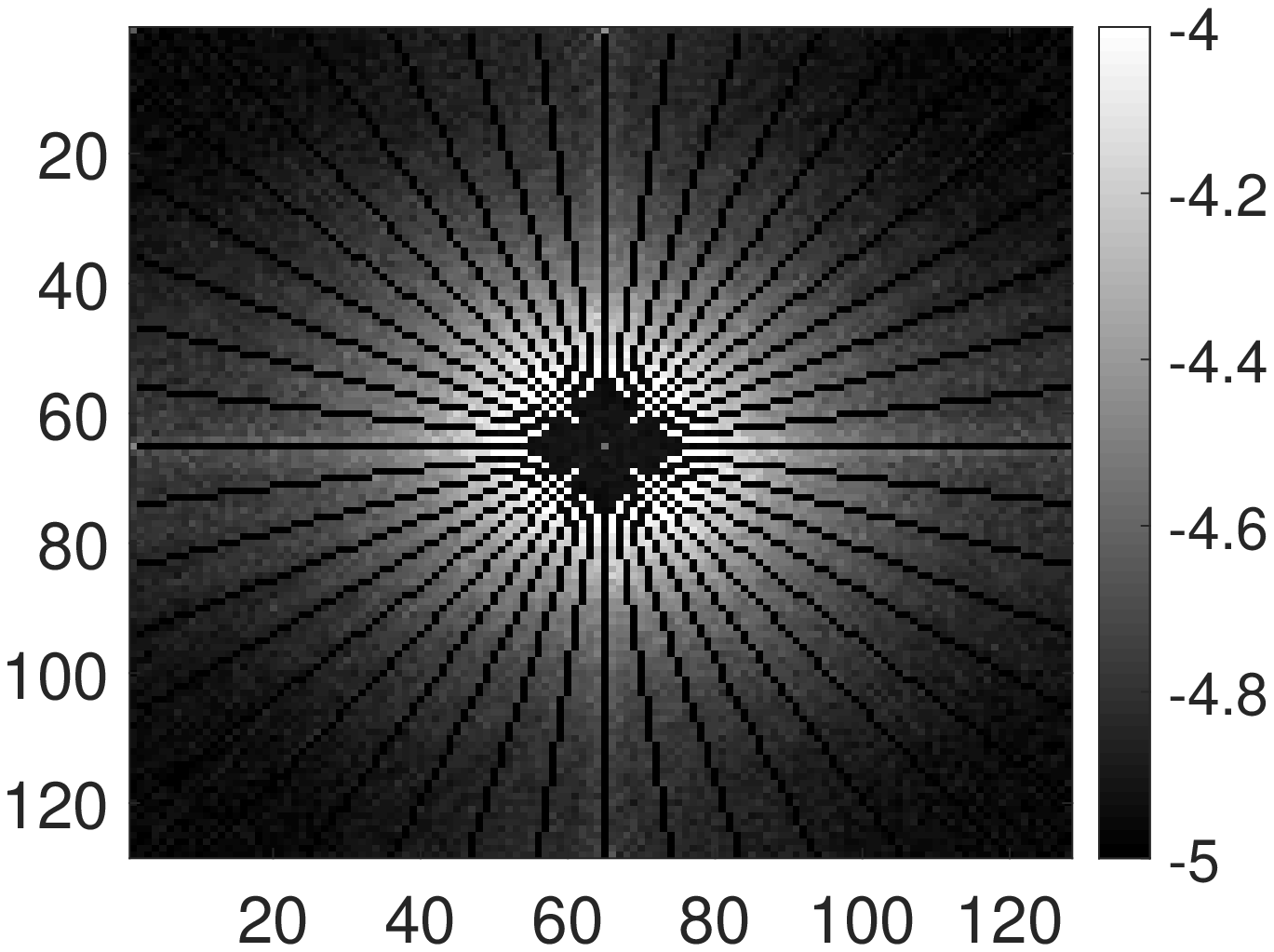}
	}
	\subfloat[SK-ROCK: std. dev. - Fourier coefs.]{
	\includegraphics[scale=.416]{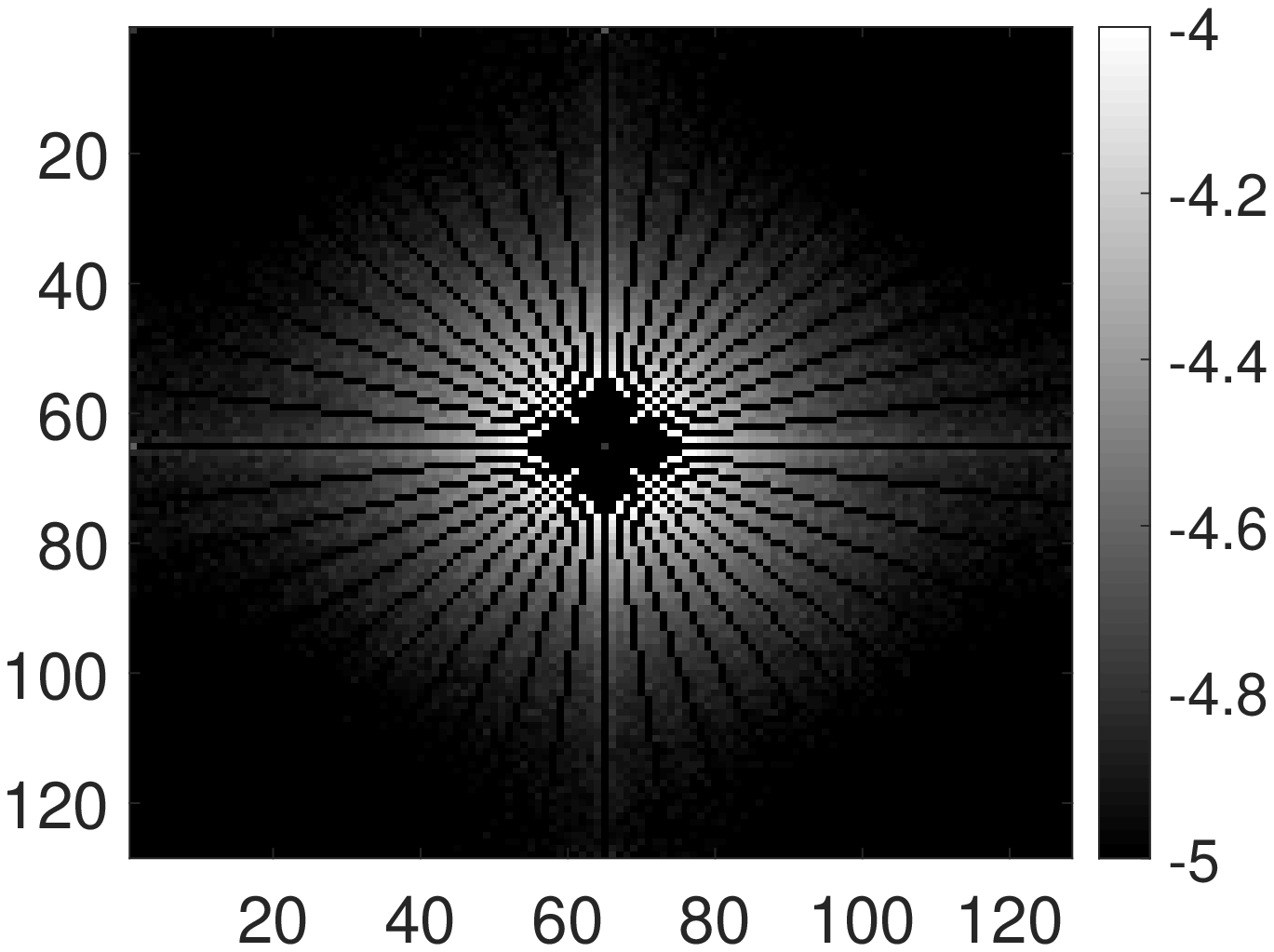}
	}
	\caption{{Tomography} experiment: {\normalfont(a)} \texttt{Shepp-Logan} phantom image ($128 \times 128$ pixels), {\normalfont(b)} tomographic observation $y$ (amplitude of Fourier coefficients in logarithmic scale). Posterior mean of $x|y$ as estimated with {\normalfont(c)} MYULA ($10^4$ samples) and {\normalfont(d)} SK-ROCK ($10^3$ samples, $s=10$).
	Standard deviations of the amplitude of the Fourier coefficients of $x|y$ as estimated with  {\normalfont(e)} MYULA ($10^4$ samples) and {\normalfont(f)} SK-ROCK ($10^3$ samples, $s=10$).}
	\label{fig:tomography_experiment_observation}
\end{figure}

Figure \ref{fig:tomography_experiment_observation} presents an experiment with the \texttt{Shepp-Logan} phantom test image of size $d=128 \times 128$ pixels, which we use to generate a noisy observation $y$ by measuring $15\%$ of the original Fourier coefficients, corrupted with additive Gaussian noise with $\sigma = 10^{-2}$ (to improve visibility, Figure \ref{fig:tomography_experiment_observation}(b) shows the amplitude of the Fourier coefficients in logarithmic scale, unobserved coefficients are depicted in black). Following on from this, we use MYULA and SK-ROCK with $s = 10$ to generate $10^4$ and $10^3$ samples respectively from $p(x|y)$ with $\lambda = 0.2 \times 10^{-4}$ {which is in the order of $L_f^{-1}$, as it is recommended in \cite[Section 3.3]{pereyraMYULA}}. {We then use these samples to compute two quantities: 1) the MMSE estimators - displayed in Figures \ref{fig:tomography_experiment_observation}(c)-(d); and 2) the (marginal) standard deviations of the amplitude of the Fourier coefficients of $x|y$, depicted in Figures \ref{fig:tomography_experiment_observation}(e)-(f) in logarithmic scale. Observe that, in this experiment, both methods deliver good and similar results with the number of samples available, with MYULA producing slightly less accurate standard deviation estimates.} More interestingly, notice from Figures \ref{fig:tomography_experiment_observation}(e)-(f) that in this tomographic experiment the uncertainty is concentrated in the unobserved medium frequencies, whereas in the deconvolution experiment uncertainty was predominant in the high-frequencies.

Moreover, to analyse the convergence properties of the two methods we compute autocorrelation functions by generating $5 \times 10^6$ samples with MYULA and $5 \times 10^6/s$ samples using SK-ROCK with $s=10$. We use said samples to determine the fastest and slowest components of each chain and measure their autocorrelation functions. Table \ref{tab:resultsTomographyExp} reports the associated ESS, which show that the SK-ROCK outperform MYULA by a factor of $20.23$ in terms of ESS for the slowest component of the chain.

\begin{figure}
	\centering
	\subfloat[$\log \pi^{\lambda}(x)$]{
		\label{subfig:tomography_logpitrace}
		\includegraphics[scale=.269]{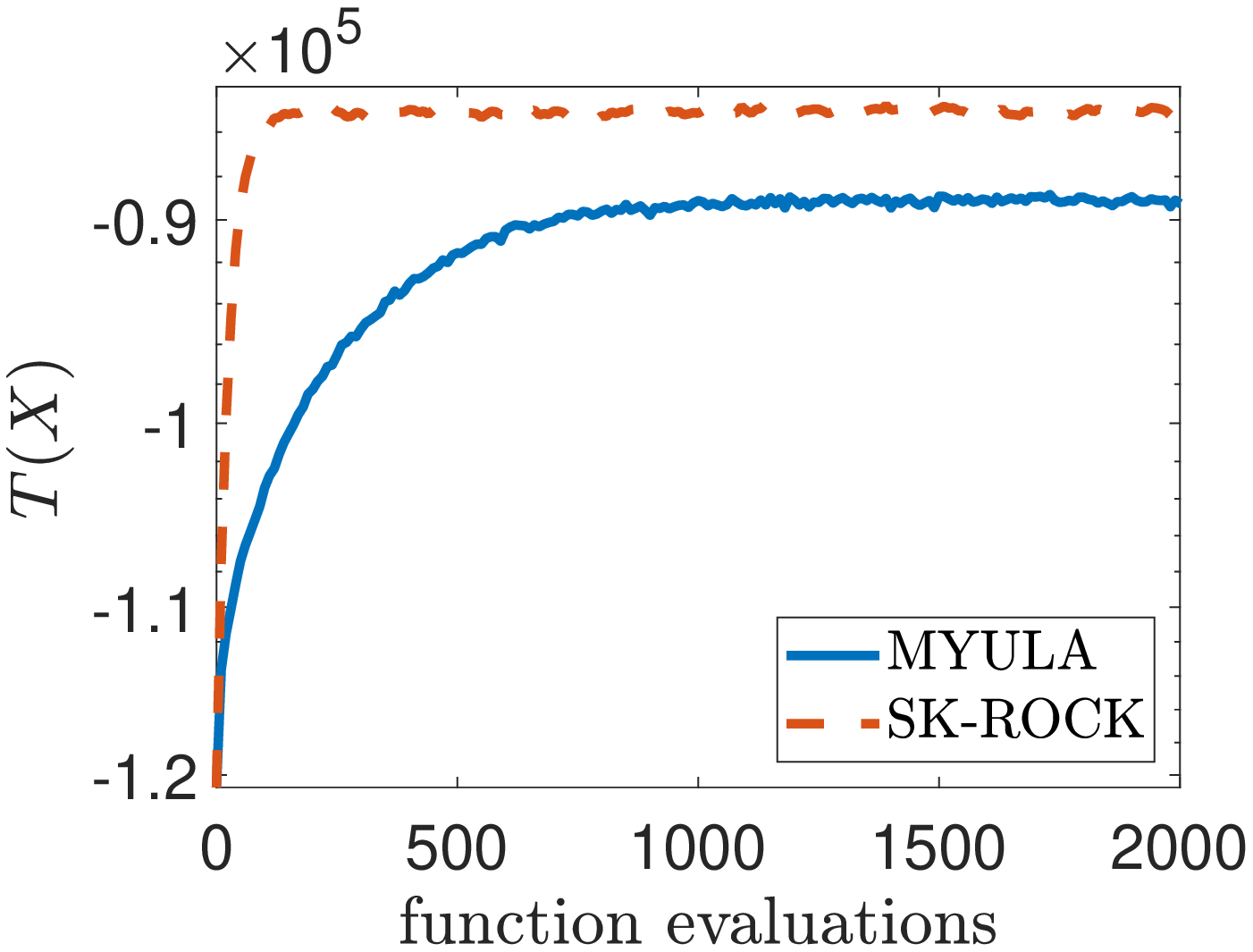}
	}
	\subfloat[$\log \pi^{\lambda}(x)$]{
		\label{subfig:tomography_chains_s10}
		\includegraphics[scale=.269]{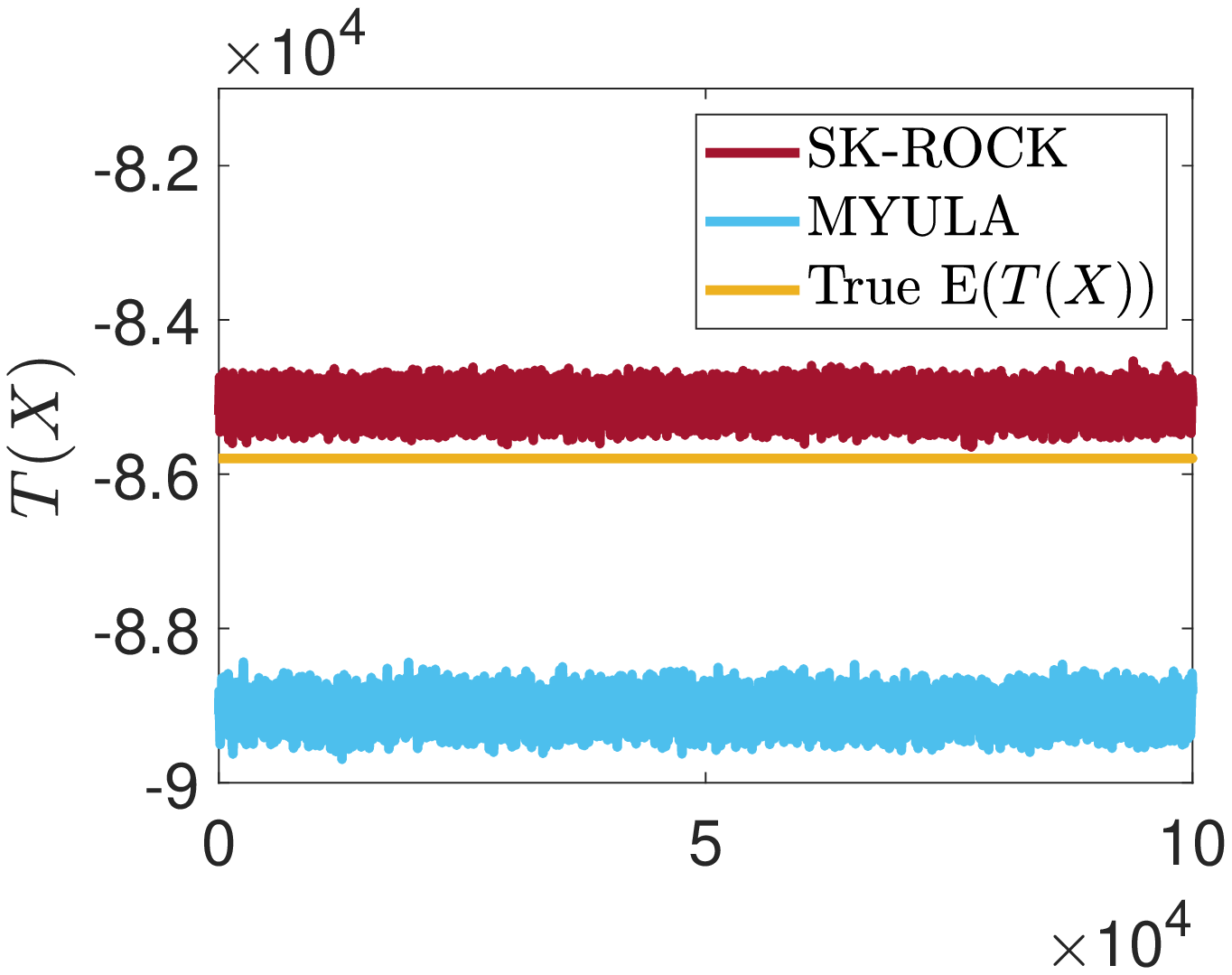}
	}
	\subfloat[ACF slow component]{
		\label{subfig:tomography_acf_slow_s10}
		\includegraphics[scale=.269]{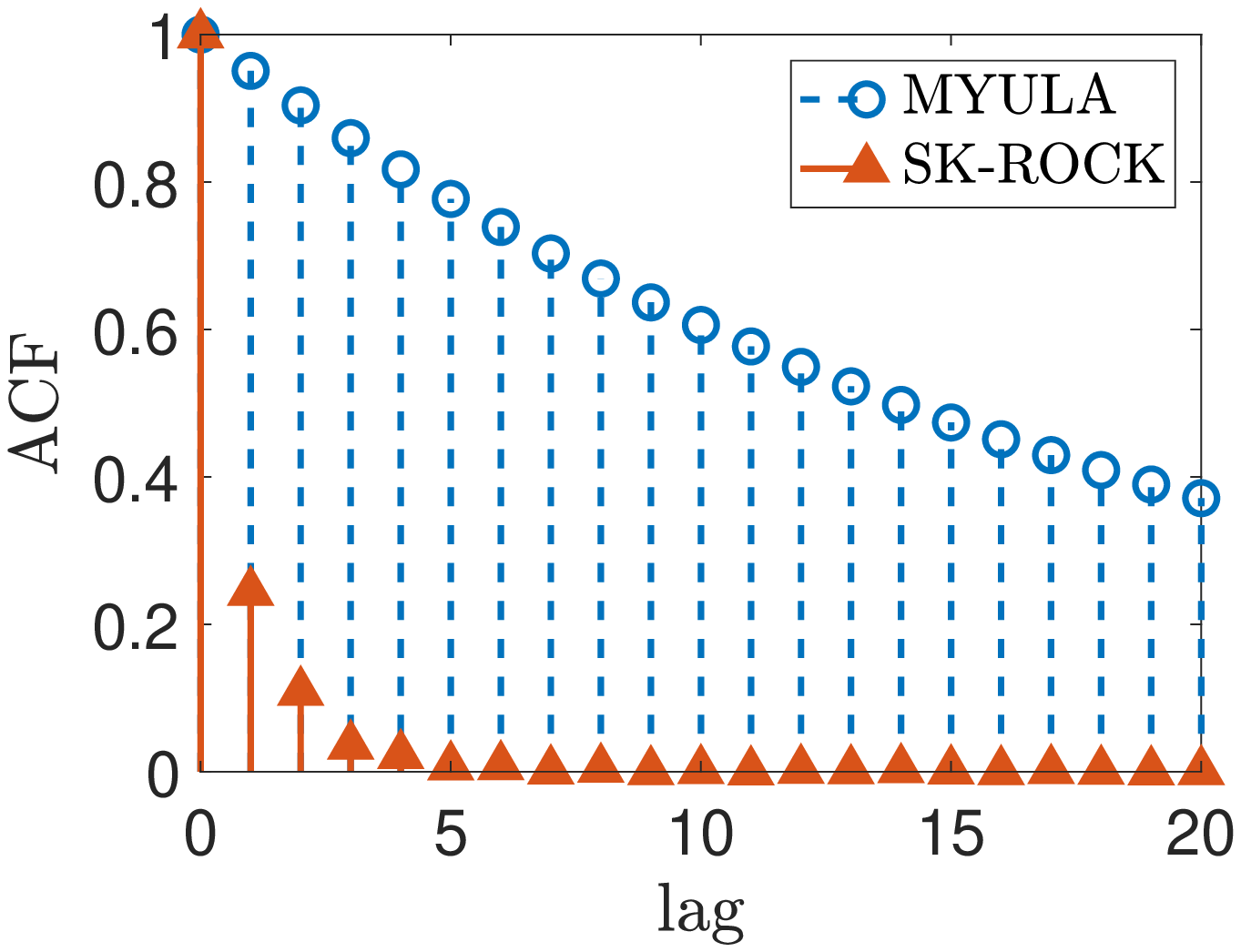}
	}
	\caption{\texttt{Tomography} experiment: {\normalfont(a)} Convergence to the {typical set of the } posterior distribution (\ref{eqn:tomography_posterior_dist}) {for  the first $3 \times 10^4$ MYULA samples and the first $3 \times 10^4 / s$ SK-ROCK ($s=10$)}. {\normalfont(b)} Last $10^5$ values of $\log \pi(x)$ from MYULA and SK-ROCK ($s=10$) chains. {\normalfont(c)} {Autocorrelation function for the slowest component}.}
	\label{fig:tomography_experiment_logPiTrace_chains_hist}
\end{figure}

These superior convergence properties can be clearly observed in Figure \ref{fig:tomography_experiment_logPiTrace_chains_hist}(c), which presents the autocorrelation plots for the slowest components of the chains. For completeness, Table \ref{tab:resultsTomographyExp} also reports the values obtained with SK-ROCK with $s=5$.

\begin{table}
	{\footnotesize
		\caption{\texttt{Tomography} experiment: Summary of the results after generating {$5 \times10^6$ samples with MYULA and $5 \times 10^6 / s$ samples} with SK-ROCK. Computing time $20$ hours per method.}  \label{tab:resultsTomographyExp}
		\begin{center}
			
			\begin{tabular}{|c|c|c|c|c|c|c|c|} \hline
				\bf Method & \bf Stepsize & \bf ESS & \bf ESS & \bf Speed-up & \bf Speed-up \\
				 & $\delta$ & \bf slow comp. & \bf fast comp. & \bf slow comp. & \bf fast comp.\\ \hline
				MYULA & $1.67 \times 10^{-5}$ & $1.31 \times 10^4$ & $1.64 \times 10^5$ & -  & -\\ 
				SK-ROCK ($s=5$)& $5.02 \times 10^{-4}$ & $5.31 \times 10^4$ & $2.56 \times 10^5$ & $4.05$  &$1.56$ \\ 
				SK-ROCK ($s=10$) & $2.30 \times 10^{-3}$ & $2.65 \times 10^5$ & $1.33 \times 10^5$ & $20.23$ & $0.81$  \\ \hline
			\end{tabular}
		\end{center}
	}
\end{table}

Finally, as in previous experiments, Figure \ref{fig:tomography_experiment_logPiTrace_chains_hist}(a) presents trace plots for the two chains during their burn-in stages;  again we can see  that SK-ROCK reaches the typical set of $x|y$ significantly faster than MYULA. Figure \ref{fig:tomography_experiment_logPiTrace_chains_hist}(b) shows the $\log\pi(x)$ trace of both methods in stationary regime, and similarly to the \textit{cameraman} and \textit{hyperspectral} experiments we have also included the entropy $\mathbb{E}(T(X))$ of the distribution calculated by  a very long run of the P-MALA algorithm, which targets (\ref{eqn:tomography_posterior_dist}) exactly. As can be seen clearly, SK-ROCK presents a lower bias than MYULA.

\section{Discussion and conclusion}
\label{sec:conclusions}
{This paper presented a new proximal MCMC method to perform Bayesian computation in inverse problems related to imaging sciences. Similarly to previous proximal MCMC methods, the methodology is derived from a Moreau-Yosida regularised overdamped Langevin diffusion process. However, instead of the conventional EM discrete approximation of the Langevin diffusion, the proposed method employs a significantly more advanced discrete approximation that has better convergence properties in problems that are ill-posed and ill-conditioned.} 

{The explicit EM approximation struggles in problems that are ill-posed or ill-conditioned because of the corresponding severe time-step restrictions. These same issues arise in the case of gradient descent and proximal gradient optimisation algorithms that also suffer from time-step restrictions. In optimisation, this has been successfully addressed by using accelerated proximal optimisation algorithms {\cite{AB17}}. The SK-ROCK approximation used in this paper achieves a similar acceleration quality. For Gaussian models, we prove rigorously the acceleration of the Markov chains in the 2-Wasserstein distance as a function of the {condition} number $\kappa$ when moderate accuracy is required. The superior behaviour of {our method} is further demonstrated with a range of numerical experiments, including non-blind image deconvolution, tomographic reconstruction, and hyperspectral  unmixing, with total-variation and $\ell_1$ priors. The generated Markov chains exhibit faster mixing, achieve larger effective sample sizes, and produce lower mean square estimation errors at equal computational budget. This allows, for example, to accurately estimate high order statistical moments and perform uncertainty quantification analyses in a more computationally efficient way.}

{Furthermore, this paper opens a number of interesting directions for future research. For example, as mentioned previously, the poor performance of EM approximations can also be mitigated by using auxiliary variable Gibbs splitting schemes \cite{Vono_TSP_2019} that have similarities with the ADMM optimisation algorithm, at the expense of additional estimation bias. Given that ADMM optimisation can be accelerated, it would be interesting to study sampling methods that combine SK-ROCK with splitting schemes to reduce that bias or achieve greater computational efficiency. Another important perspective is to theoretically analyse the non-asymptotic convergence properties of SK-ROCK for non-Gaussian log-concave models and derive bounds in total-variation and Wassertein metrics; this is highly technical and will require developing new analysis techniques. It would also be interesting to explore possible Metropolis-adjusted variants of the stochastic Runge-Kutta-Chebyshev methods discussed in this paper, and to investigate empirical Bayesian computation algorithms that combine SK-ROCK with stochastic gradient descent, which could be useful for estimating unknown model parameters such as regularisation parameters \cite{Vidal2019}.}

\appendix
\setcounter{table}{0}
\renewcommand{\thetable}{A\arabic{table}}

\section{Wasserstein distance - Gaussian process}
\label{app:w2analysis}
We begin computing the distribution $Q_n$ of the $n$ samples generated by the approximation (\ref{eqn:generalNumMethod}). We will work in the one dimensional case but the results  easily extend to higher dimensions, as can be seen later. First, we can notice that  {the solution of} (\ref{eqn:generalNumMethod}) can be expressed by the following recursive formula:
\begin{equation}\nonumber
X_n = (R_1(z))^n X_0 + \sqrt{2 \delta} \sum_{i=1}^n (R_1(z))^{n-i} (R_2(z)) \xi_i,
\end{equation}
where {$X_0$} is the initial condition of the problem. Computing expectations on both sides of the latter equation, we have:
 \begin{equation}\nonumber
\mathbb{E} (X_n) = (R_1(z))^n X_0.
\end{equation}
Then, we compute the variance as follows:
\begin{eqnarray}
\nonumber \mathbb{E} (X_n^2) - \mathbb{E} (X_n)^2 & = & 2 \delta \sum_{i=1}^n (R_1(z))^{2(n-i)} (R_2(z))^2 ,\\
\nonumber & = & 2 \delta (R_1(z))^{2n} (R_2(z))^2 \sum_{i=1}^n \frac{1}{(R_1(z))^{2i}} ,\\
\nonumber & = & 2 \delta (R_1(z))^{2n} (R_2(z))^2 \frac{1}{(R_1(z))^2} \left[ \frac{1 - \frac{1}{(R_1(z))^{2n}}}{1 - \frac{1}{(R_1(z))^2}}  \right] ,\\
\nonumber & = & 2 \delta (R_2(z))^2 \left[ \frac{(R_1(z))^{2n} -1}{(R_1(z))^2 -1} \right],
\end{eqnarray}
thus, the approximated distribution $Q_n$ of the {$n$-th sample} produced by the numerical scheme (\ref{eqn:generalNumMethod}) is defined, as follows:
\begin{equation}\nonumber
Q_n = \mathcal{N} \left( (R_1(z))^n {X_0}, 2 \delta (R_2(z))^2 \left[ \frac{(R_1(z))^{2n} -1}{(R_1(z))^2 -1} \right] \right).
\end{equation}
{We can now} compute the Wasserstein distance between the two univariate Gaussian distributions $P$ and $Q_n$:
\begin{equation}\nonumber
W_2(P ; Q_n )^2 = (R_1(z))^{2n} {X^{2}_0} + \left[ \sigma - \sqrt{2\delta} R_2(z) \left( \frac{1 - (R_1(z))^{2n}}{1 - (R_1(z))^2} \right)^{1/2} \right]^2.
\end{equation}
{As we mentioned at the beginning of this Appendix,} we can trivially extend the last result  for a $d$-dimensional Gaussian distribution i.e. let $P \sim N(0,\Sigma)$ where $\Sigma = \text{diag}(\sigma_1^2,...,\sigma_d^2)$ and $X_0=(x_0^1,...,x_0^d)^T$ and obtain the following expression for  the Wasserstein distance:
\begin{equation}\nonumber
W_2(P ; Q_n )^2 = \sum_{i=1}^d (R_1(z_i))^{2n} (x_0^i)^2 + \sum_{i=1}^d \left[ \sigma_i - \sqrt{2\delta} R_2(z_i) \left( \frac{1 - (R_1(z_i))^{2n}}{1 - (R_1(z_i))^2} \right)^{1/2} \right]^2,
\end{equation}
where $z_i = -\delta / \sigma_i^2$. This concludes the proof.

\section{Explicit bound for the Wasserstein distance}
\label{app:w2bounds}
We begin applying the triangle inequality to $W_2(P;Q_{n+1})^2$ as follows:
\begin{equation}\label{eqn:w2BoundTriangIneq}
W_2(P;Q_{n+1})^2 \leq W_2(P;\tilde{Q} )^2 + W_2(\tilde{Q} ;Q_{n+1})^2 ,
\end{equation}
where $\tilde{Q}$ is the unique invariant distribution to which (\ref{eqn:generalNumMethod}) converges when $ n \rightarrow \infty $ and it is defined as:
\[
\tilde{Q}=\mathcal{N} \left( 0, 2 \delta (R_2(z))^2 \left[ \frac{1}{1 - (R_1(z))^2} \right] \right),
\]
thus, we have that:
\begin{eqnarray}
\nonumber W_2(\tilde{Q} ;Q_{n+1})^2 & = & \sum_{i=1}^d R_1(z_i)^{2n+2} (x_0^i)^2 + \sum_{i=1}^d \left[ \left( \frac{2 \delta R_2(z_i)^2}{1 - R_1(z_i)^2} \right)^{1/2} \right. \\ \nonumber & & \left.  -  \sqrt{ 2 \delta} R_2(z_i) \left( \frac{1 - R_1(z_i)^{2n+2}}{1 - R_1(z_i)^2}\right)^{1/2} \right]^2 , \\ \label{eqn:proofBoundW2_01}
& = & \sum_{i=1}^d \left[ R_1(z_i)^{2n+2} (x_0^i)^2 + \frac{2 \delta R_2(z_i)^2}{1 - R_1(z_i)^2} \left( \sqrt{1} - \sqrt{1 - R_1(z_i)^{2n+2}} \right)^2 \right].
\end{eqnarray}
It is easy to prove the following property:
\begin{equation} \label{eqn:w2BoundPropertyProof}
\frac{1 - \sqrt{1 -x^{2n+2}}}{1 - \sqrt{1-x^{2n}}} x^2 \leq x^2 , 
\end{equation}
for $x \in (0,1)$. Thus, applying the latter in (\ref{eqn:proofBoundW2_01}) we have:
\begin{eqnarray}
\nonumber W_2(\tilde{Q} ;Q_{n+1})^2 & \leq & \sum_{i=1}^d R_1(z_i)^{2n+2} (x_0^i)^2 + \sum_{i=1}^d \frac{2 \delta R_2(z_i)^2}{1 - R_1(z_i)^2} \left( R_1(z_i)^2 \left[1 - \sqrt{1 - R_1(z_i)^{2n}} \right] \right)^2 ,\\ \nonumber
& \leq & \sum_{i=1}^d \left[ R_1(z_i)^{2n} (x_0^i)^2 + \frac{2 \delta R_2(z_i)^2}{1 - R_1(z_i)^2} \left( 1 - \sqrt{1 - R_1(z_i)^{2n}} \right)^2 \right] R_1(z_i)^2 ,\\ \nonumber & \leq &  \max_{1 \leq i \leq d} R_1(z_i)^2
 W_2(\tilde{Q} ;Q_{n})^2 .\end{eqnarray}
Thus, (\ref{eqn:w2BoundTriangIneq}) becomes:
\[
W_2(P;Q_{n+1})^2 \leq W_2(P;\tilde{Q} )^2 + \max_{1 \leq i \leq d} R_1(z_i)^2  W_2(\tilde{Q} ;Q_{n})^2 .
\]
Let:
\[
C = \max_{1 \leq i \leq d} R_1(z_i)^2,
\]
applying (\ref{eqn:w2BoundPropertyProof}) $n+1$ times, we finally have that:
\[
W_2(P;Q_{n+1})^2 \leq W_2(P;\tilde{Q} )^2 + C^{n+1} W_2(\tilde{Q} ;Q_{0})^2 ,
\]
concluding the proof.

As an attempt to minimise the bound found in the latter expression, we will try to accelerate the decay of the constant $C$ composed by $R_1(z)$ in the stochastic ROCK methods.  {This approach follows closely the approach in  \cite{EVV18}}. In particular,  in order to bound $R_1(z)$ by one, we need that $ \left| \omega_0 +\omega_1 z \right| \leq 1$, in other words we need that:
\[
-1 \leq \omega_0 - \omega_1  \frac{\delta}{\sigma_i^2} \leq 1  .
\]
Let $L:=1 / \sigma_{\min}^2$ and $\ell := 1 / \sigma_{\max}^2$, so we have that:
\[
-1 \leq \omega_0 - \omega_1  L \delta \leq \omega_0 - \omega_1  \ell \delta \leq 1,
\]
which it is the same as:
\[
-1 \leq \omega_1  \ell \delta - \omega_0 \leq  \omega_1  L \delta - \omega_0 \leq 1.
\]
Working with the first two members on the left-hand side of the latter inequality, we have that:
\begin{equation} \label{eq:safety}
\delta \geq \frac{\omega_0 - 1}{\ell \omega_1}.
\end{equation}
We choose the smallest $\delta$ to have an efficient algorithm i.e., $\delta = (\omega_0 -1) / \ell \omega_1$ and now working with the last two members on the right-hand side of the previous inequality, we have that:
\[
\kappa := \frac{L}{\ell} \leq \frac{\omega_0 +1}{\omega_0 -1} = 1 + \frac{2s^2}{\eta} ,
\]
where $\kappa$ is the condition number of our Gaussian problem. We choose the smallest $s$ to have an efficient algorithm and the latter expression determines the parameter $s$ as:
\begin{equation}
\label{eqn:bestNStagesSKROCK_App}
s = \left[ \sqrt{\frac{\eta}{2}(\kappa -1)} \; \right],
\end{equation}
where $[x]$ is the notation for the integer rounding of real numbers.


\end{document}